\documentclass{statsoc}
\usepackage[a4paper]{geometry}
\makeatletter
\def\@makefnmark{\hbox{\@textsuperscript{\normalfont\@thefnmark}}}
\renewcommand\@makefntext[1]%
    {\noindent\makebox[0pt][r]{\textsuperscript{\@thefnmark}\,}#1}
 \def\blfootnote{\gdef\@thefnmark{}\@footnotetext}
 \makeatother

 \usepackage{scrextend}
 \deffootnote{0em}{1.6em}{\thefootnotemark\ }

\usepackage{etoolbox}

\usepackage{soul}

\makeatletter
\patchcmd{\@makecaption}
  {\parbox}
  {\advance\@tempdima-\fontdimen2} 
  {}{}
\makeatother

 \usepackage[bottom]{footmisc}%
 \usepackage{amsmath}
\usepackage{graphicx}  
\usepackage{float}  
\usepackage{multirow}  
\usepackage{subfigure}  
\usepackage[utf8]{inputenc} 
\usepackage[T1]{fontenc}    
\usepackage{hyperref}       
\usepackage{url}            
\usepackage{booktabs}       
\usepackage{amsfonts}       
\usepackage{nicefrac}       
\usepackage{microtype}      
\usepackage{lipsum}		
 \usepackage{appendix}
 \usepackage{mathrsfs}
 \usepackage{amssymb}
 \usepackage{color}
  \usepackage{enumitem}
  \usepackage{algorithm}  
\usepackage{algorithmicx}  
\usepackage[noend]{algpseudocode} 
  \usepackage{epsfig}
  \usepackage{comment}
\usepackage{bm}
\usepackage{adjustbox}

 \usepackage{setspace} 
\DeclareMathOperator*{\argmin}{arg\,min}
\DeclareMathOperator*{\argmax}{arg\,max}

\def\m{\mathcal}
\def\dd{{\rm d}}
\def\mb{\mathbb}

\def\det{\mbox{\rm det}}

\newcommand{\field}[1]{\mathbb{#1}}
\newcommand{\R}{\field{R}}

\newcommand{\cO}{\mathcal{O}}

\newcommand{\Z}{\bm{Z}}
\newcommand{\X}{\bm{X}}
\newcommand{\E}{\field{E}}
\newcommand{\di}{\mathrm{d}}
\newcommand{\bbeta}{\bm{\beta}}
\newcommand{\bmu}{\bm{\mu}}

\def\II{I\negthinspace I}

\newtheorem{theorem}{Theorem}
\newtheorem{assumption}{Assumption}
\newtheorem{remark}{Remark}
\newtheorem{corollary}{Corollary}
\newtheorem{lemma}{Lemma}

\def\cod{\stackrel{\cal D}{\longrightarrow}}

\def\lf{\lfloor}
\def\rf{\rfloor}

\def\elbo{\operatorname{ELBO}}
\def\bic{{\operatorname{BIC}}}

\def\qsm{q_{S_\cM}}

\def\P{\mathbb{P}}

\def\var{{\mbox{var}}}
\def\tr{{\mbox{tr}}}
\def\diag{{\operatorname{diag}}}

\def\tmle{\widehat\theta^{\tiny \rm mle}}
\def\deltaeq{\overset{\Delta}{=}}

\def\tmstar{\theta^*_{\cM}}
\def\tb{\widetilde{b}}
\def\I{\text{I}}
\def\II{\text{II}}
\def\III{\text{III}}

\def\verts{\,||\,}
\def\cM{\mathcal{M}}
\def\qt#1{q_{\theta_{#1}}}
\def\qs#1{q_{S_{#1}}}
\newcommand{\Exp}{\E}

\title[Bayesian Model Selection via Mean-Field Variational Approximation]{Bayesian Model Selection via Mean-Field Variational \\ Approximation}
\author[Yangfan Zhang and Yun Yang]{Yangfan Zhang and Yun Yang}
\address{University of Illinois Urbana-Champaign}

\begin{document}

\begin{abstract}
This article considers Bayesian model selection via mean-field (MF) variational approximation. Towards this goal, we study the non-asymptotic properties of MF inference under the Bayesian framework that allows latent variables and model mis-specification. Concretely, we show a Bernstein von-Mises (BvM) theorem for the variational distribution from MF under possible model mis-specification, which implies the distributional convergence of MF variational approximation to a normal distribution centering at the maximal likelihood estimator (within the specified model).
Motivated by the BvM theorem, we propose a model selection criterion using the evidence lower bound (ELBO),
and demonstrate that the model selected by ELBO tends to asymptotically agree with the one selected by the commonly used Bayesian information criterion (BIC) as sample size tends to infinity. Comparing to BIC, ELBO tends to incur smaller approximation error to the log-marginal likelihood (a.k.a. model evidence) due to a better dimension dependence and full incorporation of the prior information.
Moreover, we show the geometric convergence of the coordinate ascent variational inference (CAVI) algorithm
under the parametric model framework, which provides a practical guidance on how many iterations one typically needs to run when approximating the ELBO. These findings demonstrate that variational inference is capable of providing a computationally efficient alternative to conventional approaches in tasks beyond obtaining point estimates, which is also empirically demonstrated by our extensive numerical experiments.
\end{abstract}


\section{Introduction}
Variational inference \citep[VI,][]{jordan1999introduction,bishop2006pattern} is an effective computational method for approximating complicated posterior distributions arising in Bayesian statistics. An alternative commonly adopted approach is the Markov Chain Monte Carlo (MCMC) method~\citep{hastings1970monte,gelfand1990sampling,hammersley2013monte} based on sampling, where a Markov chain is carefully constructed so that its limiting distribution matches the target distribution. Despite its popularity, MCMC is known to suffer from a number of drawbacks, including low sampling efficiency due to sample correlation and a lack of computational scalability to massive datasets. In comparison, VI turns the integration or sampling problem into an optimization problem, and can be orders of magnitude faster than MCMC for achieving the same approximation accuracy. More precisely, the target distribution in VI is approximated by a closest member in a family of tractable distributions via minimizing the Kullback-Leibler (KL) divergence, which is also equivalent to maximizing a lower bound to the logarithm of the marginal density of data, called evidence lower bound (ELBO). We refer the interested readers to~\citet{blei2017variational} for a comprehensive review on the history of VI development.


Among various approximating schemes, the mean-field (MF) approximation, which uses the approximating family consisting of all fully factorized density functions over (blocks of) the target random variables, is the most widely used and representative instance of VI that is conceptually simple yet practically powerful.
The computation of MF approximation can be realized using the coordinate ascent variational inference (CAVI) algorithm~\citep{bishop2006pattern,blei2017variational}, which iteratively optimizes each (block of) components in the factorization while keeping others fixed at their present values (see Section~\ref{sec::cavi} for more details). CAVI resembles the classical EM algorithm~\citep{dempster1977maximum} by viewing the target random variables as the unobserved latent variables. In this paper, we primarily focus on MF approximation, although our development can be analogously extended to other approximation schemes, such as Gaussian approximation and hybrid schemes that further restrict components in MF to be within certain exponential families. 

Despite the wide applications of variational inference over the past decades, it is until recent years that some general theory, trying to explain from a frequentist perspective why variational inference works so well, has been developed. Some earlier threads of theoretical research are mostly conducted in a case-by-case manner, by either explicitly analyzing the fixed point equation of the variational optimization problem, or directly analyzing the iterative algorithm for solving the optimization problem. Examples of such include Bayesian linear models~\citep{hall2011theory,hall2011asymptotic,ormerod2012gaussian}, Gaussian mixture models~\citep{titterington2006convergence,westling2015establishing}, and stochastic block models~\citep{bickel2013asymptotic,zhang2020theoretical}. 
It is a well-known fact that many VI schemes, such as MF approximation, fail to capture the dependence structure and tend to underestimate the estimation uncertainty reflected in the Bayesian posteriors~\citep{wang2005inadequacy}. This observation is first theoretically illustrated by~\citet{wang2019}, where under a local asymptotic normality (LAN) assumption, a Bernstein von-Mises type theorem \citep[i.e.~asymptotic normal approximation,][]{van2000}
is proven for the variational (approximated) posterior. 
However, the crucial LAN assumption used in~\cite{wang2019} implicitly assumes the estimation consistency for the model parameter, which still requires a case-by-case verification. In addition, it is not clear that how the asymptotic mean and covariance matrix implied by their LAN assumption relate to characteristics of the statistical model in a general sense. \citet{wang2019variational} further generalize the results in~\cite{wang2019} from well-specified models to mis-specified models.

\begin{figure}[htb]
    \centering
    \includegraphics[width=0.9\textwidth]{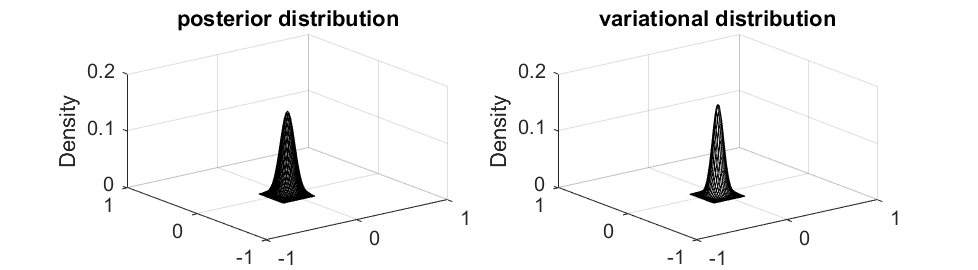}
    \caption{Plots of a two-dimensional posterior density and its mean-field approximation. }
    \label{fig:VIplot}
\end{figure}

Since VI is generally incapable of accurately approximating the target posterior distributions, another line of research focuses on the estimation consistency of point estimators obtained from variational posteriors (e.g.~using the expectation) under general settings. For example, a number of recent works~\citep{alquier2020concentration,pati2018statistical,yang2020alpha,zhang2020convergence} provide general conditions under which VI leads to consistent parameter estimation; moreover, they derive convergence rates of the point estimators that are often minimax-optimal (up to logarithmic terms) in both regular parametric and infinite-dimensional non-parametric models. 
In addition, some works~\citep{alquier2020concentration,alquier2016properties,cherief2018consistency} derive risk bounds for VI under model mis-specifided settings; while \cite{yang2020alpha} also studies the risk bounds for VI with fractional posteriors. In a nutshell, this point estimation consistency (first order information) can be attributed to the heavy penalty on the tails in the KL divergence that forces the variational distribution to concentrate around the true parameter at the optimal rate; however, the local shape of the obtained variational posteriors around the true parameter (second order information) can be far away from that of the true posterior (see Fig.~\ref{fig:VIplot} for an illustration).

Through a more refined analysis,~\cite{wei2019} show that under some mildly stronger smoothness conditions without assuming consistency, the discrepancy between a point estimator from MF approximation and the maximum likelihood estimator (MLE) is of higher-order compared to the root-$n$ estimation error of MLE for regular parametric models. As a consequence, there is essentially no loss of efficiency in using a VI point estimator for parameter estimation, in terms of asymptotically attaining the Cram\'{e}r-Rao lower bound. In addition, they show that MF variational posterior is close to a multivariate normal distribution centered at MLE with a diagonal precision matrix whose diagonal elements are equal to corresponding elements in the Fisher information at the true parameter. They further propose a consistent variational weighted likelihood bootstrap method for uncertainty quantification for the MF approximation. 

The overarching goal of the current paper is to carry out further methodological and theoretical investigations complimenting existing findings by looking into the model selection and algorithmic convergence aspects of MF variational approximation. While there are some existing theoretical results on model selection with variational inference, they either show only an oracle inequality for the selected model \citep{cherief2019consistency}, implying that the estimation performance of the selected model is not much worse than that of the true model, or use an additional aggregation step to combine multiple variational models to achieve an optimal convergence rate \citep{ohn2021adaptive}. It remains an open problem to study whether the model selected based on ELBO maximization is consistent; that is, whether the probability of choosing the right model tends to one as sample size tends to infinity. In this paper, we propose ELBO as an alternative criterion for model selection, and show that it is asymptotically equivalent to the Bayesian information criterion \citep[BIC,][]{bic1978}. It is well-known~\citep{yang2005can} that BIC, derived as an asymptotic approximation to the log-marginal likelihood (a.k.a.~model evidence), is consistent in selecting the true model; while another commonly used Akaike information criterion~\citep[AIC,][]{aic1974}, equivalent to Mallows's $C_p$ statistic in regression analysis, is minimax-rate optimal for prediction but tends to over-estimate the model size. In this work, we find that ELBO generally leads to a better approximation to the model evidence than BIC due to the full incorporation of prior information and a better dimension dependence in the approximation error. On the other hand, our numerical studies suggest that the proposed ELBO criterion tends to be less conservative than BIC in the presence of weak signals, and can achieve comparable predictive performance as AIC with a much more parsimonious selected model. 
Furthermore, we study the algorithmic convergence of the CAVI algorithm under regular parametric models by providing generic conditions on the initialization and step size under which CAVI exhibits geometric convergence towards the exact value up to the statistical accuracy of the problem.
Specifically, we characterize the algorithmic convergence rate under two commonly adopted updating schemes; and our result provides theoretic guidance on how many iterations one typically needs to run when approximating the ELBO for model selection, so that the numerically selected model coincides with the theoretical one.

The rest of the paper is organized as follows. In Section \ref{sec::backgrd}, we review some background on the mean-field variational inference, present some new theoretical results on the mis-specified Bernstein-von Mises theorem, and describe in details the methodology of using MF approximation for model selection and the related computational aspects. Section \ref{sec::theory} contains the main theoretical findings of the paper. Section \ref{sec::numeric} includes the numerical results. We conclude the paper with a discussion in Section \ref{sec::conclude} and leave the description of motivating examples, their theoretical consequences, high-dimensional extensions and technical proofs to the Appendices.

\medskip
\noindent {\bf Notation.} We use lower-case letters (e.g.~$\pi$, $p$, $q$,\ldots) to denote densities, and capital letters (e.g.~$\Pi$, $P$, $Q$,\ldots) for the associated probability measures. We use $D(P\verts Q)$ to denote the KL-divergence between two distributions $P$ and $Q$, and $H(P,\,Q)$ for the Hellinger distance.


\section{Model Selection via Mean-Field Approximation and its Computation}
\label{sec::backgrd}
In this section, we first set up the modeling framework and review mean-field (MF) variational approximation for Bayesian inference; and then present some new theoretical results, including concentration and distributional convergence under possible model mis-specification. Motivated by the theory, we propose a new criterion based on the evidence lower bound (ELBO) as an alternative to the widely used BIC for model selection. After that, we introduce several common variants of coordinate ascent variational inference (CAVI) algorithms that implement MF for model selection. In Appendix~\ref{sec::mvteg}, we provide some concrete motivating examples and derive the closed form of updating formulas for computing ELBO.

\subsection{Variational Inference and Mean-Field Approximation}\label{sec:VI}
Let $X^n=(X_1,\ldots,X_n)$ denote i.i.d.~random observations from some unknown underlying data generating distribution $\P_0$ to be estimated. Consider a parametric family $\{\P_\theta:\theta\in\Theta\}$ that does not necessarily contain $\P_0$ (i.e.,~we allow model mis-specification).
In a general Bayesian framework, the goal is to approximate the posterior density $\pi_n$ of parameter $\theta\in\Theta$ given data $X^n$, which is obtained by combining a prior density $\pi(\theta)$ and data likelihood $p(X^n\mid \theta)$,
$$
\pi_n(\theta):\,=p(\theta\mid X^n)=\frac{p(X^n\mid \theta)\,\pi(\theta)}{p(X^n)},\quad\mbox{with}\quad  p(X^n) = \int_{\Theta}p(X^n\mid \theta)\,\pi(\theta)\, \dd\theta
$$
denoting the normalization constant.
In many problems such as the Gaussian mixture modeling, model augmentation with latent variables can greatly simplify the likelihood function evaluation; thereby facilitating posterior calculation. For concreteness, we consider $n$ local latent variables $S^n=(S_1,\ldots,S_n)$ where the $i$th latent variable $S_i\in\mathcal S$ is tied with $X_i$ for $i=1,2,\ldots,n$. Under this setting, the marginal likelihood function of data $X^n$ can be recovered by integrating the conditional density function $p(X^n\mid S^n,\,\theta)$ with respect to the latent variable distribution $p(S^n\mid \theta)$, i.e.
$$
p(X^n\mid \theta)=\int_{\mathcal S^n} p(X^n\mid \theta,\,s^n)\,p(s^n\mid \theta)\,\dd s^n.
$$
In the case where latent variables are of discrete type, we replace the integration above with a summation over $\mathcal S^n$.
We consider the common setting where the observation-latent variable pairs $\{(X_i,\,S_i)\}_{i=1}^n$ are conditionally i.i.d.~given $\theta$, that is,
$$
p(X^n \mid  S^n,\,\theta)=\prod_{i=1}^n p (X_i \mid  S_i,\,\theta) \quad \text { and } \quad p (S^n\mid \theta)=\prod_{i=1}^n p(S_i \mid \theta).
$$
For such a latent variable model, one is typically interested in making inference using the following joint posterior density over parameter $\theta$ and latent variables $S^n$,
$$
p(\theta,\,S^n\mid X^n)=\frac{p(X^n \mid  S^n,\,\theta)\,p (S^n\mid \theta)\,\pi(\theta)}{p(X^n)}.
$$
Unfortunately, in many problems the normalizing constant $p(X^n)$ involving a multivariate integration is analytically intractable and difficult to numerically approximate. Variational inference (VI) instead searches for a closest distribution $\widehat q(\theta,\,S^n)$ over $\Theta\times \mathcal S^n$ from some computationally friendly variational family, denoted by $\Gamma$, to approximate the joint posterior distribution by solving the following optimization problem,
\begin{equation}
\widehat q=\underset{q \in \Gamma}{\operatorname{argmin}}\  D\big(q(\cdot,\,\cdot) \, \big\|\, p(\cdot,\,\cdot \mid  X^n)\big).
 \label{mpvi}
\end{equation}
The following equivalent description of the objective functional above leads to an alternative way to interpret VI,
\begin{align}
&D\big(q(\cdot,\,\cdot) \, \big\|\, p(\cdot,\,\cdot \mid  X^n)\big)= \int_{\Theta\times \m S^n} \log \frac{q(\theta,s^n)}{p(\theta,s^n\mid X^n)} \, q(\theta,s^n) \,\di \theta\,\di s^n \label{eqn:ELBO_dec}\\
=&\int_{\Theta\times \m S^n} \big[\log p(X^n) + \log q(\theta,s^n)-\log p(X^n, \theta,s^n) \big]\, q(\theta,s^n) \, \di \theta\, \di s^n :\,=\log p(X^n)-L(q),\notag
\end{align}
where $\displaystyle L(q)=\int_{\Theta\times \m S^n}\big[\log p(X^n, \theta,s^n) -\log q(\theta,s^n)\big]\, q(\theta,s^n) \, \di \theta\, \di s^n$ is called evidence lower bound (ELBO) since it bounds the evidence $\log p(X^n)$ from below, due to the non-positivity of KL divergence. The same identity also illustrates that VI circumvents the need of calculating the unknown normalizing constant $p(X^n)$ since it only contributes to the objective functional as a constant that does not change the optimum.

According to identity~\eqref{eqn:ELBO_dec}, ELBO $L(q)$ equals to the evidence (i.e.~$\log p(X^n)$) if and only if $q$ and $p(\cdot,\cdot\mid X^n)$ coincide. As a direct consequence, the following two statistical tasks are equivalent: 1.~approximating the joint posterior for conducting statistical inference on $(\theta,\,S^n)$; 2.~approximating the evidence $\log p(X^n)$ for evaluating model goodness of fit or performing model selection. Computationally, we can alternatively maximize the analytically tractable ELBO functional $L$ to find the best approximation to the joint posterior within variational family $\Gamma$. In this way, we have turned the integration problem into an optimization problem. Although VI is not guaranteed to generate exact samples as MCMC does, the computational efficiency can be considerably improved, as some smart choices of the variational family make the optimization problem numerically solvable and simple. Moreover, the computational speed can be further boosted by taking advantage of modern optimization techniques such as stochastic approximation and distributed computing. 

In practice, a popular choice of $\Gamma$ is the mean-field family that contains all fully factorized densities with the form $q(\theta,s^n)=\prod_{j=1}^d q_{\theta_j}(\theta_j) \cdot \prod_{i=1}^n q_{S_i}(s_i)$ for all $\theta\in\Theta$ and $s^n=(s_1,\ldots,s_n)\in \m S^n$. 
As a representative illustrating example, the (closest) MF approximation to the multivariate normal distribution $N(\mu,\Theta^{-1})$ (with $\Theta$ being the precision matrix) is $N(\mu,[\diag(\Theta)]^{-1})$, where $\diag(A)$ denotes the diagonal matrix collecting all diagonal elements from a matrix $A$. In other words, the MF approximation keeps all the first-order information, while throwing away second-order interactions encoding the dependence structure, as we have already seen from Fig.~\ref{fig:VIplot}. A common variant relaxing the fully factorized MF approximation is the block MF approximation of the form $q(\theta,s^n)=q_\theta(\theta)\,q_{S^n}(s^n)$, where the dependence within components of $\theta$ or $S^n$ is preserved while that between the two blocks $\theta$ and $S^n$ is neglected.
For either the full or the block MF approximation, the corresponding optimization problem can be efficiently solved by a generic coordinate ascent~\citep{wright2015coordinate}, whose specializations to MF will be introduced with more details in Section~\ref{sec::cavi}.

To interpret and understand different sources of errors due to MF approximation $q(\theta,s^n)=q_{\theta}(\theta) \,q_{S^n}(s^n)$, we may further decompose ELBO into the following three terms~\citep{yang2020alpha},
\begin{align}
L(q)=&\int_{\Theta\times\m S^n} \Big[\log \big\{ p(s^n,\,X^n\mid \theta)\,\pi(\theta)\big\}-\log \big\{q_\theta(\theta)\,q_{S^n}(s^n)\big\}\Big] \, q_\theta(\theta)\,q_{S^n}(s^n)\,\di\theta\,\di s^n \notag\\
=&\ \ \underbrace{\int_\Theta \log p(X^n\mid \theta)\, q_\theta(\theta)\, \di\theta - D(q_\theta\verts \pi)}_{:\,= L_\theta(q_\theta)} \ \ -\ \  \Delta_J,\label{Eqn:ELBO_LV}
\end{align}
where the first term is an integrated marginal log-likelihood function (w.r.t.~the variational density $q_\theta$), the second term is the negative KL-divergence between $q_\theta$ and the prior, and the last term 
$$
\Delta_J=\int_\Theta\left[\log p(X^n\mid \theta) -\int_{\m S^n}\log\frac{p(X^n,\,s^n\mid \theta)}{q_{S^n}(s^n)}\,q_{S^n}(s^n)\,\di s^n \right]\,q_\theta \,\di\theta\geq 0
$$
can be interpreted as a non-negative Jensen gap introduced by approximating the marginal log-likelihood with a lower bound from Jensen's inequality. In particular, $L_\theta(q_\theta)$, which collects the first two terms in~\eqref{Eqn:ELBO_LV}, constitutes the ELBO associated with the objective functional $D\big(q(\cdot)\,\big|\big|\,\pi_n(\cdot)\big)$ in the variational inference of approximating the (marginal) posterior $p(\theta\mid X^n)$ with variational density $q_\theta(\theta)$; while the extra gap $\Delta_J$ is due to the presence of latent variables where $p(s^n\,|\,X^n,\theta)$ is approximated by a single distribution $q_{S^n}(s^n)$ for all $\theta\in\Theta$.

Decomposition~\eqref{Eqn:ELBO_LV} also reveals an interesting connection between the optimization of ELBO and the traditional regularized estimation. For example, when there is no latent variable, the Jensen gap term $\Delta_J$ vanishes; and maximizing the ELBO functional~\eqref{Eqn:ELBO_LV} over all density functions $q_\theta$ over $\Theta$ becomes finding a KL-divergence-regularized estimator over $\Gamma$: the first term in~\eqref{Eqn:ELBO_LV} reflects model goodness of fit to the data and encourages $q_\theta$ to assign all its mass towards the maximizer of log-likelihood function, i.e.~the maximum likelihood estimator; while the second regularization term $D(q_\theta\verts \pi)$ avoids measure collapse as it diverges to infinity as $q_\theta$ becomes close to a point mass measure.


\subsection{Concentration and Distributional Convergence}
\label{sec:concentration}

This subsection presents results about large-sample properties of posterior distributions and their MF variational approximations in terms of concentration towards certain point in parameter space $\Theta$ and convergence towards some distribution over $\Theta$, under the frequentist perspective assuming $\{X_i\}_{i=1}^n$ to be i.i.d.~from the data generating distribution $\mb P_{0}$.
These results are direct generalization of \cite{wei2019} from well-specified parametric models to mis-specified models. They also provide the cornerstone for showing the consistency of model selection based on ELBO in Section~\ref{sec::thmmodelselection}. 

For a well-specified model where $\mb P_0 = \mb P_{\theta^\ast}$ for some true parameter $\theta^\ast$ in parameter space $\Theta$, it is known that the posterior distribution tends to contract towards $\theta^\ast$ in view of the BvM theorem~\citep{van2000}. For posterior concentration beyond parametric models, refer to~\cite{shen2001rates,ghosal2000convergence,ghosal2007convergence}. However, in the context of model selection, we also need to consider candidate models that may not contain $\P_0$. Posterior contraction for mis-specified models is formally studied in \cite{kle2012}.
More precisely, if we denote
\begin{align}\label{eq:best_theta}
    \theta^*_{\cM}=\operatorname{argmin}_{\theta\in\Theta_{\cM}}D (\P_0\,||\,\P_{\theta})
\end{align}
as the parameter associated with the ``projection'' (relative to the KL divergence) of $\mb P_0$ to a generic (possibly mis-specified) model $\cM$ with parameter space $\Theta_{\cM}$. The BvM theorem under mis-specification \citep{kle2012} states that the posterior distribution tends to be close to a normal distribution centering at MLE $\tmle=\argmax_{\theta\in\theta_{\cM}} p(X^n\,|\,\theta)$ of the assumed model $\cM$, whose covariance matrix is related as usual to the Fisher information matrix; and the MLE $\tmle$ itself is asymptotically normal with asymptotic mean $\theta^*_{\cM}$ and a sandwiched-form covariance matrix.
Since we will fix the model $\cM$ in the following analysis, we will omit the $\cM$ in the subscripts in the rest of this subsection. For example, $\theta^*_{\cM}$ will be simply denoted as $\theta^*$, parameter space $\Theta_{\cM}$ as $\Theta$, and prior density $\pi_{\cM}$ as $\pi$. 

To begin with, we make some common regularity assumptions on the prior and log-likelihood function following \cite{wei2019}, with some appropriate generalization to the mis-specified setting. 

\begin{assumption}[Thickness of prior]\label{ass::prior}
The prior satisfies $\pi(\theta^*)>0$, and is continuously differentiable in a neighborhood of $\theta^*$, with growth controlled as
    $$
\left|\log \pi(\theta)-\log \pi\left(\theta^{*}\right)\right| \leq C\big(1+\left\|\theta-\theta^*\right\|^L\big), 
$$
for any $\theta$ and some positive constants $(C,L)$.
\end{assumption}

\begin{assumption}[Regularity of marginal likelihood]\label{ass::l} Let $\ell(\theta;x)=\log p(x\,|\,\theta)$ be the log-likelihood function at a single data point $x$, then:
\begin{enumerate}
\item $\ell(\theta;x)$ has continuous mixed derivatives up to order three with respect to components of $\theta$, and the mixed derivatives have finite fourth moments in a neighborhood of $\theta^*$ under $\P_0$. Moreover, there exists a measurable function $Z=Z(x)$ such that 
$$
\left|\frac{\partial^{3} \ell(\theta ; x)}{\partial \theta_{i} \partial \theta_{j} \partial \theta_{k}}\right| \leq Z(x)\big(1+\left\|\theta-\theta^*\right\|^{L}\big),\quad \text{for all }i,j,k\in[d],
$$
and $Z$ satisfies $\mb E_{\P_0}[e^{sZ(X)}]<\infty$ for some $s>0$;
\item The information matrix $V(\theta^*):\,=\E_{\P_0}[-\nabla^2 \ell(\theta^*,X)]$ is positive definite, and the covariance matrix 
$\E_{\P_0}\left[\nabla \ell(\theta^{*} ; X) \nabla \ell(\theta^{*} ; X)^T\right]$ of the score vector $\nabla \ell(\theta^{*} ; X)$ is invertible;
\item The Radon-Nikodym derivative $\frac{\dd\P_{\theta}}{\dd \P_{\theta^*}}$ satisfies $\mb E_{\P_0}\big[\frac{\dd\P_{\theta}}{\dd \P_{\theta^*}}\big]<\infty$ in a neighborhood of $\theta^*$. Moreover, for any $\epsilon>0$ there exists a sequence of test functions $\{\phi_n:\,n\geq 1\}$ such that
$$
\E_{\P_0^n}[\phi_n(X^n)]\le\exp(-nc) \quad \mbox{and} \quad \sup_{\|\theta-\theta^*\|\ge \epsilon}\E_{\P_0^n}\left[\frac{\P(X^n\mid\theta)}{\P(X^n\mid\theta^*)}\,\big(1-\phi_n(X^n)\big)\right]\le \exp(-nc_\epsilon),
$$
for some constants $c$ and $c_\epsilon$.
\end{enumerate}
\end{assumption}

\begin{assumption}[Regularity of conditional likelihood of latent variable]\label{ass::ls}
 The conditional log-likelihood function of the latent variable $\ell_S(\theta,X;s)=\log p(s\mid\theta,X)$ has continuous mixed derivatives with respect to components of $\theta$ up to order three, and the mixed derivatives have finite fourth moments in a neighborhood of $\theta^*$ under $\P_0$. Moreover, there exists a measurable function $Z=Z(x,s)$ with bounded second moment under $\P_0$ such that
$$
\left|\frac{\partial^{3} \ell(\theta,x; s)}{\partial \theta_{i} \partial \theta_{j} \partial \theta_{k}}\right| \leq Z(x,s)\,\big(1+\left\|\theta-\theta^*\right\|^{L}\big),\quad \text{for all }i,j,k\in[d].
$$
\end{assumption}
Assumptions~\ref{ass::prior}, \ref{ass::l}(a), \ref{ass::l}(b), and \ref{ass::ls} are some standard regularity conditions made on the prior and log-likelihood functions for proving the BvM theorem (for well-specified models). Assumption \ref{ass::l}(c) includes some additional regularity conditions and testability assumptions inherited from \cite{kle2012} to address mis-specifed models.
We denote the latent variable Fisher information matrix \citep[also called missing data Fisher information matrix in the missing data/EM algorithm literature, e.g.][]{dempster1977maximum}, $\E_{\P_0}[-\nabla^2 \ell_S(\theta^*,X)]$ of $\ell_S$ as $V_s(\theta^*)$ and denote $V_c(\theta^*)=V(\theta^*)+V_s(\theta^*)$ as the complete data information matrix at $\theta^\ast$. In this work, we only consider non-singular models by assuming the non-singularity of the Fisher information (Assumption~\ref{ass::l}(b)). The problem of model selection involving possibly singular models \citep[e.g., mixture models,][]{ho2019singularity} is not covered by our theory and will be left to future research. As a result, in Section 4.2 for numerical study on Gaussian mixture model, we only consider a well-specified model. However, we note that our assumptions are satisfied for under-specified Gaussian mixture models, as well as for other commonly used models such as the linear and generalized linear models. In these cases, the Fisher information matrix remains non-singular even when the model is mis-specified, as discussed in Section~\ref{sec::sim_glm}.

Theorem 3.1 in \cite{kle2012} states that the posterior distribution of a mis-specified model tends to be close to a normal distribution. In this paper, we find that the MF variational distribution also contracts to a normal distribution under model mis-specification. Following some recently developed techniques, we first include the following two results on the consistency and concentration of the marginal posterior of $\theta$ and its MF approximation.
The first result is a non-asymptotic version of Theorem 3.1 in \cite{kle2012} on the contraction of posterior under model mis-specification, which is inherited from Theorem 5.1 in \cite{ghosal2000convergence}.

\begin{lemma}\label{lem::post_subg}
Under Assumption \ref{ass::prior} to \ref{ass::ls}, for any $K\ge 1$, it holds with probability at least $1-CK^{-2}$ that the marginal posterior $\Pi_n$ of $\theta$ satisfies
\begin{align}\label{eqn:sub_Ga}
\Pi_n\big(\|\theta-\theta^{*}\| \geq C \varepsilon\big) \leq e^{-C n \varepsilon^{2}}, \quad \text { for all } \varepsilon \geq K\varepsilon_{n},
\end{align}
with $\epsilon_n=\frac{C\log n}{\sqrt{n}}$ for some constant $C$ sufficiently large.
\end{lemma}
Using a similar variational argument as the proof of Lemma 2 in \cite{wei2019}, the sub-Gaussian type concentration property of the posterior $p(\theta\,|\,X^n)$ is inherited by its MF approximation $\widehat Q_\theta$, as described by the following lemma.
\begin{lemma}\label{lem::VI_subg}
Suppose the posterior satisfies the sub-Gaussian tail property as displayed in Lemma~\ref{lem::post_subg}. Then under the same notation as Lemma~\ref{lem::post_subg}, there exist constants $(C_1,C_2,C_3)$ such that for any $K\ge1$, it holds with probability at least $1-C_1K^{-2}$ that the MF approximation $\widehat{Q}_\theta$ satisfies
$$
\widehat Q_\theta\big(\left\|\theta-\theta^*\right\| \geq C_{2} \varepsilon\big) \leq e^{-C_3 n \varepsilon^{2}}, \quad \text { for all } \varepsilon \geq K \varepsilon_{n}.
$$
\end{lemma}
Intuitively, the concentration of variational distribution $\widehat Q_\theta$ is a result of the strong penalty on the tail difference incurred by the log density ratio $\log \frac{q}{p}$ in the KL divergence objective~\eqref{mpvi}, which forces $\widehat Q_\theta$ to concentrate around the same region where the posterior $p(\theta\,|\,X^n)$ assigns most of its mass.
Lemma~\ref{lem::VI_subg} further leads to the following theorem on the distributional convergence of $\widehat Q_\theta$, which generalizes Theorem 1 in \cite{wei2019} from well-specified models to mis-specified models. Note that the counterpart of Theorem 1 for well-specified models in \cite{wei2019} is primarily used for studying the large-sample properties of the variational mean estimator $\widehat\theta_{VB}=\E_{\widehat Q_\theta}[\theta]$. In contrast, we extend and apply Theorem 1 to mis-specified models to investigate the large-sample properties of the ELBO for model selection.
\begin{theorem}\label{thm::main1}
Under Assumptions \ref{ass::prior} to \ref{ass::ls},
there exist constants $(C_4,C_5)$  and $C^*$ that depend on the model and $\theta^*$ such that for any $1\le K=O(\sqrt{n})$ it holds with at probability at least $1-C_4K^{-2}$ that
$$
D\big(\widehat Q_\theta \,\big\|\,  Q_{VB}^*\big) \le\frac{C_5 K^3(\log n)^{3}}{\sqrt{n}},
$$
where $Q_{VB}^*$ denotes the normal distribution $N\big(\tmle,\, [n\,\diag(V_c(\theta^*))]^{-1}\big)$.
\end{theorem}
As a direct consequence of this theorem, any reasonable point estimator (e.g.~by taking expectation) obtained from $\widehat Q_\theta$ is asymptotically the same as MLE $\tmle$ under the mis-specified model. The KL divergence bound from the theorem implies an $O(n^{-3/4})$ bound between the variational mean etimator $\widehat\theta_{VB}=\E_{\widehat Q_\theta}[\theta]$ and the MLE $\tmle$ by applying a transportation cost inequality. However, it is noteworthy that our analysis characterizing the updating dynamics of the CAVI iterative algorithm results in an improved error rate of 
$\|\widehat\theta_{VB}-\tmle\|=O(n^{-1}(\log n)^{9/2})$; see equation~\eqref{eq::VB_mean_error_bound} in the proof of Theorem~\ref{thm::caviconv}.
This suggests that there is essentially no loss of statistical efficiency by using the MF approximation to obtain a point estimator to the model parameter as compared to the frequentist likelihood based approaches in the mis-specified setting. However, as is common in the MF variational inference, uncertainty quantification from $\widehat Q_\theta$ can be misleading as the asymptotic covariance matrix 
$[\diag(V_c(\theta^*)]^{-1}$ may drastically underestimate the variability in the marginal posterior of $\theta$, whose asymptotic covariance matrix is $[V(\theta^\ast)]^{-1}$ \citep[Theorem 3.1,][]{kle2012}. The uncertainty underestimation mainly comes from two sources: 1.~ignoring the dependence among components of $\theta$ results in the diagnolization of the covariance matrix; 2.~ignoring the dependence between parameter $\theta$ and latent variables $S^n$ results in the inclusion of the extra $V_s(\theta^\ast)$ term in the precision matrix, i.e. the inverse of the covariance matrix. In particular, if there is no latent variable, then matrix $V_c$ appearing in the $Q_{VB}^*$ reduces to $V$.

\begin{remark}[Extensions to Block MF]
\label{rmk_main}
If we use the block mean-field approximation $q(\theta,s^n)=q_\theta(\theta)\,q_{S^n}(s^n)$, which preserves the within-block dependence as described in the previous subsection, instead of the fully factorized one as $q(\theta,s^n)=\prod_{j=1}^d q_{\theta_j}(\theta_j) \cdot \prod_{i=1}^n q_{S_i}(s_i)$, then Theorem~\ref{thm::main1} can be correspondingly extended where the variational posterior $Q_{VB}^*$ tends to be close to a normal with the same mean, but a different non-diagonal covariance matrix. Precisely, we consider the more general case of MF approximation using factorized densities with $K$ blocks whose respective sizes are $d_1,\ldots,d_K$, 
$$q_\theta(\theta)=q_1(\theta_1,\ldots,\theta_{d_1})\, q_2(\theta_{d_1+1},\ldots,\theta_{d_1+d_2})\,\cdots\, q_K(\theta_{d-d_K+1},\ldots,\theta_d).$$ 
We similarly write the complete data information matrix $V_c(\theta^*)$ in the corresponding block form as
$$
V_c(\theta^*)=
\begin{pmatrix}
V_{11} & \cdots & V_{1K}\\
\vdots & \ddots & \vdots\\
V_{K1} & \cdots & V_{KK}
\end{pmatrix},
$$
where $V_{k\ell}\in\R^{d_k\times d_\ell}$ denotes the $(k,\ell)$-th block.
Then $\widehat Q_\theta$ will be well-approximated by the normal distribution $N\big(\tmle,\,[nS_c(\theta^*)]^{-1}\big)$ with
$$
S_c(\theta^*)=
\begin{pmatrix}
V_{11} & 0 & \cdots & 0\\
0 & V_{22} & \cdots & 0\\
\vdots & \vdots & \ddots & \vdots\\
0 & 0 & \cdots & V_{KK}
\end{pmatrix}.
$$
In the special case of block MF over parameter $\theta$ and latent variables $S^n$, or  $q(\theta,s^n)=q_\theta(\theta)\,q_{S^n}(s^n)$, the approximating normal distribution becomes $N\big(\tmle,\,(nV_c)^{-1}\big)$, where the precision overestimation only comes from the extra latent variable information $V_S$ in decomposition $V_c=V+V_S$. The probit regression example described in Appendix~\ref{sec::mvteg} uses such a block approximation.
\end{remark}

\subsection{Bayesian Model Selection via Mean-Field Approximation}
\label{sec::modelselection}
In the previous subsection, we discussed the large-sample behaviour of MF variational inference in approximating the posterior. The normal approximation $Q_{VB}^\ast$ to $\widehat Q_\theta$ suggests that we may use $\widehat Q_\theta$, or more precisely, the accompanied ELBO $L(\widehat q_\theta\otimes \widehat q_{S^n})$, as a computationally feasible surrogate to the evidence $\log p(X^n)$ for performing model selection. Concretely, recall that the true data generating distribution is denoted by $\mb P_0$. We consider a list of candidate models $\{\mathcal{M}_\lambda\}_{\lambda\in\Lambda}$ that may or may not contain $\mb P_0$. Our target is to select a most parsimonious model from $\{\mathcal{M}_\lambda\}_{\lambda\in\Lambda}$ that is closest to $\mb P_0$. In our setting of parameter models, we use $\theta_{\cM_\lambda}$ and $\Theta_{\cM_\lambda}$ to denote the respective parameter and parameter space associated with model $\cM_\lambda$. The size (or complexity) of a model $\cM_\lambda$ is then the dimension $d_{\cM_\lambda}$ of $\Theta_{\cM_\lambda}$  (i.e.~number of parameters).

In the model selection literature, Bayesian information criterion \citep[BIC,][]{bic1978} is a commonly used criterion function for selecting a best model that balances between goodness-of-fit to data and model complexity, defined as
\begin{align}\label{eqn:BIC}
    \bic(\cM)=-2\widehat\ell_n(\cM)+d_{\cM}\log n,
\end{align}
for a generic model $\cM$ with $d_{\cM}$ number of parameters, where $\widehat\ell_n(\cM)=\max_{\theta_{\cM}\in\Theta_{\cM}}\log p(X^n\mid \theta_{\cM},\,\cM)$ denotes the maximal log-likelihood value under model $\cM$. To employ BIC for model selection, one selects the model with lowest BIC. More specifically, assume a prior distribution $\pi_{\cM_\lambda}$ is imposed to $\theta_{\cM_\lambda}$ under model $\cM_\lambda$, and $p(\cM_\lambda)$ denotes the prior probability assigned to model $\cM_\lambda$.
Using the Laplace approximation, it can be shown \citep[][also see Theorem~\ref{thm::translate}]{stoica2004model} that $-\bic(\cM_\lambda)/2$ provides a large sample approximation to the logarithm of the posterior probability $p(X^n\,|\,\cM_\lambda)$ of data $X^n$ given model $\cM_\lambda$, i.e.~the evidence of model $\cM_\lambda$. Therefore, minimizing BIC over all candidate models is asymptotically equivalent to maximizing the posterior probability $p(\cM_\lambda\,|\,X^n)$ over all $\lambda\in\Lambda$, as the impact from model prior $p(\cM_\lambda)$ is diminishing as sample size $n$ grows.

Since one perspective of variational inference as described in Section~\ref{sec:VI} is finding a best lower bound
$\mbox{ELBO}(\cM) = L_{\cM}(\widehat q_{\cM})$ to the evidence $\log p(X^n\,|\,\cM)$, where $L_{\cM}$ denotes the ELBO under a generic model $\cM$ and $\widehat q_{\cM}$ denotes the MF approximation (i.e.~solution of problem~\eqref{mpvi} that maximizes $L_{\cM}$ within $\Gamma$), it is natural to use $\mbox{ELBO}(\cM)$ to approximate model evidence $\log p(X^n\,|\,\cM)$ to conduct model selection. Interestingly, we find (c.f.~Section~\ref{sec::thmmodelselection}) that as the sample size $n$ tends to infinity, model selection via maximizing the ELBO leads to the same model chosen via minimizing the BIC, which is also the highest posterior probability model. Since BIC is capable of consistently selecting the smallest model containing $\mb P_0$ as $n\to\infty$~\citep{bic1978}, a property known as model selection consistency, we can conclude that ELBO inherits the same model selection consistency property.

Specifically, we find a precise characterization of the gap between the exact evidence $\log p(X^n\,|\,\cM)$  and the ELBO associated with MF approximation (Theorem~\ref{thm::translate}) under a large sample size $n$, which converges to a model-dependent constant $\displaystyle C^\ast(\cM) = \frac{1}{2}\log\frac{\det(\diag(V_c(\tmstar)))}{\det(V(\tmstar))}$ as $n\to\infty$. In comparison, $-\bic(\cM)/2$ also provides an asymptotically constant approximation to the evidence, where the limiting constant is $\displaystyle C_{\rm BIC}^\ast(\cM) =- \frac{1}{2}\log\det(V(\tmstar))+\frac{d_{\cM}}{2}\log(2\pi)+\log\pi_{\cM}(\tmstar)$ (Theorem~\ref{thm::translate}). The empirical results in Section~\ref{sec::numeric} also align well with these theoretical findings. Consequently, approximating the evidence by ELBO does not incur significantly larger error than that by BIC. Moreover, since the discrepancy between the negative half of BIC and ELBO is of constant order while the optimality gap, i.e. the smallest difference in the BIC values between the optimal model and any suboptimal model will be at least of order $\log n$ in order for the true model to be statistically identifiable, they tend to lead to the same selected model as $n\to\infty$.

More interestingly, a closer inspection of the two limiting constants $C^\ast(\cM)$ and $C_{\rm BIC}^\ast(\cM)$ reveals that ELBO generally leads to a better approximation to the model evidence than BIC due to the full incorporation of prior information and a better dimension dependence in the approximation error. For example, the definition~\eqref{eqn:BIC} of BIC completely ignores the prior contribution while ELBO (defined after equation~\eqref{eqn:ELBO_dec}) involves an integration of log-prior, explaining the extra $\log\pi_{\cM}(\tmstar)$ term in the BIC approximation error. In addition, due to the simple Laplace approximation, $C_{\rm BIC}^\ast(\cM)$ has an extra term explicitly dependent of model size $d_{\cM}$, making BIC less accurate in approximating the evidence in large or growing dimension problems; also see our numerical study in Section~\ref{sec::sim_glm} on variable selection in GLM.
Last but not least, our numerical studies in Section~\ref{sec::numeric} suggest that ELBO tends to be less conservative than BIC in the presence of weak signals, and can achieve comparable predictive performance as AIC \citep{aic1974} with a much more parsimonious selected model. Note that AIC is known to be minimax-rate optimal for prediction in linear regression but tends to over-estimate the model size~\citep{yang2005can}.

We conclude this section with a brief discussion on computation. Variational inference is commonly used when the posterior does not have an explicit form (e.g., in mixture models). In such settings, the MLE required for BIC calculation also does not admit a closed-form solution. In these cases, the MLE can be solved numerically using the EM algorithm or approximated by the expectation of the variational distribution, where the latter incurs an error of order at most $O(n^{-1})$ (see the remark after Theorem~\ref{thm::main1}). In either scenario, the computational cost for calculating the MLE (or BIC) is comparable to that of variational inference or CAVI, since the EM algorithm can be viewed as a degenerate CAVI in which the parameter component in the mean-field approximation is further restricted to be a point mass measure. Therefore, previous discussions (or Theorem~\ref{thm::translate}) suggest that ELBO tends to achieve better approximation accuracy than BIC with similar computational costs when estimating model evidence.
Additionally, a standard implementation of CAVI for computing the variational distribution requires $O(n)$ computational cost per iteration due to the necessity of accessing the full data set. In settings where $n$ is large, we may consider approximation methods such as stochastic gradient descent to reduce the per-iteration computational cost~\citep{titsias2014doubly,alquier2020concentration}. We would also like to note that calculating the BIC often requires accurately determining the effective dimension of the model. For some complex models with non-trivial constraints, such as Bayesian factor models (with constraints on the factor loading matrix to enforce identifiability) or latent variable models like Hidden Markov models, counting this effective dimension might not be straightforward. In contrast, calculating the ELBO is a straightforward by-product of implementing the MF variational inference, which automatically incorporates the effective dimension and eliminates the need for case-by-case analysis.



\subsection{Computation via Coordinate Ascent}
\label{sec::cavi}
Coordinate ascent is a natural and efficient algorithm for optimizing over densities taking a product form as in MF approximation~\eqref{mpvi}. 
For illustration, we focus on models without latent variables, where the mean-field family contains all factorized densities of the form $q(\theta)=q_{1}(\theta_{1})\, q_2(\theta_2)\cdots q_{d}(\theta_{d})$ for $\theta\in\mb R^d$. Otherwise, we may view the latent variables as one (block) coordinate of the parameter and derive the corresponding coordinate ascent algorithms. In general, we may also block approximation where each component is a multi-dimensional density. The key idea of coordinate ascent is to optimize over one component of $q(\theta)$ at a time while fixing the others. Let $q_{-j}(\theta_{-j})=\prod_{\ell\neq j} q_{\ell}(\theta_\ell)$ denote the joint density function of $\theta_{-j}$, all components in $\theta$ except for $\theta_j$. 
When optimizing over the $j^{\text {th }}$ component $q_j$, it would be helpful to express ELBO as a functional of $q_j$,
$$
\begin{aligned}
&L(q_j;\, q_{-j}) =  \int_{\mb R^d} q(\theta)\,\big\{\log p(X^n,\, \theta)-\log q(\theta)\big\} \, \dd\theta \\
=& \int_{\mb R} q_{j}\left(\theta_{j}\right) \bigg[\int_{\mb R^{d-1}} q_{-j}\left(\theta_{-j}\right) \log p\left(\theta_{j} \mid \theta_{-j}, \,X^n\right) \,\dd\theta_{-j}\bigg]\,\dd\theta_j-\int_{\mb R} q_{j}\left(\theta_{j}\right) \log q_j(\theta_{j}) \,\dd \theta_{j}+ C(\theta_{-j}),
\end{aligned}
$$
where constant $C(\theta_{-j})$ is independent of $\theta_j$. From this identify, we may
explicitly solve the optimizer $q_{j}^{*}\left(\theta_{j}\right):\,=\argmax_{q_j} L(q_j;\, q_{-j})$ as
\begin{equation}
q_{j}^{*}\left(\theta_{j}\right) \propto \exp \left\{\int_{\mb R^{d-1}} q_{-j}\left(\theta_{-j}\right) \log p\left(\theta_{j} \mid \theta_{-j}, \,X^n\right) \,\dd\theta_{-j}\right\}.
\label{cavi}
\end{equation}
As is often in practice, one can recognize $q_{j}^{*}$ above as coming from some parametric family, for example, certain exponential family, and determine the normalizing constant of $q_{j}^{*}$; otherwise, either particle methods~\citep{saeedi2017variational,wang2021particle} can be employed to approximate $q_{j}^{*}$ over $\mb R$, or $q_{j}$ can be further restricted to some parametric family, resulting in a hybrid variational approximation. 
To summarize, we can apply coordinate ascent to numerically solve the optimization problem in MF based on formula (\ref{cavi}) in an iterative manner; and the resulting method is known as the coordinate ascent variational inference \citep[CAVI,][]{bishop2006} in the literature. To avoid overly aggressive moves that may lead to periodic oscillation or even divergence, it is customary to introduce a step size parameter $\gamma\in(0,1]$ and define the next iterate as proportional to the weighted geometric average $(q_{j}^{*})^\gamma q_j^{1-\gamma}$ between $q_{j}^{*}$
and current iterate $q_j$. In particular, with full step size $\gamma=1$, the update is most greedy and the coordinate ascent becomes alternating maximization. As we will show in our theoretical analysis in Section~\ref{sec::thmcavi} and some numerical studies in Section~\ref{sec::numeric}, the introduction of a partial step size $\gamma\in(0,1)$ is necessary in some problems to avoid algorithmic non-convergence.

In practice, two common updating schemes are utilized in CAVI, depending on whether the components are sequentially or simultaneously updated. In order to draw a connection
with two well-known iterative algorithms for solving linear systems \citep[e.g.][]{trefethen1997numerical}, we will exchangeably call the followings as the Jacobi scheme and Gauss–Seidel scheme respectively, at iteration $t=0,1,\ldots$:

\smallskip
\noindent {\bf Parallel (Jacobi) update.} We update all $d$ components simultaneously based on the last iteration. To be more precise, we run following $d$ updates at each iteration.
    \begin{equation}
q_{j}^{(t+1)} \propto \exp \left\{\int_{\R^{d-1}} q_{-j}^{(t)}\left(\theta_{-j}\right) \log p\left(\theta_{j} \mid \theta_{-j}, \,X^n\right) \,\dd\theta_{-j}\right\},\quad j=1,\ldots,d.
\label{cavi_jacobi}
\end{equation}
    This algorithm can be run in parallel. However, the ELBO is not guaranteed to be monotonically decreasing in $t$ if the full step size is used. To guarantee the convergence, we may need to take a step size $\gamma\in(0,1]$, i.e.
        \begin{equation}
q_{j}^{(t+1)}\ \propto \exp \left\{\gamma\int_{\R^{d-1}} q_{-j}^{(t)}\left(\theta_{-j}\right)\left[\log p\left(\cdot \mid \theta_{-j}, X^n\right)\right] d \theta_{-j}\right\}[q_j^{(t)}]^{1-\gamma},\quad j=1,\ldots,d.
\label{cavi_jacobi_step}
\end{equation}
    This however may lead to a slower convergence.

\smallskip
\noindent {\bf Randomized sequential (Gauss–Seidel) update.} We update one component at each time using the most recent updates of the other components.
   For simplicity, we focus on the randomized sequential update, where a component is randomly picked to be updated. Concretely, for a random index $j(t)\sim \mbox{Unif}\,(1,\ldots,d)$, we compute
\begin{equation}
q_{j(t)}^{(t+1)} \propto \exp \left\{\int_{\mb R^{d-1}} q_{-j(t)}^{(t)}\left(\theta_{-j(t)}\right) \log p\left(\theta_{j(t)} \mid \theta_{-j(t)}, \,X^n\right) \,\dd\theta_{-j(t)}\right\}.
\label{cavi_gauss}
\end{equation}
In this way, ELBO value is guaranteed to be monotonically non-decreasing in $t$. However, due to the sequential nature, we cannot utilize parallel computation techniques to reduce the run time. 
For the sequential update, we may also consider the systematic variant where the components are updated in a pre-specified deterministic order.

\smallskip
Due to the numerical error, the computed ELBO value may differ from the theoretical ELBO value. It is therefore
necessary to analyze the convergence of CAVI to guide its implementation in practice, particularly when applied to perform model selection. In Section~\ref{sec::thmcavi}, we determine the algorithmic convergence rate of CAVI, and study the impact of various problem characteristics on the rate.
Most literature in CAVI convergence studies some special cases such as Gaussian mixture model \citep{wang2005inadequacy,titterington2006convergence}, stochastic block model \citep{zhang2020theoretical,mukherjee2018mean,sarkar2021random,yin2020theoretical} and Ising models \citep{jain2018mean,koehler2019fast,plummer2020dynamics}. Accessing the convergence and determining the accompanied rate under general settings is still an open problem. 

In Section~\ref{sec::thmcavi}, we identify a set of suitable conditions under which the $t$-th iteration $q^{(t)}$ from CAVI satisfies
\begin{align*}
    \mb E\big[L(\widehat q \,) - L(q^{(t)})\big] \leq \alpha^t  \,\mb E\big[L(\widehat q \,) - L(q^{(0)})\big] + \delta_n
\end{align*}
as long as initialization $q^{(0)}$ is in a constant neighborhood around the true parameter $\theta^\ast$,
where $\alpha$ is the algorithmic convergence rate depending on the model and the CAVI setup, and $\delta_n=\m O\big((\log n)^3/\sqrt{n}\big)$ is the usual root-$n$ statistical error, where the poly-$\log n$ factor in $\delta_n$ is due to the high probability argument in our proof.
This result can be interpreted as that under a warm initialization, the ELBO regret (the difference between the ELBO value $L(\widehat q \,)$ at the optimum $\widehat q$ and $L(q^{(t)})$) relative to the theoretical value has a geometric convergence towards zero up to a statistical error of the problem. Since $\E\big[L(\widehat q \,) - L(q^{(0)})\big]$ has a trivial bound as $\m O(nd)$, to guarantee the output ELBO value to be within $n^{-c}$ distance away from its theoretical value, our theory suggests that $\displaystyle O\big(d\log(nd)\big)$ iterations suffice (so that each component is updated $\displaystyle O\big(\log(nd)\big)$ times on average). 
Moreover, under a suitable metric, which will be the Hellinger distance $H(\cdot,\,\cdot)$, we show that the iterate $q^{(t)}$ at time $t$ from CAVI converges towards $\widehat q$ at a geometric speed up to a statistical error term $\delta_n'$ as
\begin{align*}
    \E[H(q^{(t)},\,\widehat q\,)] \leq C_0\, \alpha^{t/2} + \delta_n', 
\end{align*}
where the convergence rate is expected to be $\sqrt{\alpha}$ since the KL divergence is locally quadratic when expanded in $H(q^{(t)},\,\widehat q \,)$. This result can be used to determine the number of iterations for obtaining a root-$n$ consistent point estimator based on the output of $q_\theta$ from CAVI. 

Another interesting finding implied by our theory is that the randomized sequential update always converges
exponentially fast with full step size given a warm initialization. In contrast, in some examples with moderate dependence, the parallel update will converge only when a partial step size (i.e.~$\gamma\in(0,1)$) is used. However, for those step sizes under which both schemes converge, they tend to exhibit a similar convergence behavior.

\section{Theoretical Results}
\label{sec::theory}
In this section, we provide the main theoretical results of this paper.
In Section~\ref{sec::thmmodelselection}, we derive a non-asymptotic expansion of ELBO under MF approximation, and compare it with that of BIC. We also build an oracle inequality, showing that the model selected by ELBO has prediction performance comparable to the best model, even in the case all candidate models are mis-specified and do not contain $\mb P_0$.
In Section~\ref{sec::thmcavi}, we analyze the algorithmic convergence of the CAVI under the two aforementioned schemes, and discuss their consequences.

\subsection{Model Selection Consistency Based on ELBO}
\label{sec::thmmodelselection}
In this subsection, we show the consistency of model selection based on (penalized) ELBO.
Our first result provides non-asymptotic expansions of ELBO and BIC for approximating the model evidence $\log p(X^n\,|\,\cM)$. Recall that $\theta^\ast_{\cM}$ defined in~\eqref{eq:best_theta} denotes the parameter in model $\cM$ such that $\mb P_{\theta^\ast_{\cM}}$ best approximates the true data generating distribution $\mb P_0$. 
\begin{theorem}\label{thm::translate}
Suppose the same assumptions of Theorem \ref{thm::main1} to hold for a generic model $\cM$. 
There exists constants $C_6,C_7$ such that for any $1\le K=O(\sqrt{n})$, it holds with probability at least $1-C_6K^{-2}$ that
\begin{align}
&\qquad\big| \elbo(\cM) - \log p(X^n\mid \cM) + C^*(\cM)\,\big| \leq  \frac{C_7 K^3(\log n)^{3}}{\sqrt{n}},\label{translate1}\\
&\qquad \Big|-\frac{1}{2}\, \bic(\cM) -\log p(X^n\mid \cM) + C^*_{\rm BIC}(\cM)\,\Big| \leq \frac{C_7 K^3(\log n)^{3}}{\sqrt{n}},\label{translate2}\\
& \mbox{where} \qquad C^*(\cM)=\frac{1}{2}\log\frac{\det(\diag(V_c(\tmstar)))}{\ \det(V(\tmstar))},\quad\mbox{and} \notag\\
& \qquad \qquad \ \, C_{\bic}^*(\cM)=-\frac{1}{2}\log\det(V(\tmstar))+\frac{d_{\cM}}{2}\log(2\pi)+\log\pi_{\cM}(\tmstar).\notag
\end{align}
\end{theorem}
Note that inequalities~\eqref{translate1} and~\eqref{translate2} together imply the difference between $-\bic/2$ and ELBO to be asymptotically constant as $n\to\infty$, which we denote as 
\begin{equation}
\widetilde C^*(\cM)\deltaeq C^*(\cM)-C_{\bic}^*(\cM)=\frac{1}{2}\log\det(\diag(V_c(\tmstar)))-\frac{d_{\cM}}{2}\log(2\pi)-\log\pi_{\cM}(\tmstar).
\label{eq::tildeC}
\end{equation}
As a direct consequence of the theorem, the model selected by maximizing ELBO will lead to the same one that minimizes BIC, as long as the minimal BIC value is at least of order $\log n$ away from the second smallest value. For example, let $\m M^\ast$ denote the smallest model containing the data generating distribution $\mb P_0$.
Under the model identifiability condition that any underfitted model $\cM$ not containing $\mb P_0$ has a strictly positive ``KL gap'' $\inf_{\theta\in\Theta_{\m M}} D(\mb P_0\,||\,\mb P_\theta)\geq \varepsilon>0$ for some $\varepsilon>0$, it is not difficult to show (e.g.~see Appendix \ref{pf::bic}) that the difference $\bic(\cM) - \bic(\m M^\ast)$ between BIC values of $\m M$ and $\m M^\ast$ is at least $n\varepsilon - d_{\m M^\ast} \log n \gg \log n$ with high probability. On the other hand, for any overfitted model $\m M$ containing $\mb P_0$ and having a larger number of parameters $d_{\m M}\geq d_{\m M^\ast}+1$, we have that $|\widehat \ell_n(\cM)-\widehat \ell_n(\m M^\ast)|=o(\log n)$ holds with high probability (see Appendix \ref{pf::bic}); this further implies $\bic(\cM) - \bic(\m M^\ast) \geq (d_{\m M}-d_{\m M^\ast})\log n - o(\log n)\gtrsim \log n$. By combining two cases, we can conclude that maximizing ELBO will consistently select the best model $\m M^\ast$ as by minimizing the BIC since $\big|-\bic(\cM)/2 - \elbo(\cM)\big| \leq C$ for some constant $C>0$ when $n$ is sufficiently large.

Note that the same discussion in Remark \ref{rmk_main} also applies to Theorem \ref{thm::translate} for the block mean-field approximation, where the constant then becomes $\displaystyle C_{\rm b}^*(\cM)=\frac{1}{2}\log\frac{\mbox{\rm det}(S_c(\tmstar))}{\mbox{\rm det}(V(\tmstar))}$, which is never larger than the $C^\ast(\cM)$ in Theorem \ref{thm::translate}. This observation suggests that incorporating additional structure in the VI is always beneficial for improving the approximation accuracy given the computation is still tractable. In the other situation with no latent variables, the constant becomes $\displaystyle C_{\rm n}^*(\cM)=\frac{1}{2}\log\frac{\det(\diag(V(\tmstar)))}{\det(V(\tmstar))}$, which is precisely the KL divergence between the two normal distributions $N\big(\tmle,\,[n\,V(\tmstar)]^{-1}\big)$ and $N\big(\tmle,\,[n\,\diag(V(\tmstar))]^{-1}\big)$, where the former approximates the posterior distribution under model mis-specification~\citep{kle2012} and the latter approximates the MF solution $\widehat Q_\theta$ (Theorem \ref{thm::main1} without latent variables), respectively. This limiting gap $C_{\rm n}^*(\cM)$ will vanish if components of $\tmle$ are asymptotically independent (e.g.~the location-scale normal model in Section~\ref{sec:sim_normal}).

\begin{remark}[Discussion on model selection criteria]
For a Bayesian model with latent variables, we have another ELBO value to use for model selection, which is the parameter part $L_{\theta_{\cM}}(q_{\theta_{\cM}})$ in decomposition~\eqref{Eqn:ELBO_LV}, or
\begin{align*}
    L_{\theta_{\cM}}(q_{\theta_{\cM}})&=L(q_{\cM}) + \Delta_J(q_{\cM})=
    \int_{\Theta_{\cM}}\log\frac{p(X^n,\,\theta_{\cM})}{q_{\theta_{\cM}}(\theta_{\cM})}\, q_{\theta_{\cM}}(\theta_{\cM})\,\dd\theta_{\cM} \\
   & = \log p(X^n\,|\,\cM) - D\big(q_{\theta_{\cM}}\,\big|\big|\,\pi_n\big),
\end{align*}
where recall that $\pi_n$ denotes the marginal posterior of $\theta$ and the extra Jensen gap $\Delta_J(q_{\cM})$ is due to latent variables.
In the proof of Theorem~\ref{thm::translate}, it is shown that this Jensen gap asymptotically converges to a non-negative constant $\frac{1}{2}\,\mbox{\rm \tr}([\diag(V_c)]^{-1}V_s)$. Therefore, the parameter part $\elbo_\theta(\cM):\,= L_{\theta_{\cM}}(\widehat q_{\theta_{\cM}})$ leads to an improved approximation to the evidence (due to a smaller gap); and model selections based on $\elbo_\theta$ and $\elbo$ are asymptotically equivalent, and equivalent to that based on $\bic$. However, unlike $\elbo$ that can be efficiently computed via CAVI, $\elbo_\theta$ may not admit a closed form updating formula; and requires Monte Carlo methods to approximate.
\end{remark}

Bayesian hypothesis testing can be viewed as a special instance of model selection with two candidate models, denoted as $\cM_0$ and $\cM_1$. Decisions in Bayesian hypothesis testing or model selection between two models are typically based on the so-called Bayes factor, which is defined as 
$$
B(\cM_0,\cM_1)=\frac{p(X^n\mid \cM_0)}{p(X^n\mid \cM_1)}.
$$
Motivated by our general model selection procedure based on ELBO, we define the ELBO factor $\elbo(\cM_0)-\elbo(\cM_1)$ as a computationally-efficient surrogate to the log-Bayes factor, by noticing 
\begin{align*}
    \log B(\cM_0,\cM_1)&=\log \frac{P(X^n\mid \cM_0)}{P(X^n\mid \cM_1)}=  \elbo(\cM_0)-\elbo(\cM_1)+\m O_p(1),
\end{align*}
where the second equality is due to Theorem~\ref{thm::translate}.
We summarize the result in the following.
\begin{corollary}
Suppose the assumptions of Theorem \ref{thm::main1} hold for the two model candidates $\cM_0$ and $\cM_1$. Then we may approximate the Bayes factor $\displaystyle -\frac{1}{2}B(\cM_0,\cM_1):\,=-\frac{1}{2}\frac{P(X^n\mid \cM_0)}{P(X^n\mid \cM_1)}$ based on the $\elbo$, as we have with probability at least $1-C_6K^{-2}$ that
$$
\left|-\frac{\log B(\cM_0,\cM_1)}{2}-[\elbo(\cM_0)-\elbo(\cM_1)]-[\widetilde C^*(\cM_0)-\widetilde C^*(\cM_1)]\right|\leq  \frac{C_7 K^5(\log n)^{3}}{\sqrt{n}},
$$
for some constant $C_6,C_7$, and constants $\widetilde C^*(M_i)$ ($i=0,1$) are given by (\ref{eq::tildeC}).
\end{corollary}

Previously, we have seen that model selection based on ELBO tends to agree with that based on BIC when at least one of the candidate models contains $\mb P_0$ and it is statistically distinguishable. However, it is often in practice that none of the models in the list contains $\mb P_0$.
Another common situation is when some signals are weak, for example, the $\beta_{\operatorname{min}}$-condition \citep[c.f.][]{yang2016annal} does not hold in linear regression with moderate to large number of variables; then it is information-theoretically impossible to select all important variables. Moreover, using a smaller model may lead to better prediction performance than the true model by excluding those less significant variables; in fact, using the true model with many variables may lead to overfitting. For example, in the numerical study of variable selection in GLM in Section~\ref{sec::sim_glm}, we can see from the left two plots in Fig.~\ref{probit_model} that both BIC and ELBO tend to select a smaller model which turns out to have better prediction performance than the true model with $5$ variables.
To theoretically quantify the quality of the model selected via ELBO in the MF approximation in these situations, we prove an oracle inequality, which shows that the model selected by ELBO has prediction performance comparable to the best model in the list, even all candidate models may be mis-specified and do not contain $\m P_0$. 

\begin{theorem}\label{thm::pred}
Under the same assumptions as Theorem \ref{thm::translate}, there exists some constant $C_0>0$ such that for any $1\le K=O(\sqrt{n})$, it holds with probability at least $1-CK^{-2}$ that the model selected by $\elbo$ $\widehat \cM=\argmax_{\cM} \elbo(\cM)$ satisfies
$$
\int_{\Theta_{\widehat\cM}} D \big(\P_0\, \big|\big|\, \P_{\theta_{\widehat \cM}} \big) \, \widehat q_{\theta_{\widehat{\cM}}}(\theta_{\widehat \cM}) \,\dd\theta_{\widehat \cM}  \leq \inf _{\cM}\left\{\inf_{\theta_{\cM}\in\Theta_{\cM}} D\big(\P_0 \,\big|\big|\, \P_{\theta_{\cM}}\big)+\frac{d_{\cM}\log n}{2n}\right\}+\frac{C_0}{n}+\frac{CR^3K^3(\log n)^{3}}{\sqrt{n}},
$$
where $R$ is the total number of candidate models.
\end{theorem}

As mentioned in the introduction, \cite{cherief2019consistency} also proves an oracle inequality for the model selected by ELBO maximization in variational inference. However, their results apply only to models that do not contain latent variables, and their risk function is based on the $\alpha$-R\'{e}nyi divergence, which is generally weaker than the KL-divergence used in our results. Additionally, their result is stated only for $\alpha$-fractional posteriors under $\alpha < 1$, which requires fewer assumptions than our Assumptions~\ref{ass::l} and \ref{ass::ls}. Specifically, they do not require a testing condition such as Assumption \ref{ass::l}(c), which is one merit of considering Bayesian fractional posteriors~\citep{10.1214/18-AOS1712}.



\subsection{Convergence of CAVI Algorithms} 
\label{sec::thmcavi}
In this subsection, we address the theoretical question of analyzing the convergence of CAVI algorithms. The algorithmic convergence analysis provides a theoretical guidance on how many iterations are required to adequately approximate ELBO to achieve model selection consistency; according to Theorem~\ref{thm::translate} and the related discussions, a constant numerical error approximation to the ELBO would suffice for this purpose.
Since in the convergence analysis the model $\cM$ is fixed throughout, we will omit all $\cM$ in the subscripts when no ambiguity may arise. We let $D^{(t)}=D(q^{(t)}\verts \pi_n)-D(\widehat q \verts \pi_n)$ be a discrepancy measure between the $t$-th iterate $q^{(t)}$ from CAVI and the MF solution $\widehat q$. Note that we also have $D^{(t)}=L(\widehat q\,) - L(q^{(t)})$, which is the regret of $q^{(t)}$ relative to the theoretical ELBO value achieved by $\widehat q$.

\smallskip
\noindent{\bf Special case: Gaussian posterior without latent variables.}
We first consider the special case of a Gaussian posterior to help explain the intuition. Specifically, we assume the target posterior 
$\pi_n$ to be the density of $N(\tmle_\cM,[nV(\tmstar)]^{-1})$, which will be denoted by $\phi_n$. Since our later analysis concerning a general posterior 
$\pi_n$ will have a leading term related to its (asymptotic) normal approximation $N(\tmle_{\cM},[nV(\tmstar)]^{-1})$, we will preserve the notation $p^{(t)}(\theta_1,\ldots,\theta_d)=\prod_{j=1}^d p^{(t)}_j(\theta_j)$ for the $t$-th iterate in the CAVI, either the parallel or the randomized sequential update depending on the context, of maximizing $D(p\,||\,\phi_n)$ in the MF variational inference. To simplify the notation, we use the shorthand notation $\widehat\theta=\tmle_{\cM},V=V(\tmstar)$ and $S=\diag(V)$. Under this notation, the MF approximation is $\phi^\ast:\,=\argmin_{p=\otimes_{j=1}^dp_j }D(p\,||\,\phi_n)$, which is the density of $N(\widehat\theta,\,(nS)^{-1})$.

To simplify the analysis for the Gaussian posterior, let us assume the initialization $p^{(0)}$ to be the density of $N(\theta^{(0)},\,(nS)^{-1})$ for some initial mean vector $\theta^{(0)}\in\mb R^d$. This assumption is not overly restrictive. It is easy to verify that after each component in $p=\otimes_{j=1}^d p_j$ is updated at least once with full step size $\gamma=1$, all future iterates of $p$ will follow a normal distribution with covariance matrix $(nS)^{-1}$.
Due to this reason, let us denote $p^{(t)}$ as the density of $N(\theta^{(t)},\,(nS)^{-1})$. To describe the two updating schemes of the CAVI described in Section~\ref{sec::cavi}, let us derive the common step of updating one component $p_{\ell}$ from the following rule. Recall that the parallel scheme updates all components with indices $\ell=1,2,\ldots,d$ in one iteration; while the randomized sequential scheme randomly pick one index $\ell$. The common updating formula for $p_{\ell}$ is 
\begin{align*}
   p_\ell^{(t+1)}(\theta_{\ell})\propto\exp\Big(-\frac{\gamma\,n}{2}\, \mb E_{p_{-\ell}^{(t)}} \big[(\theta-\widehat\theta\,)^TV(\theta-\widehat\theta\,)\big]\Big) \cdot \big[p_\ell^{(t)}(\theta_{\ell})\big]^{1-\gamma},
\end{align*}
where $\gamma\in(0,1]$ denotes the step size.
Denote the bias at iteration $t$ as $b^{(t)}=\theta^{(t)}-\widehat\theta$. 
Then it is straightforward to translate the preceding display into an updating rule for $b^{(t)}$ as
\begin{equation}
b^{(t+1)}_\ell=(1-\gamma)\,b^{(t)}_\ell-\gamma\,\sum_{k\not=\ell}\frac{V_{\ell k}}{V_{\ell\ell}}\, b_k^{(t)}.
\label{eq::bconv}
\end{equation}
Parallel update applies (\ref{eq::bconv}) to $d$ components simultaneously, and we may write the equivalent matrix form as $b^{(t+1)}=A_\gamma b^{(t)}$ for $A_\gamma=(I-\gamma S^{-1}V)$. On the contrary, sequential update applies (\ref{eq::bconv}) to one component at each iteration, and we may directly study the decrease in ELBO as it is quadratic (c.f. Appendix~\ref{pf::gauss_cavi} for more details). It is worth noting that the parallel and the randomized sequential schemes using the preceding updating rule respectively coincide with the Jacobi and the Gauss-Seidel methods (with partial step size) for solving the system of $d$ linear equations $V x = 0$.
To analyze the convergence of the algorithm, let us study the evolution of the objective functional $D(p^{(t)}\verts \phi^*) = \frac{n}{2}\,[b^{(t)}]^TSb^{(t)}$, whose convergence implies the convergence of the algorithm. For the Gaussian posterior, we have $D^{(t)}=D(p^{(t)}\verts \pi_n)-D(\phi^\ast \verts \pi_n)=\frac{n}{2}\,[b^{(t)}]^TVb^{(t)}$, which is equivalent to $D(p^{(t)}\verts \phi^*)$ up to some multiplicative constant. We summarize the result in the following lemma, whose proof is deferred to Appendix \ref{pf::gauss_cavi}.

\begin{lemma}[Gaussian posterior without latent variables]\label{lem::gauss_cavi}
Suppose the true posterior is $N(\widehat\theta,\,(nV)^{-1})$. If the step size $\gamma$ is chosen so that the $\alpha$ defined below belongs to $(0,1)$, then $p^{(t)}$ converges exponentially fast to $\phi^*$:
$$
\E[D(p^{(t)}\verts \phi^*)]\le C\,D^{(t)}\leq C\,\alpha^t D^{(0)},\qquad t\geq 1,
$$
for some constant $\alpha\in(0,1)$ depending on the updating scheme.
In particular, we have $\alpha=1-c_s\gamma(2-\gamma)/d$ for the randomized sequential update; and $\alpha=c_p(\gamma)$ for the parallel update, where
\begin{equation}
c_s=\max_{\|b\|=1}\frac{b^TV S^{-1}Vb}{b^TVb}\quad\mbox{and}\quad c_p(\gamma)=\max_{\|b\|=1}\frac{b^T(I-\gamma S^{-1}V)V(I-\gamma S^{-1}V)b}{b^TVb}.
\label{geometric_c}
\end{equation}
\end{lemma}
Without parallel computing, the computational complexity of $d$ iterations of the sequential scheme is comparable to that of the parallel scheme. Therefore, to fairly compare the overall computational complexities, we may define one iteration in the sequential scheme as $d$ updates, whose effective contraction rate is then $(1-c_s\gamma(2-\gamma)/d)^d$, which is roughly $e^{-c_s\gamma(2-\gamma)}$ for a large $d$. Some detailed comparison of the contraction rates is deferred to the end of this subsection.

\begin{remark}[Necessity of partial step size]
From Lemma~\ref{lem::gauss_cavi}, we can see that the randomized sequential update always converges exponentially fast with full step size, i.e.~$\alpha <1$ for $\gamma =1$. In fact, as $\alpha=1-c_s\gamma(2-\gamma)/d$ in randomized sequential update and $c_s>0$ due to  non-singularity of $V$, we may prefer taking $\gamma=1$ to obtain fastest convergence rate. In contrast, the parallel update may require a partial step size to guarantee the convergence. This is not an artifact of our proof technique --- consider the following counterexample with $\gamma=1$: we take $d=3$, $V=\frac{1}{3}\bm{I}_3+\frac{2}{3} \bm{1}_3\cdot\bm{1}_3$ where $\bm{I}_3$ is the identity matrix and $\bm{1}_3$ the all-one vector. If initialized at $b^{(0)}=(1,1,1)$, the parallel scheme leads to $b^{(t)}=(-\frac{4}{3})^t\,b^{(0)}$, and therefore both $b^{(t)}$ and the ELBO diverges. 
\end{remark}

\smallskip
\noindent{\bf General case I: non-Gaussian posterior without latent variables.}
Recall that the CAVI output at $t$-th iteration for a general posterior $\pi_n$ is denoted as $q^{(t)}$, and the associated CAVI output for its large sample normal approximation $N(\tmle_{\cM},[nV(\tmstar)]^{-1})$ (density denoted as $\phi_n$) is $p^{(t)}$. In addition, the MF approximation to $\phi_n$ is $N(\widehat\theta,\,(nS)^{-1})$, whose density is denoted as $\phi^\ast$; and $\widehat q$ is the MF approximation to $\pi_n$.
Recall that $D^{(t)}=D(q^{(t)}\verts \pi_n)-D(\widehat q\verts \pi_n)=L(\widehat q \,) - L(q^{(t)})$ denotes the regret of $q^{(t)}$ relative to the theoretical ELBO value achieved by $\widehat q$.
We prove the following theorem, describing the convergence of $D^{(t)}$ and $q^{(t)}$, by applying perturbation analysis for generalizing Lemma \ref{lem::gauss_cavi} to a general posterior. 
\begin{theorem}[General posteriors without latent variables]\label{thm::caviconv}
Under Assumptions \ref{ass::prior} and \ref{ass::l}, there exist positive constant $C$, such that if the support of initial distribution is in some small neighborhood around $\theta^\ast$, or $supp(q^{(0)})\subseteq B_{\theta^*}(\delta)=\{\theta:\|\theta-\theta^*\|\le \delta\}$ for some sufficiently small constant $\delta>0$, then we have for any $1\le K=O(\sqrt{n})$ it holds with probability at least $1-CK^{-2}$ that
\begin{equation}
\E[D^{(t)}] \le C\,\alpha^t D^{(0)}+\frac{C K^3(\log n)^3}{\sqrt{n}},\quad t\geq 1,
\label{mainineq3}
\end{equation}
where the contraction rate $\alpha$ is given in Lemma~\ref{lem::gauss_cavi}.
Moreover, under the same high probability event, there exists some constant $C_0$ depending on the initialization $q^{(0)}$ such that
\begin{equation}
\E[H(q^{(t)},\,\widehat q \,)]\le \alpha^{t/2}C_0+C\sqrt{\frac{K^3(\log n)^{3}}{\sqrt{n}}}.
\label{eqn:mainineq4}
\end{equation}
\end{theorem}
The proof of the theorem is quite technical and provided in Appendix \ref{pf::caviconv}, where the expression of $C_0$ is also included. As we can see, the general case inherits the same contraction rate $\alpha$ from the Gaussian case. However, the convergence is now only up to the $\m O\big(\log^3(n)/\sqrt{n}\big)$ statistical accuracy; and only a local geometric convergence can be proved. The local convergence is not an artifact of our proof, as the numerical example in Section~\ref{sec::community} shows that the fast geometric convergence indeed only occurs after the iterate enters the contraction basin as a local neighborhood around $\theta^\ast$.  We note that the $\log ^3(n) / \sqrt{n}$ in our error bound is larger than the $\sqrt{\log (n)} / \sqrt{n}$ error bound proved in \cite{alquier2020concentration} for the gradient-based method for MF inference. However, their result is only stated for the MF approximation to $\alpha$-fractional posteriors with $\alpha < 1$, which requires fewer assumptions and tends to be technically less involved to prove. It would be interesting to explore if the same error bound also holds for the MF approximation to the usual posterior without tempering.

 In view of Theorem \ref{thm::main1}, it suffices to run $O\big(d\log(nd)\big)$ iterations to guarantee model selection consistency by using the approximated ELBO value produced from the CAVI algorithm. 
The first inequality in the theorem states that $D(q^{(t)}\verts \pi_n)$ converges to $D(\widehat q\verts \pi_n)$ exponentially fast up to a statistical error term of order $\m O(\log^c n/\sqrt{n})$. However, the convergence of $D(q^{(t)}\verts \pi_n)$ to $D(\widehat q\verts \pi_n)$ (which is roughly equal to the constant $C^*(\cM)$, up to the same statistical error) does not imply the convergence of $q^{(t)}$ to $\widehat q$. Therefore, we also provide the second inequality that implies the convergence of $q^{(t)}$ towards $\widehat q$, where the loss function used is the Hellinger distance.
We use the weaker Hellinger distance to characterize the convergence of $q^{(t)}$ by applying the triangle inequality because KL-divergence is not a metric and hence triangle inequality is not applicable. 

\smallskip
\noindent{\bf General case II: non-Gaussian posterior with latent variables.}
When there are latent variables, we may apply similar arguments to analyze the convergence of the CAVI algorithm. The (parallel) updating formula in CAVI with latent variables, for example, becomes
\begin{align*}
\qs i^{(t+1)}\left(s_i\right) &\propto \exp \left\{\int_{\R^{d}} q_\theta^{(t)}\left(\theta\right) \log p\left(s_i \mid \theta, X_i\right) \,\dd\theta\right\}, \quad i=1,\ldots,n,\\
q_{\theta_j}^{(t+1)}(\theta_j)&\propto \exp\bigg\{\int_{\mb R^{d-1}}\left(\int_{\m S^n} q^{(t+1)}_{S^n}(s^n)\,\log p(\theta\,|\,X^n,s^n)\,\dd s^n \right) \,q^{(t)}_{\theta_{-j}}(\theta_{-j})\,\dd\theta_{-j}\bigg\},\quad j=1,\ldots,d.
\end{align*}
We will follow a similar analysis based on Taylor expansions on $\theta$, first analyzing $\qs i ^{(t+1)}$ by utilizing the concentration property of $q_\theta^{(t)}$ and then plugging it into $\qt j^{(t+1)}$. This perturbation analysis helps us identify the leading terms in the CAVI updates, resulting in a key updating formula~\eqref{eqn:leading_up} (in Appendix~\ref{sec::cavi_latent}) describing the ``noiseless'' case dynamic, similar to the updating formula~\eqref{eq::bconv} for the Gaussian posterior without latent variables as obtained earlier.
With this key updating formula, we can then proceed with the analysis as in the case without latent variables. First, we study the CAVI for the leading ``noiseless'' Gaussian (or quadratic) case. We then employ perturbation analysis to analyze the CAVI for the original problem. The randomized sequential update can be interpreted and analyzed in a similar way. This leads to the following theorem, whose proof is provided in Appendix~\ref{sec::cavi_latent}.
Recall that we used $V=V(\theta^\ast)$ to denote the observed data Fisher information, $V_s = V_s(\theta^\ast)$ for the missing data Fisher information, and $V_c=V+V_s$ for the complete data Fisher information. Additionally, $S$ and $S_c$ are the diagonal matrices corresponding to the diagonal parts of $V$ and $V_c$, respectively.

\begin{theorem}[General posteriors with latent variables]\label{thm::caviconv2}
Under Assumptions \ref{ass::prior}, \ref{ass::l} and \ref{ass::ls}, the conclusions of Theorem~\ref{thm::caviconv} remain true, albeit with different contraction rates. In particular, we have $\alpha = 1-c_s(\gamma)\gamma/d$ for the randomized sequential update; and $\alpha = c_p(\gamma)$ for the parallel update, where
$$
c_s(\gamma)=\max_{\|b\|=1}\frac{b^TVS_c^{-1}(2S_c-\gamma S) S_c^{-1}V b}{b^TVb} \quad\mbox{and}\quad c_p(\gamma)=\max_{\|b\|=1}\frac{b^T(I-\gamma S_c^{-1}V)V(I-\gamma S_c^{-1}V)b}{b^TVb}.
$$
\end{theorem}
By comparing the contraction rates of the CAVI algorithm with latent variables (Theorem~\ref{thm::caviconv2}) to those without latent variables (Theorem~\ref{thm::caviconv}), we can observe that the presence of latent variables generally slows down the convergence of CAVI. 
Specifically, this reduction in convergence rate is characterized by a factor related to the ratio between the observed data information and the complete data information, or the matrix operator norm of $S_c^{-1} S$, where diagonalization is due to the mean-field approximation on $q_\theta$; see the sequential update case in Appendix~\ref{sec::cavi_latent} for a more concrete calculation. Our convergence result for the CAVI algorithm is consistent with existing findings on the algorithmic contraction rate of the EM algorithm~\citep{dempster1977maximum} --- the convergence slows down as the missing data information $V_s$ increases, making the latent variables harder to learn, which in turn affects the algorithmic convergence of the estimation of $\theta$.



We end this section with a discussion on the efficiency comparison of the two updating schemes using an illustrating example without latent variables.
We define an epoch for the sequential scheme to be $d$ iterations so that each coordinate is updated once (on average for the randomized sequential update), and an epoch for the parallel scheme as one iteration.
We compare the (averaged) decrease in ELBO for the two methods after one epoch.
We consider $V=(1-a)I_d+a1_d\cdot 1_d^T$. For the sequential update, $\Exp[D^{(t)}]$ has a relative decrease by at least $1-\big(1-\frac{\gamma(2-\gamma)\,(1-a)}{d}\big)^d$, which will be $1-\big(1-(1-a)/d\big)^d$ at $\gamma=1$ after one epoch. For the parallel update, we need $\gamma((d-1)a+1)<2$ to avoid divergence. Taking $\gamma=\frac{2}{(d-2)a+2}$ leads to a largest relative decrease by $1-\left(1-\frac{2-2a}{da-2a+2}\right)^2$. When $a$ is close to $1$, we can see that the decrease from the parallel update tends to be $4/d$ of that from the sequential update after one epoch. As $a$ tends to zero, the relative decrease from the parallel update is about $1-(da/2)^2\approx 1$, which is roughly $1.58$ times larger (better) than the $1-e^{-1}$ decrease from the sequential update when $d$ is large. Therefore, efficiency-wise, the parallel update is generally preferred in low dependence situations ($da$ small); while the sequential update is preferred in high dependence and high dimensional situations.
Implementation-wise, the parallel update may benefit from parallel computing; the sequential update always converges while the parallel update may need fine tuning the step size.

\section{Numerical Study}\label{sec::numeric}

This section includes numerical experiments on assessing the algorithmic convergence and the performance of using ELBO for model selection. Detailed descriptions about some of the considered examples are provided in Appendix~\ref{sec::mvteg}.

\subsection{Location-Scale Normal Model}\label{sec:sim_normal}
\begin{figure}[htb]
    \centering
    \includegraphics[width=15cm]{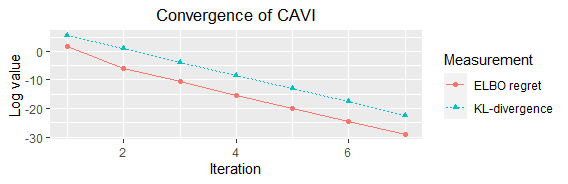}
    \caption{Convergence of CAVI for location-scale normal model.}
    \label{normal_conv}
\end{figure}

We consider the location-scale normal model $N(\mu,\sigma^2)$ with parameter $\theta=(\mu,\sigma^2)$ as described in Appendix~\ref{sec::mvteg}. We generate sample from $N(100,100^2)$ with sample size $n=10$. We take the prior as $\mu\sim N(0,100^2)$ and $\sigma^2\sim \mbox{IG}(1/100,1/100)$. 
To assess the convergence of the CAVI algorithm, we plot the logarithms of the ELBO regret $L(\widehat q \,)-L(q^{(t)})$ and the KL divergence $D(q^{(t)}\verts \widehat q \,)$ between $q^{(t)}$ and $\widehat q$ respectively versus iteration count $t$. Note that the posterior has  asymptotically independent components in this setting, so the two updating regimes lead to similar result, and we present sequential update for illustration. Here the optimal variational distribution $\widehat q$ is approximately obtained from the CAVI output after sufficiently many iterations.
The results in~Fig.~\ref{normal_conv} show that both measurements are nearly linearly decreasing in the log-scale, implying the geometric convergence. The two curves are nearly parallel, which implies the same convergence rate. Note that the Fisher information $V$ (equal to $V_c$ without latent) in this example is diagonal. Therefore, the leading constant term of the contraction rate $\alpha$ from Theorem \ref{thm::caviconv} vanishes and the remainder $\m O(1/\sqrt{n})$ term characterizes the contraction rate. 

\begin{figure}[htb]
    \centering
    \includegraphics[width=15cm]{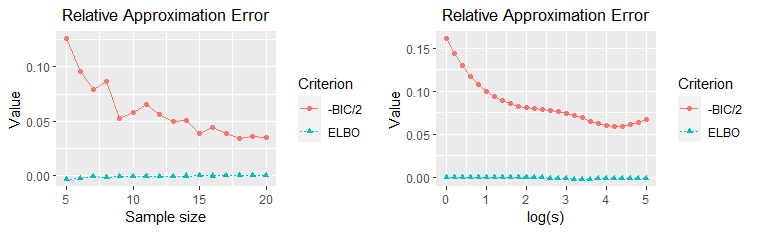}
    \caption{Relative approximation error of ELBO and -BIC/2 for normal parameters, with different sample sizes or different prior parameters.}
    \label{normal_gap_both}
\end{figure}

In terms of the approximation accuracy of ELBO and BIC, the diagonal information matrix implies that the limiting constant gap $C^*(\cM)$ between evidence and ELBO equals 0, and they only differ by a statistical error term. On the other hand, the constant gap $C_{\bic}^*(M)$ between evidence (or ELBO) and $-\bic/2$  is generally non-zero. We can also see this difference via a numerical comparison, by plotting the relative approximation errors of ELBO and BIC under different sample sizes, defined as $\displaystyle\frac{\elbo-\log p(X\mid\cM)}{\big|\log p(X\mid\cM)\big|}$ and $\displaystyle\frac{-\bic/2-\log p(X\mid\cM)}{\big|\log p(X\mid\cM)\big|}$ respectively. The evidence is calculated based on $10^6$ Monte Carlo replicates. The result is included in the left panel of Fig.~\ref{normal_gap_both}.
To assess the impact of prior information on the approximation accuracy, we run an extra simulation study by fixing the sample size $n$ to be $10$, and taking the prior as $\mu\sim N(0,s^2)$ and $\sigma^2\sim \mbox{IG}(s^{-1},s^{-1})$ with varying $s$ over the grid $\exp(m)$ for $m=0,0.2,\ldots 5$. We report the result in the right panel of Fig.~\ref{normal_gap_both}. As expected, smaller $s$ leads to larger approximation error of BIC, as BIC completely neglects the prior contribution in the evidence.

\subsection{Gaussian Mixture Model}\label{sec:sim_GMM}
\begin{figure}[htb]
    \centering
    \includegraphics[width=15cm]{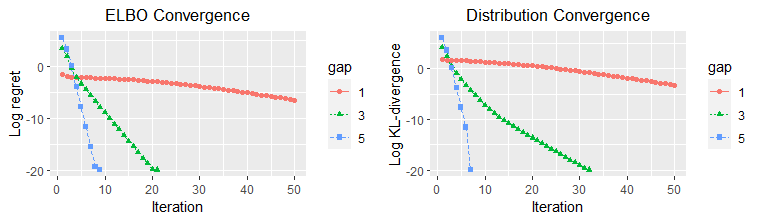}
    \caption{Convergence of CAVI in GMM with gap $\Delta=\{1,3,5\}$.}
    \label{gmm_conv}
\end{figure}

We consider the Gaussian mixture model (GMM) with $K=3$ clusters centered at $\Delta\cdot(-1,0,1)$ (a detailed description is provided in Appendix~\ref{sec::mvteg}). We take the gap (or cluster separation) $\Delta\in\{1,3,5\}$, and the prior parameter $\sigma=2$ in the i.i.d.~$N(0,\sigma^2)$ prior for cluster centers (we choose relatively smaller $\sigma$ so that $\widetilde C^*(\cM_0)$ is more distinguishable in the plot for different $\Delta$). We take the sample size $n$ over the grid $\lf\exp(m)\rf$ for $m=4,4.1,\ldots,7.9,8$, and calculate $-\bic/2-\elbo$ in comparison with theoretical values $ \widetilde C^*(\cM_0)=\Delta^2/\sigma^2+K(\log\sigma^2-\log K)/2$. We first focus on the GMM with a correctly specified number of clusters $K=3$. We can see the simulated values are fairly close to the theoretical limit $\widetilde C^*(\cM_0)$ from Fig.~\ref{gmm_conv}.
\begin{figure}[htb]
    \centering                              
    \includegraphics[width=15cm]{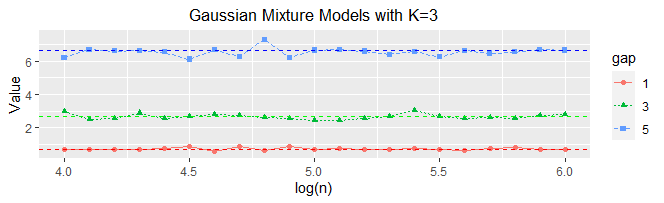}
    \caption{Comparison between simulated $-\bic/2-\elbo$ and theoretical values $\widetilde C^*(\cM_0)$ over three different gaps $\Delta$. The dashed horizontal lines refer to theoretical values.}
    \label{gmm_gap}
\end{figure}

\begin{figure}[htb]
    \centering
    \includegraphics[width=15cm]{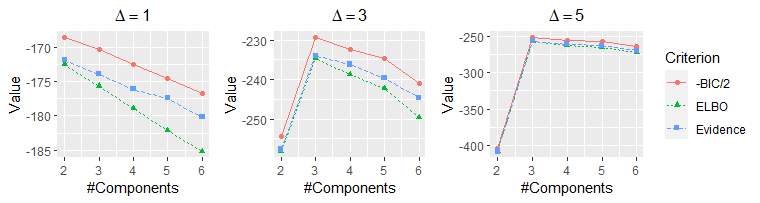}
    \caption{ELBO, BIC and evidence values versus number of components in GMM.}
    \label{gmm_model}
\end{figure}

We also assess the convergence of the sequential CAVI algorithm by plotting the logarithm of the ELBO regret and $D(q^{(t)}\verts \widehat q \,)$ under $\Delta\in\{1,3,5\}$, $n=100$ and $\sigma=10$. The numerical results from Fig.~\ref{gmm_conv} show the geometric convergence of both measurements as they are nearly linearly decreasing in the log-scale, which is consistent to the prediction from Theorem~\ref{thm::caviconv}. The slopes of two measurements are also close under the same $\Delta$, implying the same convergence rate. Moreover, since a large gap $\Delta$ leads to a lower posterior dependence, our Theorem~\ref{thm::caviconv} predicts a smaller contraction rate $\alpha$ (faster convergence), which is also consistent with the numerical results.
Next, we compare the use of BIC and ELBO for evidence approximation. We consider two settings: 1.~prior standard deviation $\sigma=s$ for $s=\exp(m)$ with $m=0,0.2,\ldots,4$ under fixed sample size $n=100$; 2.~sample size $n=\lf\exp(m)\rf$ with $m=4,4.1,\ldots,6$ with fixed $\sigma=10$. We take the gap $\Delta=1$. As we can see from Fig.~\ref{gmm_model}, the approximation error of BIC is quite sensitive to the prior choice; while the approximation of ELBO is much more accurate and stable. However, both approximations become accurate as we increase the sample size.
\begin{figure}[htb]
    \centering
    \includegraphics[width=15cm]{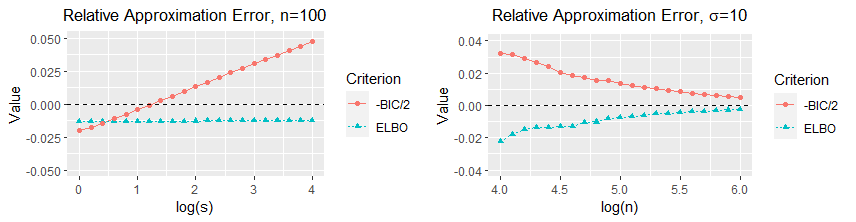}
    \caption{Approximation errors of ELBO and BIC in GMM under $\Delta=1$.}
    \label{gmm_error}
\end{figure}

Note that GMM has degenerate Fisher information when the number of clusters $K$ is misspecified, which leads to a slower $n^{-1/4}$ rate of convergence when $K$ is over-specified \citep{chen1995optimal}. With known mixing weights, the overfitted models are still degenerate. For example, the two component GMM with equal weights is degenerate when true model is a single Gaussian \citep{dwivedi2020singularity}. 
Although our theory no longer applies in such a situation, it is still interesting to compare the model selection performance based on BIC, ELBO and the true evidence.
For simplicity, we assume the underlying mixing weights to be known and fixed in our simulation setting. We set $\Delta\in\{1,3,5\}$, sample size $n=100$ and prior with $\sigma=10$. The results are summarized in the Fig.~\ref{gmm_model}.
As we can see, the three model selection criteria result in the same selected models across all settings.
When the gap $\Delta=1$ is relatively small, all criteria select the smaller model with two components. As we increase the gap to $\Delta\in\{3,5\}$, all criteria select the true model. In terms of approximation accuracy for the evidence, we can see that ELBO tends to be more accurate than BIC, particularly for smaller models under stronger signal-to-noise ratios where the Fisher information is non-singular near the mis-specifed MLE.

\subsection{Generalized Linear Model}
\label{sec::sim_glm}
\begin{figure}[htb]
    \centering
    \includegraphics[width=15cm]{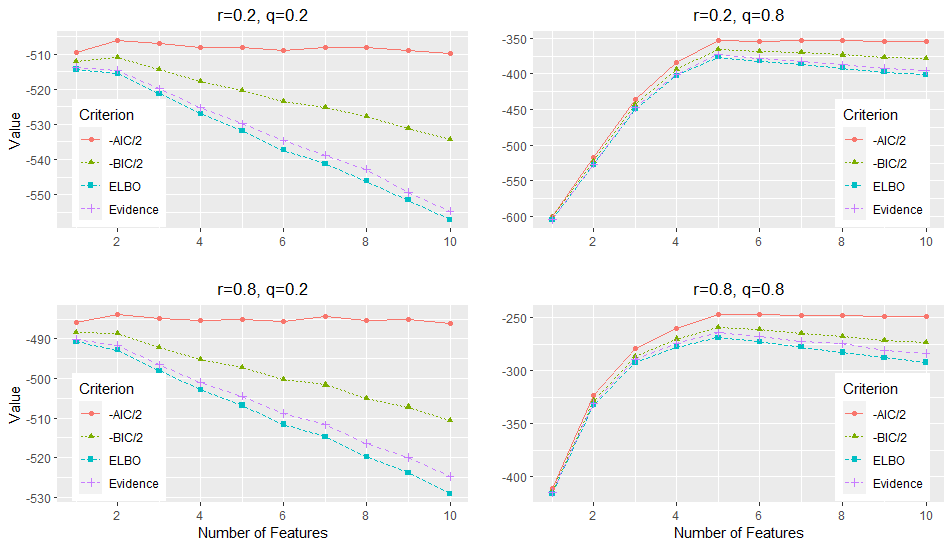}
    \caption{ELBO, AIC, BIC and evidence values versus number of variables in probit regression.}
    \label{probit_model}
\end{figure}
In this section, we consider variable selection in probit regression as a representative instance of generalized linear model (GLM), whose detailed description is deferred to Appendix~\ref{sec::mvteg}.
The simulation setting is as follows. We generate $p$-dim feature vectors $X_i$ i.i.d.~from $N(0,\Sigma)$. The covariance matrix $\Sigma(r)=(\sigma_{ij})\in\R^{p\times p}$ is of AR(1) structure so that $\sigma_{ij}=r^{|i-j|}$. We take $(n,d_{\cM_0})=(1000,10)$, and the $p$-dim (regression) coefficient $\bbeta$ has the $5$ nonzero terms: $\beta_j=q^{j-1}\cdot1(j\le5),\, j=1,\ldots,10$. The prior of $\bbeta$ is the centered multivariate normal distribution with covariance matrix $\Sigma_0=\sigma^2 I_{d_\cM}$ where $\sigma=10$. To assess the variable selection performance, we compare ELBO with AIC, BIC, and the true evidence (calculated based on $10^4$ Monte Carlo replicates).
As we can see from Fig.~\ref{probit_model}, when the signal is relatively strong ($q=0.8$), all methods except AIC correctly select the true model; and AIC selects a larger model (with $6$ features) when the dependence is strong ($r=0.8$). Moreover, the AIC curve is quite flat and the discrepancy between AIC values under different models are small. Under the weak signal case ($q=0.2$), although the proposed criterion and BIC fail to select the true model, they still select the same model as that based on evidence (i.e.~highest posterior model). Under all four settings, ELBO is much more accurate than BIC in terms of the approximation for the evidence.

\begin{figure}[htb]
    \centering
    \includegraphics[width=15cm]{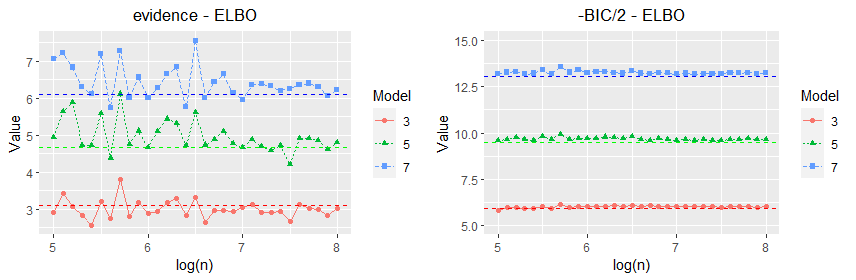}
    \caption{Comparison between simulated and theoretical values for $C^*(\cM)$ and $\widetilde C^*(\cM)$ under the models with $\{3,5,7\}$ included variables. The dashed horizontal lines refer to theoretical values.}
    \label{probit_gap}
\end{figure}

In addition, to compare the numerical gaps with the theoretical constants $C^*(\cM)$ and $\widetilde C^*(\cM)$ (which together also characterize $C_{\bic}^*(\cM)$) in Theorem \ref{thm::translate} and \eqref{eq::tildeC}, Fig.~\ref{probit_gap} includes their simulated values for the models including $3$, $5$ and $7$ features (recall the true model contains $5$ features), respectively, with the sample size $n$ over the grid of $\lf\exp(m)\rf$ for $m=5$, $5.1$, \ldots , $7.9$, $8$. In this study, we set $q=0.8$ and $r=0.8$. We estimate $\bbeta^*_{\cM}$ for the underfitted model (with 3 features) based on $10^7$ samples, so that the estimated $\widehat\bbeta_{\cM}$ can be treated as $\bbeta^*_{\cM}$. As we can see from Fig.~\ref{probit_gap}, the simulated values are fairly close to the theoretical ones, which justifies our theory that evidence$-\elbo$ and $-\bic/2-\elbo$ are equal to $C^*(\cM)$ and $\widetilde C^*(\cM)$ up to some higher-order terms. We note that the simulated values of $C^*(\cM)$ have higher fluctuations, which might be partially caused by the high variance of the Monte Carlo approximation to the evidence.

\begin{figure}[htb]
    \centering
    \includegraphics[width=15cm]{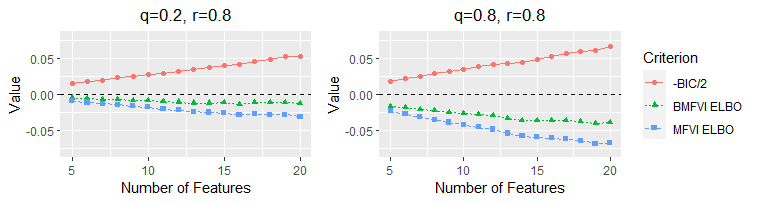}
    \caption{Relative approximation errors of ELBO and -BIC/2 under weak ($q=0.2$) and strong ($q=0.8$) signal regimes. BMFVI: block MF;  MFVI: fully factorized MF. }
    \label{probit_error}
\end{figure}

According to Theorem~\ref{thm::translate}, the difference between -BIC/2 and evidence has a term proportional to the dimension $d_\cM$, which is consistent with the trend of the gap observed from Fig.~\ref{probit_model}. 
On the other hand, as we used block MF (BMFVI) for probit regression, the approximation error of ELBO only comes from the difference between $V_c$ and $V$ characterized by $\displaystyle\frac{1}{2}\log\frac{\det(V_c)}{\det(V)}$, which grows linearly in $d_{\cM}$ with slope independent of $\sigma_0$. To formally demonstrate the impact of $d_{\cM}$ on approximation the evidence, we plot the approximation errors of BIC and ELBO for $d_\cM=5,\ldots,20$ in Fig.~\ref{probit_error}. In the plot, we also include the results for the fully factorized MF (MFVI), whose limiting approximation error $\displaystyle\frac{1}{2}\log\frac{\det(\diag(V_c))} {\det(V)}$ tends to be more sensitive to $d_{\cM}$ than BMFVI, as expected. 
 We can see that ELBO with BMFVI has overall best approximation accuracy, while the approximation error of BIC is largest due to its strong dimension dependence. On the other hand, the ELBO from MFVI has larger approximation error than that from BMFVI, due to the additional independence imposed among parameter components. 
 Therefore, we should take advantage of the block mean field structure and implement BMFVI in practice whenever computationally tractable.

To assess the prediction performance of various criteria in the presence of weak signals (i.e.~selection consistency may not be achievable) in the context of Theorem \ref{thm::pred}, we consider a more challenging setting where $d_{\cM_0}=100$, $\beta_j=0.8^j$ for $1\le j\le d_{\cM_0}$, and the same covariance matrix $\Sigma(r)$ with $r\in\{0.2,0.8\}$ for the random feature vectors. 
Although now the true model includes all $100$ important features, signals decay geometrically fast; as a result, both BIC and ELBO would select much smaller models. Although model selection consistency does not hold under this setting, we use numerical study to compare the prediction performance of model selection based on AIC, BIC and ELBO. In this example, we include AIC as the benchmark corresponding to the theoretically minimax-optimal criterion for prediction~\citep{aic1974}.
We generate $10^4$ data, and randomly select $n=200,500,1000$ of them as training data, and the remaining as test data. Then we select the model using the training data with AIC, BIC and ELBO, and evaluate their prediction performance on the test data. We compare the classification errors and logistic losses of ELBO and BIC evaluated on the test data. Note that the logistic loss is defined as $-n_{\text{test}}^{-1}\sum_{i=1}^{n_{\text{test}}}\big[y_i\log \widehat p_i+(1-y_i)
\log (1-\widehat p_i)$\big], where $\widehat p_i$ is the fitted probability on the $i$-th test data with the selected model. The logistic loss is the log-likelihood function evaluated on the test data, and approximates the population-level KL-divergence between the true and the fitted prediction distributions up to some fixed constant. We report the average classification errors along with their standard deviations, and the median of logistic losses from 100 Monte Carlo replicates. 
From Table~\ref{table:GLM}, the prediction accuracy of ELBO is overall comparable to AIC, and outperforms BIC. However, AIC tends to include more variables than ELBO for slightly reducing its classification error; while due to the tendency of overfitting the data by including too many variables, AIC may incur slightly larger logistic loss on the test data.
Overall, ELBO seems to be a better balance between AIC and BIC by selecting a parsimonious model with comparable prediction accuracy as AIC.
\begin{table}
\caption{\label{table:GLM}Prediction comparison among ELBO, AIC and BIC in probit regression.}
\centering
\footnotesize
 \hspace{-.2cm}
 \begin{adjustbox}{max width=\textwidth}
\begin{tabular}{|c|c|c|c|c|c|c|c|c|c|c|}
\hline
\multirow{2}{*}{$r$} &
  \multirow{2}{*}{$n$} &
  \multicolumn{3}{c|}{classification error in \%} &
  \multicolumn{3}{c|}{logistic loss} &
  \multicolumn{3}{c|}{average model size} \\ \cline{3-11} 
   &  & ELBO & AIC & BIC & ELBO & AIC & BIC & ELBO & AIC & BIC  \\ \hline
\multirow{3}{*}{0.2} & 200 & 21.5 (1.6) & 20.7 (1.6) & 24.1 (2.7) & 0.421 & 0.430 & 0.451 & 5.2 (0.9) & 7.8 (1.6) & 3.6 (0.9)\\ \cline{2-11} 
&  500  & 18.8 (0.7) & 18.5 (0.8) & 20.0 (1.1) & 0.363 & 0.366 & 0.381 & 7.4 (0.9) & 9.8 (1.5) & 5.7 (0.8)\\ \cline{2-11} 
& 1000 & 18.0 (0.5) & 17.8 (0.5) & 18.7 (0.6) & 0.346 & 0.346 & 0.356 & 8.7 (0.9) & 11.2 (1.4) & 7.2 (0.8)  \\ \hline
\multirow{3}{*}{0.8} & 200 & 13.3 (1.0) & 13.0 (1.0) & 14.6 (1.5) & 0.276 & 0.282 & 0.289 & 3.8 (0.9) & 4.8 (1.2) & 2.7 (0.6) \\ \cline{2-11} 
&  500  & 11.4 (0.6) & 11.2  (0.7) & 12.5 (0.9) & 0.229 & 0.227 & 0.248 & 5.5 (0.9) & 6.9 (1.2) & 4.0 (0.6) \\ \cline{2-11} 
& 1000 & 10.7 (0.5) & 10.5 (0.5) & 11.3 (0.5) & 0.215 & 0.211 & 0.225 & 7.0 (0.9) & 8.6 (1.0) & 5.4 (0.8)  \\ \hline
\end{tabular}
\end{adjustbox}
\end{table}

\begin{figure}[htb]
    \centering
    \includegraphics[width=15cm]{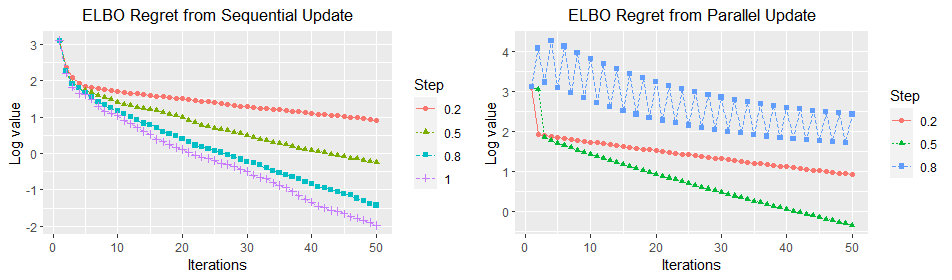}
    \caption{ELBO convergence in probit regression under sequential and parallel updating schemes.}
    \label{probit_elbo_thm3}
\end{figure}

Lastly, we assess the algorithmic convergence of the two CAVI algorithms for implementing MFVI. We take the true model with $n=100$ samples and $d_{\cM_0}=10$ features with $\bbeta^*=0.1\cdot \bm{1}_{10}$, and feature covariance matrix $\Sigma=(\sigma_{ij})_{i,j=1}^p$ with $\sigma_{ij}=0.9+0.1\times 1(i=j)$. The prior standard deviation is $\sigma_0=1$ and $\bbeta$ initialized at $\bbeta^{(0)}=\bm{0}_{10}$.
Fig.~\ref{probit_elbo_thm3} reports the logarithm of ELBO regret with different step sizes from parallel and randomized sequential updates. 
As we can see, the sequential update is always convergent and has faster convergence rate with a larger step size; moreover, the log ELBO regret has a linear decay, imply the geometric convergence of ELBO.  On the contrary, the parallel update only converges under smaller step sizes --- the ELBO converges extremely slow due to oscillations or even diverges to $\infty$ under larger or full step size (hence not presented), and $\gamma=0.8$ tends to be the threshold under which CAVI converges fast without oscillations. Consistent with our theory, the parallel CAVI has geometric convergence for small step sizes $\gamma=0.2$ and $0.5$. In addition, for both updating schemes, the slope (equal to $\log\alpha$) of the log regret is roughly proportional to the step size in absolute value (except for $\gamma=0.8$ in the parallel update).


\subsection{Community Detection}
\label{sec::community}
In this example, we consider the model selection problem of selecting the number of communities in the community detection described in Appendix~\ref{sec::mvteg}.
We set the true underlying data generating stochastic block model (SBM) to have $5$ communities. The five communities have equal number of members. The connectivity matrix $\bm B$ is generated as $\bm B_{ab}\overset{iid}{\sim} \mbox{Unif}(0,0.4)$ if $i\neq j$ and $\bm B_{aa} = 0.6$. For candidate models, we consider $K=2,\ldots,10$ communities. We take the priors on $\bm B_{ab}$ as Beta(1,1) and on $Z_i$ as Categorical($1/K,\ldots,1/K$). We calculate the ELBO using the parallel update in CAVI with full step size, since we find that the algorithm always converges. In terms of BIC, since we cannot find the exact MLE, we approximate MLE by the variational mean from $\widehat q_\theta$ and obtain the BIC correspondingly. 

\begin{figure}[htb]
    \centering
    \includegraphics[width=15cm]{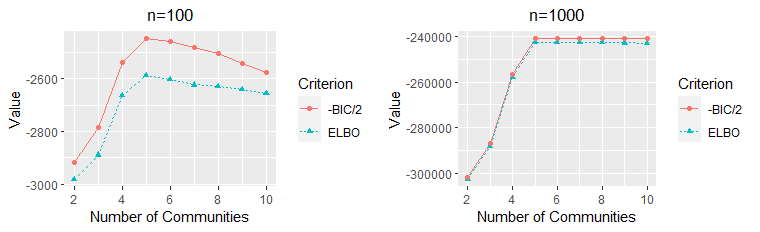}
    \caption{ELBO and BIC from parallel CAVI in SBM.}
    \label{community_models}
\end{figure}

For SBM the Fisher information is infeasible as is for MLE. However, we can visually find $\widetilde C^*(\cM)$, i.e. the gap between -BIC/2 and ELBO via the numerical differences. Note that it is no longer a ``constant'' gap since we have growing number of parameters in SBM. We have included the ELBO and BIC values obtained with $n=100$ and 1000 nodes in Fig.~\ref{community_models}. Both criteria select the true model under both settings, but the gap between two criteria is comparable to the gap between different models, particularly for the overfitted model. This is due to the growing number of parameters.

\begin{figure}[htb]
    \centering
    \includegraphics[width=15cm]{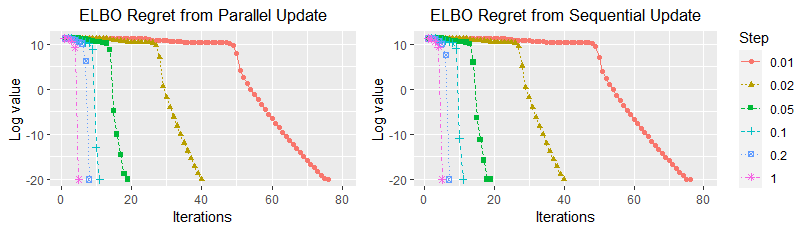}
    \caption{ELBO and BIC from parallel CAVI in SBM with $n=1000$.}
    \label{community_elbo_conv}
\end{figure}

We plot in Fig.~\ref{community_elbo_conv} the regret of ELBO versus iteration counts from both parallel and sequential CAVI algorithms under the true model, with different step sizes. 
We can see that two updating regimes have overall similar performance. The ELBO decays slowly in the first several iterations, and converges exponentially fast afterwards. The convergence speed (slope) is roughly proportional to the step size and the two updating regimes have almost identical speed under small step sizes. This is consistent with our findings in Theorem \ref{thm::caviconv} that CAVI has geometric convergence given a good initialization, as the ELBO in SBM starts to converge exponentially fast after entering the local contraction basin around the true parameter.


\subsection{Real Data Application}
In this section, we apply and compare the three model selection criteria, namely ELBO, AIC and BIC, in four classification datasets: Adult, Taiwan Company Bankruptcy (TRB), Musk and Car Insurance Claims (CIC). The first three are publicly available on the UCI machine learning repository, and the CIC data is available at kaggle.

The original Adult dataset contains 14 features, including 6 continuous and 8 categorical ones. Instead, we use the pre-processed version on the LibSVM repository containing 123 binary features and 32,561 samples. The response  variable is a binary one encoding whether the individual has salary $\ge$50K.
The TRB dataset contains 95 features and 6,819 samples. We remove one constant feature and use the remaining to predict whether the company will go bankruptcy,  based on the business regulations of the Taiwan Stock Exchange. Since the data is highly unbalanced as only 220 data points have response one (referring to bankruptcy), we repeat each of them for additional 10 times so the derived dataset contains 9,019 samples. This is equivalent to a weighted probit regression.
For the Musk dataset about molecules, we use its second version that contains 6598 samples and 166 features. The goal is to predict whether the molecules are musks or not.
The original CIC dataset contains 10,000 samples and 17 features. We remove the samples containing missing values and apply one-hot encoding to the categorical features, and the pre-processed data contains 8,149 samples and 26  features.

 We compare the prediction performance in terms of classification error, logistic loss and selected model size, similar to the numerical study in Section~\ref{sec::sim_glm}. For each dataset, we randomly select $1000$ samples as the training data, which is used to conduct model selection with ELBO, AIC and BIC. The selected models are fitted using the same $1000$ training data and evaluated on the remaining data as the test data. This procedure is repeated $100$ times, and we report the average classification errors and median logistic losses, and also the average sizes of the models selected by different criteria. As we can see from Table~\ref{table:real_data}, AIC has the overall best performance in terms of prediction as expected, which is consistent with the optimality of AIC in prediction accuracy \citep{yang2005can}. On the other hand, the performance of the proposed ELBO criterion tends to be in the between of AIC and BIC, with slightly worse prediction performance than AIC but better than BIC. The model size selected by ELBO also lies between AIC and BIC.
\begin{table} 
\caption{\label{table:real_data}Comparison between ELBO, AIC and BIC in real data applications.}
\centering
\footnotesize
 \hspace{-.2cm}
 \begin{adjustbox}{max width=\textwidth}
\begin{tabular}{|c|c|c|c|c|c|c|c|c|c|}
\hline
  \multirow{2}{*}{Data} &
  \multicolumn{3}{c|}{classification error in \%} &
  \multicolumn{3}{c|}{logistic loss} &
  \multicolumn{3}{c|}{average model size} \\ \cline{2-10} 
   & ELBO & AIC & BIC & ELBO & AIC & BIC & ELBO & AIC & BIC  \\ \hline
  Adult & 18.6 (1.3) & 18.2 (0.8) & 20.0 (1.9) & 0.369 & 0.366 & 0.384 & 7.0 (1.3) & 11.0 (1.8) & 3.7 (0.8) \\ \hline 
  TRB & 14.9 (0.7) & 14.4 (0.7) & 15.5 (0.8) & 0.328 & 0.353 & 0.317 & 5.8 (1.7) & 8.7 (1.9) & 3.4 (0.7) \\ \hline
  Musk & 10.2 (1.1) & 8.6 (0.7) & 12.2 (1.3) & 0.265 & 0.243 & 0.290 & 10.6 (3.1) & 17.7 (3.5) & 5.5 (1.3) \\ \hline
  CIC & 17.0 (0.8) & 16.9 (0.6) & 17.9 (1.6) & 0.333 & 0.330 & 0.349 & 10.2 (1.2) & 11.0 (1.1) & 8.0 (1.0) \\ \hline
 \end{tabular}
  \end{adjustbox}
\end{table}

\section{Conclusion}
\label{sec::conclude}
In summary, we proved a non-asymptotic BvM-type theorem for the MF variational distribution, stating that the variational distribution concentrates to a normal distribution centered at MLE with diagonal covariance matrix. Motivated by this normal approximation, we proposed a model selection criterion based on evidence lower bound (ELBO), and studied the model selection consistency. We found that BIC, ELBO and evidence tend to differ from each other by some constant gaps; and ELBO tends to provide a better approximation to evidence than BIC due to a better dimension dependence and the full incorporation of the prior information. To characterize prediction efficiency, we also proved an oracle inequality for the variational distribution from the model selected by ELBO. Lastly, we showed the geometric convergence of two commonly used CAVI algorithms for solving the optimization problem in MF approximation; and the analysis suggested that it suffices to run $O(d\log(nd))$ iterations of CAVI algorithm to approximate the ELBO in order to achieve model selection consistency.

Next, we discuss some possible future directions.
First, note that we have assumed the non-singularity of the Fisher information to develop the theory. 
It would be interesting to study the model selection based on ELBO for singular models, such as the over-specified Gaussian mixture model numerically examined in Section~\ref{sec:sim_GMM}, and study the connection between ELBO and the so-called singular BIC as discussed in \cite{drton2017bayesian} and \cite{watanabe2010asymptotic}. Although our current theory on the consistent estimation (up to a constant) of model evidence using ELBO applies only to under-specified GMM with a non-singular Fisher information, our numerical results for GMM suggest that the ELBO is still capable of identifying the true model. On the theoretical front, for a singular model $\mathcal{M}$, it is proved (see, for instance, ~\cite{drton2017bayesian} and ~\cite{watanabe2010asymptotic}) that the model evidence $\log p(X^n | \mathcal{M})$ can be approximated by the singular BIC, or $\text{sBIC}(\mathcal M) = -\widehat \ell_n(\mathcal M) + \lambda (\mathcal M) \log n - [m(\mathcal M) - 1] \log\log n$ up to a bounded constant. Here, $\lambda(\mathcal M)$ is called the learning rate or real log-canonical threshold of model $\mathcal{M}$ which summarizes the effective model dimensionality, and $m(\mathcal M)$ represents the multiplicity of $\lambda(\mathcal M)$. For regular models with invertible Fisher information matrices, $\lambda(\mathcal M)$ simplifies to $d_{\mathcal M}/2$ and $m(\mathcal M)=1$. As such, sBIC reduces to half of the usual BIC defined in~\eqref{eqn:BIC}. Recent work~\cite{bhattacharya2020evidence} demonstrates that the ELBO from mean-field variational inference accurately captures the leading $\lambda(\mathcal M)\log n$ term for any singular model in its canonical or normal-crossing configuration. However, while Hironaka’s theorem in algebraic geometry guarantees the existence of an analytical map (or reparametrization map) for any singular model that can locally transform the posterior distribution into a normal-crossing form, it is typically challenging to explicitly construct such an analytical map. In fact, the determination of $\lambda(\mathcal M)$ and $m(\mathcal M)$ for a generic singular model $\mathcal M$ is still an unresolved issue. As such, investigating whether the ELBO from MF variational inference can recover the leading $\lambda(\mathcal M) \log n$ term for a general singular model not in its normal-crossing from is an interesting question to explore.
Second, in more complicated models, the latent variables and sample points may not be one-on-one. For example, if we treat the assignment variables as latent variables instead of unknown parameters in mixed membership community models, then we have $n$ latent variables and $\binom{n}{2}$ data. Our theories do not cover these settings and we also leave them to future study.
Finally, it would be interesting to extend the current development to models with dependent latent variables, such as hidden Markov models or state space models.

\section*{Acknowledgement}
We are grateful to two anonymous reviewers for their constructive comments, which have led to substantial improvements of this paper. Y. Yang's research was supported in part by NSF DMS-2210717.

\bibliographystyle{rss}
\bibliography{ref}

\begin{thebibliography}{59}
\expandafter\ifx\csname natexlab\endcsname\relax\def\natexlab#1{#1}\fi
\expandafter\ifx\csname url\endcsname\relax
  \def\url#1{\texttt{#1}}\fi
\expandafter\ifx\csname urlprefix\endcsname\relax\def\urlprefix{URL: }\fi

\bibitem[{Alquier and Ridgway(2020)}]{alquier2020concentration}
Alquier, P. and Ridgway, J. (2020) Concentration of tempered posteriors and of their variational approximations.
\newblock \textit{The Annals of Statistics}, \textbf{48}, 1475--1497.

\bibitem[{Alquier et~al.(2016)Alquier, Ridgway and Chopin}]{alquier2016properties}
Alquier, P., Ridgway, J. and Chopin, N. (2016) On the properties of variational approximations of {Gibbs} posteriors.
\newblock \textit{The Journal of Machine Learning Research}, \textbf{17}, 8374--8414.

\bibitem[{Bhattacharya et~al.(2015)Bhattacharya, Pati, Pillai and Dunson}]{bhattacharya2015dirichlet}
Bhattacharya, A., Pati, D., Pillai, N.~S. and Dunson, D.~B. (2015) Dirichlet--{Laplace} priors for optimal shrinkage.
\newblock \textit{Journal of the American Statistical Association}, \textbf{110}, 1479--1490.

\bibitem[{Bhattacharya et~al.(2020)Bhattacharya, Pati and Plummer}]{bhattacharya2020evidence}
Bhattacharya, A., Pati, D. and Plummer, S. (2020) Evidence bounds in singular models: probabilistic and variational perspectives.
\newblock \textit{arXiv preprint arXiv:2008.04537}.

\bibitem[{Bhattacharya et~al.(2019)Bhattacharya, Pati and Yang}]{10.1214/18-AOS1712}
Bhattacharya, A., Pati, D. and Yang, Y. (2019) {Bayesian fractional posteriors}.
\newblock \textit{The Annals of Statistics}, \textbf{47}, 39 -- 66.

\bibitem[{Bickel et~al.(2013)Bickel, Choi, Chang and Zhang}]{bickel2013asymptotic}
Bickel, P., Choi, D., Chang, X. and Zhang, H. (2013) Asymptotic normality of maximum likelihood and its variational approximation for stochastic blockmodels.
\newblock \textit{The Annals of Statistics}, \textbf{41}, 1922--1943.

\bibitem[{Bishop(2006{\natexlab{a}})}]{bishop2006pattern}
Bishop, C.~M. (2006{\natexlab{a}}) Pattern recognition.
\newblock \textit{Machine learning}, \textbf{128}.

\bibitem[{Bishop(2006{\natexlab{b}})}]{bishop2006}
--- (2006{\natexlab{b}}) \textit{Pattern recognition and machine learning}.
\newblock springer.

\bibitem[{Blei et~al.(2017)Blei, Kucukelbir and McAuliffe}]{blei2017variational}
Blei, D.~M., Kucukelbir, A. and McAuliffe, J.~D. (2017) Variational inference: {A} review for statisticians.
\newblock \textit{Journal of the American statistical Association}, \textbf{112}, 859--877.

\bibitem[{Chen(1995)}]{chen1995optimal}
Chen, J. (1995) Optimal rate of convergence for finite mixture models.
\newblock \textit{The Annals of Statistics}, 221--233.

\bibitem[{Chen and Chen(2008)}]{chen2008}
Chen, J. and Chen, Z. (2008) Extended {Bayesian} information criteria for model selection with large model spaces.
\newblock \textit{Biometrika}, \textbf{95}, 759--771.

\bibitem[{Ch{\'e}rief-Abdellatif(2019)}]{cherief2019consistency}
Ch{\'e}rief-Abdellatif, B.-E. (2019) Consistency of {ELBO} maximization for model selection.
\newblock In \textit{Symposium on Advances in Approximate Bayesian Inference}, 11--31. PMLR.

\bibitem[{Ch{\'e}rief-Abdellatif and Alquier(2018)}]{cherief2018consistency}
Ch{\'e}rief-Abdellatif, B.-E. and Alquier, P. (2018) Consistency of variational {Bayes} inference for estimation and model selection in mixtures.
\newblock \textit{Electronic Journal of Statistics}, \textbf{12}, 2995--3035.

\bibitem[{Dempster et~al.(1977)Dempster, Laird and Rubin}]{dempster1977maximum}
Dempster, A.~P., Laird, N.~M. and Rubin, D.~B. (1977) Maximum likelihood from incomplete data via the {EM} algorithm.
\newblock \textit{Journal of the Royal Statistical Society: Series B (Methodological)}, \textbf{39}, 1--22.

\bibitem[{Drton et~al.(2017)Drton, Plummer, Claeskens, Rousseau, Robert, Imai, Kuroki, Friel, McKeone, Oates et~al.}]{drton2017bayesian}
Drton, M., Plummer, M., Claeskens, G., Rousseau, J., Robert, C.~P., Imai, T., Kuroki, M., Friel, N., McKeone, J., Oates, C. et~al. (2017) A {Bayesian} information criterion for singular models.
\newblock \textit{Journal of the Royal Statistical Society. Series B (Statistical Methodology)}, 323--380.

\bibitem[{Dwivedi et~al.(2020)Dwivedi, Ho, Khamaru, Wainwright, Jordan and Yu}]{dwivedi2020singularity}
Dwivedi, R., Ho, N., Khamaru, K., Wainwright, M.~J., Jordan, M.~I. and Yu, B. (2020) Singularity, misspecification and the convergence rate of {EM}.
\newblock \textit{The Annals of Statistics}, \textbf{48}, 3161--3182.

\bibitem[{Gelfand and Smith(1990)}]{gelfand1990sampling}
Gelfand, A.~E. and Smith, A.~F. (1990) Sampling-based approaches to calculating marginal densities.
\newblock \textit{Journal of the American statistical association}, \textbf{85}, 398--409.

\bibitem[{Ghosal et~al.(2000)Ghosal, Ghosh and Van Der~Vaart}]{ghosal2000convergence}
Ghosal, S., Ghosh, J.~K. and Van Der~Vaart, A.~W. (2000) Convergence rates of posterior distributions.
\newblock \textit{Annals of Statistics}, 500--531.

\bibitem[{Ghosal and Van Der~Vaart(2007)}]{ghosal2007convergence}
Ghosal, S. and Van Der~Vaart, A. (2007) Convergence rates of posterior distributions for noniid observations.
\newblock \textit{The Annals of Statistics}, \textbf{35}, 192--223.

\bibitem[{Hall et~al.(2011{\natexlab{a}})Hall, Ormerod and Wand}]{hall2011theory}
Hall, P., Ormerod, J.~T. and Wand, M.~P. (2011{\natexlab{a}}) Theory of {Gaussian} variational approximation for a {Poisson} mixed model.
\newblock \textit{Statistica Sinica}, 369--389.

\bibitem[{Hall et~al.(2011{\natexlab{b}})Hall, Pham, Wand and Wang}]{hall2011asymptotic}
Hall, P., Pham, T., Wand, M.~P. and Wang, S.~S. (2011{\natexlab{b}}) Asymptotic normality and valid inference for {Gaussian} variational approximation.
\newblock \textit{The Annals of Statistics}, \textbf{39}, 2502--2532.

\bibitem[{Hammersley(2013)}]{hammersley2013monte}
Hammersley, J. (2013) \textit{Monte {Carlo} methods}.
\newblock Springer Science \& Business Media.

\bibitem[{Han and Yang(2019)}]{wei2019}
Han, W. and Yang, Y. (2019) Statistical inference in mean-field variational {Bayes}.
\newblock \textit{arXiv preprint arXiv:1911.01525}.

\bibitem[{Hastings(1970)}]{hastings1970monte}
Hastings, W.~K. (1970) Monte {Carlo} sampling methods using {Markov} chains and their applications.

\bibitem[{Hirotugu(1974)}]{aic1974}
Hirotugu, A. (1974) A new look at the statistical model identification.
\newblock \textit{IEEE Transactions on Automatic Control}, \textbf{19}, 716--723.

\bibitem[{Ho and Nguyen(2019)}]{ho2019singularity}
Ho, N. and Nguyen, X. (2019) Singularity structures and impacts on parameter estimation in finite mixtures of distributions.
\newblock \textit{SIAM Journal on Mathematics of Data Science}, \textbf{1}, 730--758.

\bibitem[{Ishwaran and Rao(2005)}]{ishwaran2005spike}
Ishwaran, H. and Rao, J.~S. (2005) Spike and slab variable selection: frequentist and {Bayesian} strategies.
\newblock \textit{The Annals of Statistics}, \textbf{33}, 730--773.

\bibitem[{Jain et~al.(2018)Jain, Koehler and Mossel}]{jain2018mean}
Jain, V., Koehler, F. and Mossel, E. (2018) The mean-field approximation: {Information} inequalities, algorithms, and complexity.
\newblock In \textit{Conference On Learning Theory}, 1326--1347. PMLR.

\bibitem[{Jordan et~al.(1999)Jordan, Ghahramani, Jaakkola and Saul}]{jordan1999introduction}
Jordan, M.~I., Ghahramani, Z., Jaakkola, T.~S. and Saul, L.~K. (1999) An introduction to variational methods for graphical models.
\newblock \textit{Machine learning}, \textbf{37}, 183--233.

\bibitem[{Kleijn et~al.(2012)Kleijn, Van~der Vaart et~al.}]{kle2012}
Kleijn, B. J.~K., Van~der Vaart, A.~W. et~al. (2012) The {Bernstein-von-Mises} theorem under misspecification.
\newblock \textit{Electronic Journal of Statistics}, \textbf{6}, 354--381.

\bibitem[{Koehler(2019)}]{koehler2019fast}
Koehler, F. (2019) Fast convergence of belief propagation to global optima: {Beyond} correlation decay.
\newblock \textit{Advances in Neural Information Processing Systems}, \textbf{32}.

\bibitem[{Mukherjee et~al.(2018)Mukherjee, Sarkar, Wang and Yan}]{mukherjee2018mean}
Mukherjee, S.~S., Sarkar, P., Wang, Y. and Yan, B. (2018) Mean field for the stochastic blockmodel: {Optimization} landscape and convergence issues.
\newblock \textit{Advances in neural information processing systems}, \textbf{31}.

\bibitem[{Ohn and Lin(2021)}]{ohn2021adaptive}
Ohn, I. and Lin, L. (2021) Adaptive variational {Bayes}: {Optimality}, computation and applications.
\newblock \textit{arXiv preprint arXiv:2109.03204}.

\bibitem[{Ormerod and Wand(2012)}]{ormerod2012gaussian}
Ormerod, J.~T. and Wand, M.~P. (2012) Gaussian variational approximate inference for generalized linear mixed models.
\newblock \textit{Journal of Computational and Graphical Statistics}, \textbf{21}, 2--17.

\bibitem[{Pati et~al.(2018)Pati, Bhattacharya and Yang}]{pati2018statistical}
Pati, D., Bhattacharya, A. and Yang, Y. (2018) On statistical optimality of variational {Bayes}.
\newblock In \textit{International Conference on Artificial Intelligence and Statistics}, 1579--1588. PMLR.

\bibitem[{Plummer et~al.(2020)Plummer, Pati and Bhattacharya}]{plummer2020dynamics}
Plummer, S., Pati, D. and Bhattacharya, A. (2020) Dynamics of coordinate ascent variational inference: {A} case study in 2d {Ising} models.
\newblock \textit{Entropy}, \textbf{22}, 1263.

\bibitem[{Saeedi et~al.(2017)Saeedi, Kulkarni, Mansinghka and Gershman}]{saeedi2017variational}
Saeedi, A., Kulkarni, T.~D., Mansinghka, V.~K. and Gershman, S.~J. (2017) Variational particle approximations.
\newblock \textit{The Journal of Machine Learning Research}, \textbf{18}, 2328--2356.

\bibitem[{Sarkar et~al.(2021)Sarkar, Wang and Mukherjee}]{sarkar2021random}
Sarkar, P., Wang, Y.~R. and Mukherjee, S.~S. (2021) When random initializations help: a study of variational inference for community detection.
\newblock \textit{J. Mach. Learn. Res.}, \textbf{22}, 22--1.

\bibitem[{Schwarz et~al.(1978)}]{bic1978}
Schwarz, G. et~al. (1978) Estimating the dimension of a model.
\newblock \textit{The Annals of Statistics}, \textbf{6}, 461--464.

\bibitem[{Shen and Wasserman(2001)}]{shen2001rates}
Shen, X. and Wasserman, L. (2001) Rates of convergence of posterior distributions.
\newblock \textit{The Annals of Statistics}, \textbf{29}, 687--714.

\bibitem[{Stoica and Selen(2004)}]{stoica2004model}
Stoica, P. and Selen, Y. (2004) Model-order selection: a review of information criterion rules.
\newblock \textit{IEEE Signal Processing Magazine}, \textbf{21}, 36--47.

\bibitem[{Titsias and L{\'a}zaro-Gredilla(2014)}]{titsias2014doubly}
Titsias, M. and L{\'a}zaro-Gredilla, M. (2014) Doubly stochastic variational {Bayes} for non-conjugate inference.
\newblock In \textit{International conference on machine learning}, 1971--1979. PMLR.

\bibitem[{Titterington and Wang(2006)}]{titterington2006convergence}
Titterington, D. and Wang, B. (2006) Convergence properties of a general algorithm for calculating variational {Bayesian} estimates for a normal mixture model.
\newblock \textit{Bayesian Analysis}, \textbf{1}, 625--650.

\bibitem[{Trefethen and Bau~III(1997)}]{trefethen1997numerical}
Trefethen, L.~N. and Bau~III, D. (1997) \textit{Numerical linear algebra}, vol.~50.
\newblock Siam.

\bibitem[{Van~der Vaart(2000)}]{van2000}
Van~der Vaart, A.~W. (2000) \textit{Asymptotic statistics}, vol.~3.
\newblock Cambridge university press.

\bibitem[{Wang and Titterington(2005)}]{wang2005inadequacy}
Wang, B. and Titterington, D.~M. (2005) Inadequacy of interval estimates corresponding to variational {Bayesian} approximations.
\newblock In \textit{International Workshop on Artificial Intelligence and Statistics}, 373--380. PMLR.

\bibitem[{Wang and Blei(2019{\natexlab{a}})}]{wang2019variational}
Wang, Y. and Blei, D. (2019{\natexlab{a}}) Variational {Bayes} under model misspecification.
\newblock \textit{Advances in Neural Information Processing Systems}, \textbf{32}, 13357--13367.

\bibitem[{Wang and Blei(2019{\natexlab{b}})}]{wang2019}
Wang, Y. and Blei, D.~M. (2019{\natexlab{b}}) Frequentist consistency of variational {Bayes}.
\newblock \textit{Journal of the American Statistical Association}, \textbf{114}, 1147--1161.

\bibitem[{Wang et~al.(2021)Wang, Chen, Liu and Kang}]{wang2021particle}
Wang, Y., Chen, J., Liu, C. and Kang, L. (2021) Particle-based energetic variational inference.
\newblock \textit{Statistics and Computing}, \textbf{31}, 1--17.

\bibitem[{Watanabe and Opper(2010)}]{watanabe2010asymptotic}
Watanabe, S. and Opper, M. (2010) Asymptotic equivalence of {Bayes} cross validation and widely applicable information criterion in singular learning theory.
\newblock \textit{Journal of machine learning research}, \textbf{11}.

\bibitem[{Westling and McCormick(2015)}]{westling2015establishing}
Westling, T. and McCormick, T.~H. (2015) Establishing consistency and improving uncertainty estimates of variational inference through {M}-estimation.
\newblock \textit{arXiv preprint arXiv:1510.08151}, \textbf{1}.

\bibitem[{Wright(2015)}]{wright2015coordinate}
Wright, S.~J. (2015) Coordinate descent algorithms.
\newblock \textit{Mathematical Programming}, \textbf{151}, 3--34.

\bibitem[{Yang(2005)}]{yang2005can}
Yang, Y. (2005) Can the strengths of {AIC} and {BIC} be shared? {A} conflict between model indentification and regression estimation.
\newblock \textit{Biometrika}, \textbf{92}, 937--950.

\bibitem[{Yang et~al.(2020)Yang, Pati and Bhattacharya}]{yang2020alpha}
Yang, Y., Pati, D. and Bhattacharya, A. (2020) $\alpha$-variational inference with statistical guarantees.
\newblock \textit{The Annals of Statistics}, \textbf{48}, 886--905.

\bibitem[{Yang et~al.(2016{\natexlab{a}})Yang, Wainwright and Jordan}]{yang2016computational}
Yang, Y., Wainwright, M.~J. and Jordan, M.~I. (2016{\natexlab{a}}) On the computational complexity of high-dimensional {Bayesian} variable selection.
\newblock \textit{The Annals of Statistics}, \textbf{44}, 2497--2532.

\bibitem[{Yang et~al.(2016{\natexlab{b}})Yang, Wainwright, Jordan et~al.}]{yang2016annal}
Yang, Y., Wainwright, M.~J., Jordan, M.~I. et~al. (2016{\natexlab{b}}) On the computational complexity of high-dimensional {Bayesian} variable selection.
\newblock \textit{The Annals of Statistics}, \textbf{44}, 2497--2532.

\bibitem[{Yin et~al.(2020)Yin, Wang and Sarkar}]{yin2020theoretical}
Yin, M., Wang, Y.~R. and Sarkar, P. (2020) A theoretical case study of structured variational inference for community detection.
\newblock In \textit{International Conference on Artificial Intelligence and Statistics}, 3750--3761. PMLR.

\bibitem[{Zhang and Zhou(2020)}]{zhang2020theoretical}
Zhang, A.~Y. and Zhou, H.~H. (2020) Theoretical and computational guarantees of mean field variational inference for community detection.
\newblock \textit{The Annals of Statistics}, \textbf{48}, 2575--2598.

\bibitem[{Zhang and Gao(2020)}]{zhang2020convergence}
Zhang, F. and Gao, C. (2020) Convergence rates of variational posterior distributions.
\newblock \textit{The Annals of Statistics}, \textbf{48}, 2180--2207.

\end{thebibliography}

\newpage
\pagenumbering{arabic}
\setcounter{page}{1}
\appendix
\begin{center}
{\bf\Large {\bf\large Supplement to ``Bayesian Model Selection via Mean-Field Variational Approximation''}}
\end{center}
Appendix A describes a number of motivating examples in Bayesian statistics where we apply variational inference to approximate the posterior with closed form updating formulas. Appendix B applies the theory from Section~\ref{sec::theory} to some of the motivating examples in Appendix A and discusses the consequences. Appendix C discuss extensions of the BIC and the proposed ELBO based model selection criterion to high-dimensional problems.
Appendix D includes the proofs of the main results in the paper. The rest and technique results are left to Appendix E. To ease the notation in the proofs, we use $C$ to denote a generic constant whose value may vary line by line; we may also omit the subscript $\cM$ that refers to the model candidate, when there is no ambiguity.

\section{Motivating Examples}
\label{sec::mvteg}
In this section, we present several representative examples. For the CAVI algorithm, our discussion primarily centers on a full step size of $\gamma=1$. Extensions involving different step sizes can be formulated in a straightforward manner.

\smallskip
\noindent {\bf Location-scale normal distribution.}
In the first example, we consider Bayesian inference on the mean and variance parameters in the location-scale normal distribution family. Suppose we have i.i.d.~samples $X^n=(X_1,\ldots,X_n)$ from $N(\mu,\sigma^2)$, with priors on parameter $\theta=(\mu,\sigma^2)$ given by
$$
\mu\sim N(\mu_0,\sigma_0^2),\quad \sigma^2\sim\mbox{IG}(a,b),
$$
where $\mbox{IG}(a,b)$ stands for the inverse Gamma distribution with shape parameter $a$ and rate parameter $b$.
We approximate the posterior $p(\mu,\sigma^2\mid X)$ by a factorized distribution with form
$$
q(\mu,\sigma^2)=q_\mu(\mu)\, q_{\sigma^2}(\sigma^2).
$$
It can be shown that the optimal densities $q_{\mu}^\ast$ and $q_{\sigma^2}^\ast$ within the CAVI iterations taking the forms as 
$$
\begin{aligned}
q_\mu^\ast(\mu \mid \sigma^2, X)& \sim N\left(\frac{n\bar{X} \E_q[\sigma^{-2}]+\mu_0 \sigma_0^{-2}}{n\E_q[\sigma^{-2}]+\sigma_0^{-2}}, \frac{1}{n\E_q[\sigma^{-2}]+\sigma_0^{-2}}\right); \\
q_{\sigma^2}^\ast(\sigma^{2} \mid \mu, X)&\sim \mathrm{IG}\left(a+\frac{n}{2}, b+\frac{1}{2}\Big(\|X-\E_q[\mu] \mathbf{1}_{n}\|^2+n\var_q(\mu)\Big)\right).
\end{aligned}
$$
Since the optimal distribution follows $\widehat q_\mu\sim N(m,s^2)$ and $\widehat q_{\sigma^2}\sim \mbox{IG}(A,B)$,
we can then apply CAVI algorithm to solve the optimization problem in an iterative manner, by updating $m,s^2$ and $B$ ($A\equiv a+n/2$ is fixed). Since the mean of Gamma($a,b$) equals $a/b$, we have the following updating formulas.
\begin{align*}
s^2=\,&\frac{1}{n(a+n/2)\,/\,B+\sigma_0^{-2}};\\
m=\,&\big(n\overline X(a+n/2)\,/\,B+\mu_0\sigma_0^{-2}\big)s^2;\\
B=\,&b+\frac{1}{2}\Big(\|X-m\mathbf{1}_n\|^2+ns^2\Big).
\end{align*}
We may find the optimal ELBO as
\begin{align*}
  L(\widehat q \,)=&\frac{1}{2}-\frac{n}{2}\log(2\pi)+\frac{1}{2}\log\frac{s^2}{\sigma_0^2}-\frac{(m-\mu_0)^2+s^2}{2\sigma_0^2}\\
  &+a\log b-A\log B+\log\Gamma(A)-\log\Gamma(a). 
\end{align*}

\smallskip
\noindent {\bf Generalized linear model.}
We turn our attention to a more complex example: the generalized linear model (GLM). In GLM, the conditional mean of the response $Y$ given $X\in\R^d$ equals
$$
\E[Y\mid X]=g^{-1}(X^T\bbeta),
$$
where $g$ is the link function. For example, $g(p)=\log \frac{p}{1-p}$ in logistic regression and $g(\mu)=\log\mu$ in Poisson regression.

When $g$ is equal to the identity function, the GLM reduces to ordinary linear regression. For linear regression model, we typically use normal and inverse Gamma priors for mean and variance parameters, which are conjugate in the sense that the posteriors are also from the same family (see the first example). However, we do not have standard conjugate prior for GLM in general, and we may use variational inference to approximate the posterior that can be computationally expensive to sample from.

For concreteness, we consider the probit regression, where $g^{-1}$ equals the cdf of standard normal distribution, denoted by $\Phi$. Therefore, we have
$$
Y_i\mid X_i\overset{i.i.d.}{\sim}\mbox{Ber}(\Phi(X_i^T\bbeta)). 
$$
We impose a normal prior $\bbeta\sim N(\bm{0},\Sigma_0)$. To facilitate computation, it is common to augment the model
by introducing $n$ latent variables $\Z:\,=Z^n=(Z_1,\ldots,Z_n)$ with latent distribution
$$
Z_i\mid \bbeta\overset{i.i.d.}{\sim}N(X_i^T\bbeta,1),
$$
so that $p(Y_i\mid Z_i)=I(Z_i\ge 0)^{Y_i}I(Z_i<0)^{1-Y_i}$. Under the model augmentation, we can write the logarithm of the joint posterior distribution over the parameter-latent pair $(\bbeta, Z^n)$ as
\begin{align*}
\log p(Z^n,\bbeta\,|\,Y^n)=&\sum_{i=1}^n[Y_i\log I(Z_i\ge 0)+(1-Y_i)\log I(Z_i<0)]\\
&-\bbeta^T\Sigma_0^{-1}\bbeta/2-\|\Z-\X\bbeta\|^2/2+const,    
\end{align*}
where $\X$ is the $n$-by-$d$ design matrix whose $i$th row is $X_i$.

With CAVI updating formula (\ref{cavi}), we have
$$
q^*_{Z_i}(Z_i\mid \bbeta)\sim
\begin{cases}
N_+(X_i^T\E_q[\bbeta],1)& Y_i=1,\\
N_-(X_i^T\E_q[\bbeta],1), & Y_i=0,
\end{cases}
$$
where $N_+$ and $N_-$ denote the normal distributions truncated to positive and negative part, respectively. For $\bbeta$, we have
$$
q^*_{\bbeta}(\bbeta\mid Z^n)\sim N([\X^T\X+\Sigma_0^{-1}]^{-1}\X^T\E_q[\Z],[\X^T\X+\Sigma_0^{-1}]^{-1}).
$$
Note that both distributions are normal or truncated normal, with fixed variance. As a result, we may simply update $\bmu_{\bbeta}:\,=\E[\bbeta]$ and $\bmu_{\Z}:\,=\E[\Z]$ as follows.
\begin{align*}
\bmu_{\bbeta}=&(\X^T\X+\Sigma_0^{-1})^{-1}\X^T\bmu_Z,\\
\bmu_{Z_i}=&X_i^T\bmu_{\bbeta}+\frac{\phi(X_i^T\bmu_{\bbeta})}{\Phi(X_i^T\bmu_{\bbeta})^{Y_i}[\Phi(X_i^T\bmu_{\bbeta})-1]^{1-Y_i}},
\end{align*}
where $\phi$ is the pdf of standard normal distribution.
The optimal ELBO has an explicit form as
$$
L(\widehat q \,)=\sum_i[Y_i\log\Phi(\X\bmu_{\bbeta})+(1-Y_i)\log(1-\Phi(\X\bmu_{\bbeta}))]-\frac{1}{2}\bmu_{\bbeta}^T\Sigma_0^{-1}\bmu_{\bbeta}-\frac{1}{2}\log\det(\Sigma_0\X^T\X+I_d).     
$$
As discussed in the previous subsection, one can use this ELBO value as the criterion to conduct variable selection by selecting a subset of variables that maximizes it.

Note that we have used the BMFVI approximation for the probit regression model. We can always implement MFVI, though, by further imposing the independence between parameters. Define $\bm A=\X^T\X+\Sigma_0^{-1}\in\R^{d\times d}$, and $x_j$ denotes $j$th column (feature) of $\X$. In the fully factorized MF, the updating rule of $\bbeta$ in CAVI is given by
$$
q^*_{\beta_j}(\beta_j\mid\bbeta_{-j}, Z^n)\sim N(A_{jj}^{-1}(\bm A_{j,-j}^T\bbeta_{-j}+\mu_{\Z}^Tx_j),A_{jj}^{-1}),
$$
while the distribution of $Z_i$ is updated in the same way as BMFVI.
The optimal ELBO value is then
\begin{align*}
&\sum_i[Y_i\log\Phi(\X\bmu_{\bbeta})+(1-Y_i)\log(1-\Phi(\X\bmu_{\bbeta}))]-\frac{1}{2}\bmu_{\bbeta}^T\Sigma_0^{-1}\bmu_{\bbeta}\\&-\frac{1}{2}\log\det(\Sigma_0)-\frac{1}{2}\log\det(\diag(\X^T\X+\Sigma_0^{-1})). 
\end{align*}
We note that the fully factorized MF would introduce additional approximation error, and hence we suggest using BMFVI for approximating the posterior in practice, whenever such an algorithm is computationally feasible. 

\smallskip
\noindent {\bf Community detection.}
\cite{zhang2020theoretical} apply variational inference to the following stochastic block model (SBM) for communication detection. Consider an undirected and unweighted network with adjacency matrix denoted by $\bm A\in\{0,1\}^{n\times n}$ such that $\bm A=\bm A^T$.
We assume the network to have no self-loops and hence $\bm A_{ii}\equiv 0$. Each edge $\bm A_{ij}$ ($i<j$) follows Bernoulli distribution with probability $\P_{ij}$. In SBM, the connectivity probability $\P_{ij}$ only depends on the communities which the nodes $i$ and $j$ belong to. We assume that there are $k$ communities, and the connectivity probability matrix for them is denoted by $\bm B\in\R^{K\times K}$, where the elements $\bm B_{ab}$ refers to the connectivity probability between communities $a$ and $b$. Let the assignment (latent) variables $Z_i$ denote the community membership ($Z_i=\ell$ if node $i$ belongs to community $\ell$) of node $i$. Then we can rewrite the connectivity probability as $\P_{ij}=\bm B_{Z_iZ_j}$.

In community detection problem, we hope to recover the  assignment variables $Z_i$, with observed adjacency matrix $\bm A$. We can write the distribution of $\bm A$ in terms of the parameters $\bm B$ and the assignment variables $Z_i$ as 
$$
\P(\bm A\mid \Z,\bm B)=\prod_{i<j}\bm B_{Z_iZ_j}^{\bm A_{ij}}(1-\bm B_{Z_iZ_j})^{1-\bm A_{ij}}.
$$
We take the following priors on $Z_i$ and $\bm B$, with $Z_i$ following categorical distribution with hyperparameters $\{\pi_{i\cdot}\}_{a=1}^k$ such that $\sum_{a=1}^k \pi_{i,a}=1$, and $\bm B_{ab}$ from Beta priors such that $\bm B_{ab}\sim$Beta$(\alpha_{ab},\beta_{ab})$ for $1\le a\le b\le K$.  In view of Bayesian framework, the inference on $Z^n$ depends on the posterior distribution $\P(Z^n,\bm B\mid \bm A)$. Unfortunately, the posterior is computationally infeasible due to the difficulty of calculating the marginal probability $p(\bm A)$.

With mean-field variational inference we approximate the posterior $p(Z^n,\bm B\mid \bm A)$ with $q(Z^n,\bm B)=\prod_{i=1}^nq(Z_i)\prod_{1\le a\le b\le K}q_{ab}$. It can be shown that the above priors are conjugate to the SBM, so the variational distribution has form 
$$
q(Z^n,\bm B)=\prod_{i=1}^n q_{\phi_{i\cdot}}(Z_i)\times\prod_{1\le a\le b\le K} q_{\alpha_{ab},\beta_{ab}}(\bm B_{ab}),
$$
where $q_{\phi_{i\cdot}}$ is the cdf of categorical distribution with probabilities $\{\phi_{i,a}\}_{a=1}^k$ and $q_{\alpha,\beta}$ is the cdf of Beta($\alpha,\beta$). The CAVI can be derived accordingly as follows.

Note that the likelihood function is
$$
\begin{aligned}
\P(\bm A\mid Z^n,\bm B)=&\prod_{1\le a, b\le K}
\prod_{i<j}\left[\bm B_{ab}^{\bm A_{ij}}(1-\bm B_{ab})^{1-\bm A_{i j}}\right]^{ Z_{i, a} Z_{j, b}} .
\end{aligned}
$$
Multiplying the above with the prior, we have
$$
\begin{aligned}
\P(\bm A,Z^n,\bm B)=&\prod_{1\le a, b\le K}
\prod_{i<j}\left[\bm B_{ab}^{\bm A_{ij}}(1-\bm B_{ab})^{1-\bm A_{i j}}\right]^{ Z_{i, a} Z_{j, b}} \times\prod_{i=1}^n\pi_{i,Z_i}^{pri}\\
&\times \prod_{1\le a<b\le K}\left[\frac{\Gamma\left(\alpha_{ab}^{\mathrm{pri}}+\beta_{ab}^{\mathrm{pri}}\right)}{\Gamma\left(\alpha_{ab}^{\mathrm{pri}}\right) \Gamma\left(\beta_{ab}^{\mathrm{pri}}\right)} \bm B_{ab}^{\alpha_{ab}^{\mathrm{pri}}-1}(1-\bm B_{ab})^{\beta_{ab}^{\mathrm{pri}}-1}\right].
\end{aligned}
$$
Using the CAVI algorithm, by calculating $p(\bm B\mid Z^n,\bm A)$ and $p(Z^n\mid \bm B,\bm A)$, we have the updating rule for $\bm B_{ab}\sim$ Beta$(\alpha_{ab},\beta_{ab})$ as
$$
\alpha_{ab}=\alpha_{ab}^{pri}+\sum_{i\not=j}\pi_{i,a}\pi_{j,b}\bm A_{ij},\quad \beta_{ab}=\beta_{ab}^{pri}+\sum_{i\not=j}\pi_{i,a}\pi_{j,b}(1-\bm A_{ij}),\quad\forall a\not=b,
$$
and 
$$
\alpha_{aa}=\alpha_{aa}^{pri}+\sum_{i<j}\pi_{i,a}\pi_{j,a}\bm A_{ij},\quad \beta_{aa}=\beta_{aa}^{pri}+\sum_{i<j}\pi_{i,a}\pi_{j,a}(1-\bm A_{ij}),\quad \forall a.
$$
In addition, $Z_i$ follows a categorical distribution with parameter $\{\pi_{i,a}\}_{a=1}^K$, whose updating rule is
\begin{align*}
\pi_{i,a}^{new}\propto\pi_{i,a}^{pri}\exp\left\{\sum_{b=1}^K\sum_{j\not=i}\pi_{j,b}\Big(\bm A_{ij}\E_q[\log\bm B_{ab}]+(1-\bm A_{ij})\E_q[\log(1-\bm B_{ab})\,]\Big)\right\}.
\end{align*}
Correspondingly, the optimal ELBO value can be expressed as
\begin{align*}
L(\widehat q \,)=&\sum_{1\le a\le b\le K }\Bigg[\log\frac{\text{Beta}(\alpha_{ab},\beta_{ab})}{\text{Beta}(\alpha_{ab}^{pri},\beta_{ab}^{pri})}-(\alpha_{ab}-\alpha_{ab}^{pri})\psi(\alpha_{ab})-(\beta_{ab}-\beta_{ab}^{pri})\psi(\beta_{ab})\\
&~~~~~~~~+(\alpha_{ab}+\beta_{ab}-\alpha_{ab}^{pri}-\beta_{ab}^{pri})\psi(\alpha_{ab}+\beta_{ab})\Bigg]+\sum_{1\le i\le n}\sum_{1\le a\le K}\pi_{i,a}\log\frac{\pi_{i,a}^{pri}}{\pi_{i,a}}\\
&+\sum_{1\le i<j\le n}
\sum_{1\le a, b\le k}\pi_{i,a}\pi_{j,b}\big((\psi(\alpha_{ab})-\psi(\beta_{ab}))\bm A_{ij}+\psi(\alpha_{ab})-\psi(\alpha_{ab}+\beta_{ab})\big),
\end{align*}
where $\psi$ denotes the first order derivative of the logarithm of the Gamma function.

\smallskip
\noindent {\bf Gaussian mixture model.}
Our last example is the Gaussian mixture model (GMM), where the underlying data generating model for data $X^n=(X_1,\ldots,X_n)$ is a mixture of $K$ Guassian components, given by $\P_\mu=\sum_{k=1}^K\frac{1}{K}N(\mu_k,1)$, where the parameter vector is $\mu=(\mu_1,\cdots,\mu_k)$. 
To facilitate the computation, it is a common practice of introducing a latent variable $c_i$ for each $X_i$, where $$c_i\sim\operatorname{Categorical}(1/K,\ldots,1/K),$$ 
so that the conditional distribution of $X_i$ given $(c_i,\mu)$ is $$X_i\mid c_i,\mu\sim N(c_i^T\mu,1).$$ 
We impose a common normal prior on the parameters, i.e.~$\mu_k\overset{i.i.d.}{\sim} N(0,\sigma^2), k=1,\ldots,K$. In the mean-field variational inference, we approximate the joint posterior $p(\{c_i\}_{i=1}^n,\mu\mid X^n)$ with a factorized distribution $\prod_{k=1}^Kq_{\mu_k}\prod_{i=1}^nq_{c_i}$. It turns out that the normal prior is a conjugate prior in the sense that the variational distributions of $\mu_k$ and $c_i$ are also normal and categorical, respectively. Therefore, we may write
$$
q_{\mu_k}\sim N(m_k,s_k^2),\quad q_{c_i}\sim \text{Categorical}(\phi_{i1},\ldots,\phi_{iK}),.
$$
for $i=1,\ldots,n$ and $k=1,\ldots,K$.
Due to the conditional conjugacy, the MF approximation can be efficiently computed by optimizing the parameters of these distributions. We again apply the coordinate ascent algorithms to find the variational parameters governing the distributions, whose updating formula is given by
\begin{align*}
\phi_{ik}\propto&\exp\left\{X_im_k-\frac{s_k^2+m_k^2}{2}\right\},\quad
m_k=\frac{\sum_i X_i\phi_{ik}}{\sigma^{-2}+\sum_i\phi_{ik}},\quad \mbox{and}\quad s_k^2=\frac{1}{\sigma^{-2}+\sum_i\phi_{ik}}.
\end{align*}
The optimal ELBO value also admits a closed form expression as
\begin{align*}
L(\widehat q \,)=&\sum_{i,k}X_i\phi_{ik}m_k-\frac{1}{2}\sum_{i,k}\phi_{ij}(m_k^2+s_k^2)-\sum_{i,k}\phi_{ik}\log\phi_{ik}-\frac{1}{2\sigma^2}\sum_k(m_k^2+s_k^2)\\
&+\frac{1}{2}\sum_k\log s_k^2-\frac{n}{2}\log2\pi-\frac{\sum_iX_i^2}{2}-n\log K+\frac{K}{2}(1-\log\sigma^2).
\end{align*}
The ELBO function across iterations can also serve as a convergence monitor, providing a basis for determining a stopping rule.

\section{Applications to Examples}
\label{sec::app_eg}
In this section, we apply Theorem~\ref{thm::translate} to approximate the model evidence using ELBO in various examples. 

\smallskip
\noindent {\bf Generalized linear model.}
We have seen how VI and CAVI can be used to approximate the posteriors arising from generalized linear models (GLM). For the problem of model selection in GLM, there are exponentially (in the total number $p$ of features) many model candidates.
Each model is characterized by the subset of features included in the model. If we use $X_\cM$ to denote the columns of features included in model $\cM$, the GLM for model $\cM$ is given by
$$
\E[Y\,|\,X_\cM]=g^{-1}(X_\cM^T \bbeta_\cM).
$$
For probit regression, we have $Y\,|\,X_\cM\sim \mbox{Ber}(\Phi(X_\cM^T \bbeta_\cM))$, where $\Phi$ is the cdf of the standard normal distribution. The constant gaps $C^*(\cM)$ can be calculated explicitly for each model. Since we have the Fisher's information
$$
V(\tmstar)=\E\left[\frac{\phi^2(X_{1,\cM}^T\bbeta_\cM)}{\Phi(X_{1,\cM}^T\bbeta_\cM)(1-\Phi(X_{1,\cM}^T\bbeta_\cM))}X_{1,\cM}^TX_{1,\cM}\right],
$$
where $X_{i,\cM}$ denotes the $i$-th row (observation) of $X_\cM$. Since $Z_{i,\cM}\,|\,X_{i,\cM},\bbeta\sim N(X_{i,\cM}^T\bbeta,1)$, we have $V_c(\tmstar)=\E[X_{1,\cM}^TX_{1,\cM}]$, and therefore
$$C^*(\cM)=\frac{1}{2}\log\frac{\det(V_c(\tmstar))}{\det(V(\tmstar))},\quad C_{\bic}^*(\cM)=-\frac{1}{2}\log\det\big(\Sigma_0V(\tmstar)\big)-\frac{1}{2}(\bbeta_\cM^*)^T\Sigma_0^{-1}\bbeta_\cM^*.$$
A sufficient condition for the consistency of ELBO is $C^*(\cM)=o(\log n)$ for any model $\cM$, which is true if 
$p\ll n$.

\smallskip
\noindent {\bf Community detection.}
\cite{zhang2020theoretical} study the theoretical properties of VI assuming the number of communities $k$ known. In practice, $k$ is often unknown, and we may apply the ELBO-based criterion to select the parameter $k$. This becomes a model selection problem. Although the MLE is not analytically solvable for the community detection problem, making it impossible to derive a closed-form expression for the Fisher information or the constant gaps predicted by Theorem~\ref{thm::translate}, both ELBO and BIC are numerically computable for different model candidates. We empirically demonstrate through simulation (see Section~\ref{sec::community}) that model selection consistency via ELBO is still achieved.

\smallskip
\noindent {\bf Gaussian mixture model.}
Since the Gaussian mixture model is degenerate under over-parametrization (more components than actually have), we only focus on well-specified model $\cM=\cM_0$ to apply Theorem~\ref{thm::translate}.
Since $X_i\,|\,c_i,\mu\sim N(c_i^T\mu,1)$ and $c_i$ does not depend on $\mu$, we can show that the complete data Fisher information takes the diagonal form as $V_c=\diag(1/K,\ldots,1/K)$.
Therefore, we have
$$
\widetilde C^*(\cM)=\frac{\|\mu\|^2}{2\sigma^2}+\frac{K}{2}(\log\sigma^2-\log K).
$$
However, the Fisher information $V$ of $X_i$ does not take an explicit form, and we have to use Monte Carlo method to approximate it and $C^*(\cM)$. The corresponding results are presented in Section~\ref{sec:sim_GMM}, where the simulated values of $-$BIC$/2-$ELBO closely align with the theoretical values (computed via the Monte Carlo method). Moreover, although Theorem~\ref{thm::translate} does not apply to over-parametrized models, our numerical results suggest that ELBO minimization still selects the true model.

\section{Model Selection in High-Dimensional Problems}\label{sec:high_dim_BIC}
In model selection problems that may evolve a large number of candidate models, such as variable selection in high-dimensional regression, the usual BIC with the $\log n$ penalty no longer leads to consistent model selection. The inconsistency comes from the fact that BIC only approximates the log-(marginal-)likelihood $\log p(X^n\,|\,\cM)$ according to Theorem~\ref{sec::thmmodelselection}; however, the intrinsic penalty on model complexity in Bayesian inference due to the integration over the parameter space (i.e.~the $d_{\cM}\log n$ term) is too weak to rule out all over-sized models and a sparsity inducing prior is critical in controlling the model complexity~\citep{ishwaran2005spike,bhattacharya2015dirichlet,yang2016computational}. In particular, the negative log-prior on each model $\log \pi(\cM)$ effectively serves as the model complexity penalty, leading to the so-called extended BIC \cite[see e.g.][]{chen2008}, defined as follows,
$$
\bic_{\pi}(\cM)=\bic(\cM)-2\log \pi(\cM),
$$
which now approximates negative twice of the log posterior probability $-2\log p(\cM\,|\,X^n)$. For example, in 
 high-dimensional linear regression with $p$ features where $p\gg n$, we may choose $\pi(\cM)\propto p^{-\kappa \, d_{\cM}}$ that exponentially penalizes a model with $d_{\cM}$ variables; and the extra penalty due to the prior becomes $2\kappa\log p\cdot  d_{\cM}$, which imposes a much stronger penalty to the number of parameters than BIC.
Model selection consistency based on the extended BIC can be proved in a number of settings where the usual BIC is not adequate \cite[see e.g.][]{chen2008}. 
Correspondingly, we may consider the following objective that also penalizes ELBO by the prior,
$$
\cO(\cM)=\elbo(\cM)+\log{\pi(\cM)}=L(\widehat q_\cM)+\log{\pi(\cM)}.
$$
The penalization term makes the criterion suitable for more settings in the same manner as the extended BIC by approximating $\log p(\cM\,|\,X^n)$ from below (up to a model independent constant). As a direct consequence of Theorem~\ref{thm::translate}, the following corollary shows that the difference between the prior penalized ELBO is close to the extended BIC up to a constant.
\begin{corollary}\label{col::translate_gamma}
Under the same assumptions as Theorem \ref{thm::translate}, we have
$$
|-\bic_\pi(\cM)/2-\cO(\cM)-\widetilde C^*(\cM)|\leq \frac{C_7 K^5(\log n)^{3}}{\sqrt{n}},
$$
where $\widetilde C^*(\cM)$ is given in \eqref{eq::tildeC}.
\end{corollary}

\section{Proofs of Main Results}
In this section, we present proofs of the main results and some claims made in the paper, while deferring some technical results to the next section.

\subsection{Proof of Lemma \ref{lem::VI_subg}}
Recall that Lemma \ref{lem::post_subg} implies a sub-Gaussian tail bound for the marginal posterior distribution of $\theta$. We will use the optimality of the variational distribution  $\widehat Q=\widehat Q_\theta \otimes \widehat Q_{S^n}$ to show that $\widehat Q_\theta$ satisfies the same bound as well. Define a class of probability measures in the mean-field family,
$$
Q_\lambda=\left.(1-\lambda) \widehat{Q}\right|_{A_{n, \ell}^c}+\left.\lambda \widehat{Q}\right|_{A_{n, \ell}}, \quad \text { for all } \lambda \in[0,1],
$$
where we have defined $Q_A(B)=Q(A\cap B)/Q(A)$ and $A_{n,\ell}=\{\theta: |\theta_\ell-\theta^*_\ell|\ge C'\epsilon\}$ for some constant $C'$ to be determined later. In particular, we have $\widehat Q=Q_{\widehat\lambda}$ for $\widehat\lambda=\widehat Q(A_{n,\ell})$ and hence the optimality of $\widehat Q$ imples $\widehat\lambda=\arg\min_{\lambda\in[0,1]}D(q_\lambda\verts p(\cdot,\cdot\mid X^n))$. Note that we have the following decomposition for the objective function as
\begin{equation}
\begin{aligned}
& D\left(q_\lambda\verts p\left(\cdot,\cdot\mid X^n\right)\right)=\int_{\Theta \times \m S^n} \log \left(\frac{q_\lambda (\theta, s^n)}{p(\theta,s^n \mid X^n)}\right)\left[\left.(1-\lambda) \widehat Q\,\right|_{A_{n, \ell}^c}+\left.\lambda \widehat Q\,\right|_{A_{n, \ell}}\right] (\dd\theta,\dd s^n) \\
=&\left.(1-\lambda) \int_{A_{n, \ell}^c} \log \left(\frac{\left.(1-\lambda) \,\widehat q\,\right|_{A_{n, \ell}^c}( \theta,s^n)}{p\left(\theta, s^n\mid X^n\right)}\right) \widehat{Q}\,\right|_{A_{n, \ell}^c}(\dd \theta, \dd s^n)\\
&\qquad\qquad\qquad\qquad\qquad\qquad\qquad\qquad\qquad+\left.\lambda \int_{A_{n,\ell}} \log \left(\frac{\left.\lambda\, \widehat q\,\right|_{A_{n,\ell}}(\theta,s^n)}{p\left(\theta, s^n\mid X^n\right)}\right) \widehat{Q}\,\right|_{A_{n,\ell}}(\dd \theta, \dd s^n)\\
=& \left.(1-\lambda) \int_{A_{n,\ell}^c} \log \left(\frac{\left.\widehat q\,\right|_{A_{n, \ell}^c}(\theta,s^n)}{\left.p\,\right|_{A_{n, \ell}^c}(\theta,s^n\mid X^n)}\right) \widehat{Q}\,\right|_{A_{n, \ell}^c}(\dd \theta, \dd s^n)\\
&\qquad\qquad\qquad\qquad\qquad\qquad\qquad\qquad\qquad+\left.\lambda \int_{A_{n,\ell}} \log \left(\frac{\left.\widehat q\,\right|_{A_{n, \ell}}(\theta,s^n)}{\left.p\,\right|_{A_{n,\ell}}(\theta,s^n\mid X^n)}\right) \widehat Q\,\right|_{A_{n,\ell}}(\dd \theta, \dd s^n) \\
& +(1-\lambda) \log \left(\frac{1-\lambda}{P \left(A_{n,\ell}^c \mid X^n\right)}\right)+\lambda \log \left(\frac{\lambda}{P\left(A_{n,\ell} \mid X^n\right)}\right) \\
=: & (1-\lambda) d_{n 1}+\lambda d_{n 2}+D\left(\operatorname{Ber}(\lambda) \| \operatorname{Ber}\left(P(A_{n,\ell} \mid X^n)\right)\right),
\end{aligned}
\label{eq::kl_lambda}
\end{equation}
Letting its derivative with respect to $\lambda$ at $\widehat\lambda$ to be zero, as $\widehat\lambda$ is the minimizer, we have
$$
\widehat{\lambda}=\frac{\beta_n \exp \left(-d_{n 2}\right)}{\beta_n \exp \left(-d_{n 21}\right)+\left(1-\beta_n\right) \exp \left(-d_{n 1}\right)} \leq \frac{\beta_n \exp \left(-d_{n 2}\right)}{\left(1-\beta_n\right) \exp \left(-d_{n 1}\right)} \leq \frac{\beta_n}{1-\beta_n} \exp \left(d_{n 1}\right),
$$
where we denoted $\beta_n=P(A_{n,\ell} \mid X^n)$. Lastly, by plugging in $\lambda=\widehat\lambda$ as the minimizer in \eqref{eq::kl_lambda}, we obtain
$$
D(\hat q\verts p(\cdot,\cdot\mid X^n)) \ge D(\operatorname{Ber}(\widehat\lambda) \| \operatorname{Ber}(\beta_n)).
$$
By selecting $C'$ large enough such that $\beta_n \leq \Pi_n\left(\left\|\theta-\theta^*\right\| \geq C' \varepsilon\right) \leq e^{-C C'^2 n \varepsilon^2} \leq n^{-CC'^2} \leq 1 / 2$, we further obtain
$$
d_{n1}\le 2\beta_n\exp(d_{n1}) \le 2\beta_n\exp\left\{\frac{D(\hat q\verts p(\cdot,\cdot\mid X^n))}{1-\widehat\lambda}\right\}.
$$
By Lemma 4 in \cite{wei2019}, we have with probability at least $1-CK^{-2}$ that 
$$
D(\operatorname{Ber}(\widehat\lambda) \| \operatorname{Ber}(\beta_n)) \le D(\hat q\verts p(\cdot,\cdot\mid X^n)) \le CK^2\log n .
$$
On the other hand, note that
$$
D(\operatorname{Ber}(\widehat\lambda) \| \operatorname{Ber}(\beta_n)=\widehat\lambda\log\frac{\widehat\lambda}{\beta_n}+(1-\widehat\lambda)\log\frac{1-\widehat\lambda}{1-\beta_n}\ge -\log2+\widehat\lambda CC'^2\log n.
$$
Taking $C'$ large enough leads to $\widehat\lambda < 1/2$. Therefore, we conclude that
$$
\widehat Q_\theta(|\theta_\ell-\theta^*_\ell|\ge C'\epsilon)=\widehat\lambda\le 2\exp\left(-CC'^2n\varepsilon^2+2CK^2n\epsilon_n^2\right)\le \exp(-CC'^2n\epsilon^2/2),
$$
by taking $C'$ large enough. Finally, a simple union bounds over $\ell=1,2,\ldots,d$ leads to the desired inequality.

\subsection{Proof of Theorem \ref{thm::main1}}
\label{pf::main1}
First we introduce the following useful result, which is Lemma 5 in \cite{wei2019}. We use $\mu_Q$ to denote the mean of any probability measure $Q$, and $Q_{(\mu)}$ to denote a translation of $Q$ such that its expectation equals to $\mu$.
\begin{lemma}
\label{lem::kltri_diag}
Let $Q_\theta=\otimes_{j=1}^dQ_j$ be a probability measure from mean-field family, $\mu\in\R^d$ be a vector and $\Sigma\in\R^{d\times d}$ be a positive definite matrix. Then we have
\begin{align*}
&D(Q_\theta\verts N(\mu,\Sigma^{-1}))-D(N(\mu,[\diag(\Sigma)]^{-1})\verts N(\mu,\Sigma^{-1}))\\
=&\sum_{j=1}^dD(Q_{(\mu_j),j}\verts N(\mu_j,\Sigma_{jj}^{-1}))+\frac{1}{2}(\mu_{Q_\theta}-\mu)^T\Sigma(\mu_{Q_\theta}-\mu).
\end{align*}
Moreover, we have
$$
D(Q\verts N(\mu,[\diag(\Sigma)]^{-1})\le \frac{\max_j\Sigma_{jj}}{\lambda_{\min}(\Sigma)}[D(Q\verts N(\mu,\Sigma^{-1}))-D(N(\mu,[\diag(\Sigma)]^{-1})\verts N(\mu,\Sigma^{-1})].
$$
\end{lemma}
This inequality plays the role of a weaker version of the triangular inequality for KL-divergence when the second distribution is a normal.
 In particular, when $\Sigma$ is the identity matrix, the inequality in the lemma becomes a triangular inequality for KL-divergence as the ratio becomes one.

We also introduce the following lemma that generalizes Lemma 7 in \cite{wei2019} from well-specified models to mis-specified models. In addition, we improve the dimension-dependent exponent $(d+3)$ in the logarithmic term $(\log n)^{d+3}$ in \cite{wei2019} to a universal constant.  A proof is deferred to Appendix \ref{pf::kltri}.
\begin{lemma}
\label{lem::kltri}
Under Assumptions \ref{ass::prior} and \ref{ass::l}, for any $1\le K=O(\sqrt{n})$, it holds with at least probability $1-CK^{-2}$ that for any probability measure $Q_\theta$ in the mean-field family satisfying the same sub-Gaussian tail decay as in Lemma \ref{lem::post_subg}, we have
$$
\left|D\left(Q_{\theta} \| \Pi_{n}\right)-D\left(Q_{\theta} \| N\left(\tmle,\left[n V(\theta^*)\right]^{-1}\right)\right)\right| \leq \frac{C K^3(\log n)^{3}}{\sqrt{n}}.
$$
\end{lemma}
Since both $\widehat Q_\theta$ and $Q^*_{VB}\deltaeq N(\tmle,[n\diag(V_c)]^{-1})$ satisfy the sub-Gaussian property, the lemma can be applied to them.

Now let us proceed to the proof of the theorem.
Define the ``profile'' KL divergence $F_n(Q_\theta)=\min_{Q_{S^n}} D(Q_\theta\otimes Q_{S^n} \verts  P(\cdot\,,\cdot\,|\,X^n))$ so that the variational distribution $\widehat Q_\theta$ for the parameter minimizes this functional, or $\widehat Q_\theta=\argmin_{Q_\theta}F_n(Q_\theta)$. The following lemma characterizes an asymptotic expression of $F_n$. Note that we have defined $\Sigma_Q$ as the variance matrix for any probability measure $Q$, which is diagonal if $Q$ belongs to the mean-field family. The result is directly adopted from Appendix B.11 in \cite{wei2019}, whose proof is based on the Taylor expansion and the following moment bound due to the sub-Gaussian property of $Q_\theta$ (as in Lemma \ref{lem::post_subg}), i.e. for some constant $D>0$
\begin{equation}
\E_{Q_\theta}[\|\theta-\theta^*\|^m]\le C(DK\epsilon_n)^m,\quad  m=1,2,\ldots.
\label{eq::mmt_bd}
\end{equation}
\begin{lemma}
\label{lem::profile_kl}
Under Assumptions \ref{ass::prior} to \ref{ass::ls}, for any $1\le K=O(\sqrt{n})$, it holds with at least probability $1-CK^{-2}$ that for any probability measure $Q_\theta$ in the mean-field family satisfying the same sub-Gaussian tail decay as in Lemma \ref{lem::post_subg}, we have
$$
\left|F_n(Q_\theta)-D(Q_\theta\verts \Pi_n)-\frac{n}{2}\tr(\Sigma_{Q_\theta} V_s(\theta^*))\right|\le \frac{C K^{3}(\log n)^{3}}{\sqrt{n}}.
$$
\end{lemma}
Note that when there is no latent variable, the ``profile'' KL divergence coincides with $D(Q_\theta\verts \Pi_n)$. Therefore, this lemma illustrates that the third term is coming from the mean-field approximation that ignores the dependence between the latent variables and parameters. This term will eventually contributes to the shape (asymptotic covariance matrix) of the variational distribution $\widehat Q_\theta$.

The two preceding lemmas imply that the minimizer $\widehat Q_\theta$ of $F_n$ is essentially the minimizer of $D(Q_\theta\verts N(\tmle,[nV_c(\theta^*)]^{-1}))$, up to some higher-order term of order $\alpha_n={K^3(\log n)^{3}}/{\sqrt{n}}$. To be more precise, for any probability measure $Q_\theta$ in the mean-field family satisfying the same sub-Gaussian tail decay as in Lemma \ref{lem::post_subg}, we have
\begin{equation}
\left|F_n(Q_\theta)-D\left(Q_{\theta} \| N\left(\tmle,\left[n V(\theta^*)\right]^{-1}\right)\right)-\frac{n}{2}\tr(\Sigma_{Q_\theta} V_s(\theta^*))\right|\le 2\alpha_n
\label{eq::profile_1}
\end{equation}
Also note that after straightforward calculation, we have
\begin{equation}
\begin{aligned}
&D\left(Q_{\theta} \verts N\left(\tmle,\left[n V_c(\theta^*)\right]^{-1}\right)\right)-D\left(Q_{\theta} \verts N\left(\tmle,\left[n V(\theta^*)\right]^{-1}\right)\right)\\
=&\frac{n}{2}\tr(\Sigma_{Q_\theta} V_s(\theta^*))+\frac{n}{2}(\mu_{Q_\theta}-\tmle)^T V_s(\theta^*)(\mu_{Q_\theta}-\tmle)+\frac{1}{2}\log\det(V_c^{-1}V),
\end{aligned}
\label{eq::kl_diff}
\end{equation}
where we recall that $\mu_Q$ denotes the mean of any probability measure $Q$.

Now by applying (\ref{eq::profile_1}) and (\ref{eq::kl_diff}) to $\widehat Q_\theta$ and $Q^*$, both satisfying the sub-Gaussian tail property, along with the optimality of $\widehat Q_\theta$, we can conclude that
\begin{equation}
\begin{aligned}
&D\left(\widehat{Q}_{\theta} \verts N\left(\tmle,[n V_c(\theta^{*})]^{-1}\right)\right)\\
\le& D\left(Q_{VB}^* \verts N\left(\tmle,\left[n V_c(\theta^*)\right]^{-1}\right)\right)+\frac{n}{2}\left(\mu_{\widehat Q_\theta}-\tmle\right)^{T} V_s\left(\theta^{*}\right)\left(\mu_{\widehat{Q}_{\theta}}-\tmle\right)+4\alpha_n.
\end{aligned}
\label{eq::klbound}
\end{equation}
Recall $Q_{VB}^*=N(\tmle,[\diag(V_c)]^{-1})$. By applying Lemma \ref{lem::kltri_diag} with $(\mu,\Sigma)=(\tmle,V_c)$ and combining it with above, we can get
$$
\frac{n}{2}\left(\mu_{\widehat Q_\theta}-\tmle\right)^{T} V\left(\theta^{*}\right)\left(\mu_{\widehat{Q}_{\theta}}-\tmle\right)\le 4\alpha_n.
$$
The positive definiteness of $V(\theta^*)$ further implies that
\begin{equation}
 \frac{n}{2}\left(\mu_{\widehat Q_\theta}-\tmle\right)^{T} V_s\left(\theta^{*}\right)\left(\mu_{\widehat{Q}_{\theta}}-\tmle\right)\le C\alpha_n,
 \label{eq::quad_vs}
\end{equation}
for some constant $C>0$ depending on the information matrices.
Finally, apply again Lemma \ref{lem::kltri_diag}, we can conclude that
\begin{align*}
&D(\widehat Q_\theta\verts Q_{VB}^*)\\
\le&C\left[D\left(\widehat{Q}_{\theta} \verts N\left(\tmle,[n V_c(\theta^{*})]^{-1}\right)\right)
-D\left(Q_{VB}^* \verts N\left(\tmle,\left[n V_c(\theta^*)\right]^{-1}\right)\right)\right]\\
\le & \frac{n}{2}\left(\mu_{\widehat Q_\theta}-\tmle\right)^{T} V_s\left(\theta^{*}\right)\left(\mu_{\widehat{Q}_{\theta}}-\tmle\right)+C\alpha_n\le C'\alpha_n,
\end{align*}
which is the desired result.

\subsection{Proof of Theorem \ref{thm::translate}}
\label{pf::translate}
Since we will demonstrate that the theorem holds for each fixed model 
$\m M$, we omit all subscripts of $\m M$ in the following proof for the sake of simplicity in notation.

From (\ref{eq::profile_1}), (\ref{eq::kl_diff}) and (\ref{eq::klbound}), we have
\begin{align*}
F_n(\widehat Q_\theta) \le &\,D\left(\widehat Q_{\theta} \verts N\left(\tmle,\left[n V(\theta^*)\right]^{-1}\right)\right)+\frac{n}{2}\tr(\Sigma_{\widehat Q_\theta} V_s(\theta^*))+ 2\alpha_n\\
=&\,D\left(\widehat Q_\theta \verts N\left(\tmle,\left[n V_c(\theta^*)\right]^{-1}\right)\right)-\frac{1}{2}\log\det(V_c^{-1}V)\\
&\,-\frac{n}{2}(\mu_{\widehat Q_\theta}-\tmle)^T V_s(\theta^*)(\mu_{\widehat Q_\theta}-\tmle)+2\alpha_n\\
\le &\, D\left(Q_{VB}^* \verts N\left(\tmle,\left[n V_c(\theta^*)\right]^{-1}\right)\right)-\frac{1}{2}\log\det(V_c^{-1}V)+6\alpha_n\\
=&\,\frac{1}{2}\log\frac{\det(\diag(V_c))}{\det(V_c)}+\frac{1}{2}\log\frac{\det(V_c)}{\det(V)}+6\alpha_nC^*(\cM)+6\alpha_n.
\end{align*}
On the other hand, we have by applying (\ref{eq::profile_1}) and (\ref{eq::kl_diff}) that
\begin{align*}
F_n(\widehat Q_\theta) \ge &\,D\left(\widehat Q_{\theta} \verts N\left(\tmle,\left[n V(\theta^*)\right]^{-1}\right)\right)+\frac{n}{2}\tr(\Sigma_{\widehat Q_\theta} V_s(\theta^*))-2\alpha_n\\
=&\,D\left(\widehat Q_\theta \verts N\left(\tmle,\left[n V_c(\theta^*)\right]^{-1}\right)\right)-\frac{1}{2}\log\det(V_c^{-1}V)\\
&\,-\frac{n}{2}(\mu_{\widehat Q_\theta}-\tmle)^T V_s(\theta^*)(\mu_{\widehat Q_\theta}-\tmle)-2\alpha_n\\
\ge &\, D\left(Q_{VB}^* \verts N\left(\tmle,\left[n V_c(\theta^*)\right]^{-1}\right)\right)-\frac{1}{2}\log\det(V_c^{-1}V)-C\alpha_n\\
=&\,C^*(\cM)-C\alpha_n,
\end{align*}
where we have used (\ref{eq::quad_vs}) in the last inequality and the fact that $Q_{VB}^*$ minimizes $D(Q\verts  N\big(\tmle,$ $[n V_c(\theta^*)]^{-1}\big)$. Therefore, we conclude that
\begin{equation}\label{eqn:ine1}
\left|D(\widehat Q_\theta\otimes\widehat Q_{S^n}\verts p(\cdot\,,\cdot\,|\,X^n))-C^*(\cM)\right|=\left|F_n(\widehat Q_\theta)-C^*(\cM)\right|\le C\alpha_n
\end{equation}
holds with probability at least $1-CK^{-2}$. It remains to note that
$$
\log p(X^n\,|\,\cM)-\elbo(\cM)=D(\widehat Q_\theta\otimes\widehat Q_{S^n}\verts p(\cdot\,,\,\cdot\,|\,X^n)),
$$
which together with~\eqref{eqn:ine1} implies the first claimed result.

The second result arises from the Laplace approximation for BIC, which implies
$$\log p(X^n\,|\,\cM)=-\frac{\bic(\cM)}{2}+\log\pi_\cM(\tmstar)+\frac{d_\cM}{2}\log(2\pi)-\frac{1}{2}\log\det(V)+o_p(1).$$ 
In fact, this result can be proven using an argument similar to that used in the proof of Theorem \ref{thm::translate} based on a Taylor expansion,  which leads to an explicit error bound as follows,
$$\left|\log p(X^n\,|\,\cM)+\frac{\bic(\cM)}{2}-\log\pi_\cM(\tmstar)-\frac{d_\cM}{2}\log(2\pi)+\frac{1}{2}\log\det(V)\right|\le C\alpha_n.$$ 

\subsection{Proof of Theorem \ref{thm::pred}}
\label{pf::pred}
Since we select the model $\widehat \cM$ with largest ELBO value, we have for any candidate model $\cM$, $\elbo(\widehat \cM)\ge \elbo(\cM)$. Therefore, we have that for each model $\cM$, it holds with probability at least $1-CR^{-1}K^{-2}$ (by replacing constant $K$ with $\sqrt{R}K$ in all the results therein, where recall $R$ is the total number of candidate models) that
$$
-\bic(\widehat \cM)/2-\widetilde C^*(\widehat \cM) \ge -\bic( \cM)/2-\widetilde C^*(\cM)-CR^3\alpha_n, 
$$
which, combined with Theorem \ref{thm::translate}, implies
$$
\ell_n(\tmle_{\widehat \cM};X^n)\ge \ell_n(\tmle_{\cM};X^n)+\frac{d_{\widehat \cM}-d_\cM}{2}\log n+\widetilde C^*(\widehat \cM)-\widetilde C^*(\cM)-CR^3\alpha_n.
$$
A simple application of triangle inequality leads to
\begin{align*}
&\left|\frac{1}{n}\ell_n(\tmle_{\cM};X^n)-\E_{\P_0}[\ell(\theta_{\cM}^*;X)]\right|\\
\le&\left|\frac{1}{n}\ell_n(\theta_{\cM}^*;X^n)-\frac{1}{n}\ell_n(\tmle_{\cM};X^n)\right|+\left|\frac{1}{n}\ell_n(\theta_{\cM}^*;X^n)-\E_{\P_0}[\ell(\theta_{\cM}^*;X)\right|.
\end{align*}
In addition, Assumption \ref{ass::l} implies that 
$$
\left|\frac{1}{n}\ell_n(\theta_{\cM}^*;X^n)-\frac{1}{n}\ell_n(\tmle_{\cM};X^n)\right|\le \frac{1}{n}\sum_i|Z(X_i)|\cdot\|\tmle_{\cM}-\theta_{\cM}^*\|^2,
$$
which is trivially bounded by $CR^3\alpha_n$ with probability at least $1-CRK^{-2}$. 
More over, by the Markov inequality, we have with probability at least $1-CR^{-1}K^{-2}$,
$$
\left|\frac{1}{n}\ell_n(\theta_{\cM}^*;X^n)-\E_{\P_0}[\ell(\theta_{\cM}^*;X)\right|\le \frac{C\sqrt{R}K}{\sqrt{n}}.
$$

Putting all pieces together, we obtain that for any model $\cM$,
$$
\E_{\P_0}[\ell(\theta_{\widehat \cM}^*;X)] \ge \E_{\P_0}[\ell(\theta_{\cM}^*;X)]+\frac{d_{\widehat \cM}-d_\cM}{2n}\log n+\frac{\widetilde C^*(\widehat \cM)-\widetilde C^*(\cM)}{n}-CR^3\alpha_n,
$$
and hence
$$
D(\P_0\verts\P_{\theta_{\widehat \cM}^*}) \le D(\P_0\verts\P_{\theta_{\widehat \cM}^*})+\frac{d_\cM-d_{\widehat \cM}}{2n}\log n+\frac{\widetilde C^*(\cM)-\widetilde C^*(\widehat \cM)}{n}+CR^3\alpha_n.
$$
It remains to note that
\begin{align*}
&\left|\int D(\P_0\verts\P_{\theta_{\widehat \cM}})\di \widehat q_{\theta_{\widehat \cM}}(\theta_{\widehat \cM})-D(\P_0\verts\P_{\theta_{\widehat \cM}^*})\right|\\
=&\,\left|\int\E_{\P_0}\left[\frac{\ell(\theta_{\widehat \cM}^*;X)}{\ell(\theta_{\widehat \cM};X)}\right]\di Q_{\theta_{\widehat \cM}}(\theta_{\widehat \cM})\right|
\le C\int \|\theta_{\widehat \cM}-\theta_{\widehat \cM}^*\|^2\di Q_{\theta_{\widehat \cM}}(\theta_{\widehat \cM})\le C'RK^2\epsilon_n^2,
\end{align*}
where we have used (\ref{eq::mmt_bd}) and the fact that
$$
\E_{\P_0}\left[\frac{\ell(\theta_{\cM}^*;X)}{\ell(\theta_{\cM};X)}\right]\le C\|\theta_{\cM}-\theta_{\cM}^*\|^2
$$
holds for any model $\cM$ with some constant $C$. Then, we can conclude that for any model $\cM$, it holds with probability at least $1-CR^{-1}K^2$ that
\begin{align*}
&\int D(\P_0\verts\P_{\theta_{\widehat \cM}})\,\di \widehat q_{\theta_{\widehat \cM}}(\theta_{\widehat \cM})\\
\le &\,D(\P_0\verts\P_{\theta_{\cM}^*})+\frac{d_\cM-d_{\widehat \cM}}{2n}\log n+\frac{\widetilde C^*(\cM)-\widetilde C^*(\widehat \cM)}{n}+CR^3\alpha_n\\
\le &\, D(\P_0\verts\P_{\theta_{\cM}^*})+\frac{d_\cM}{2n}\log n+\frac{C_0}{n}+CR^3\alpha_n,
\end{align*}
where $\displaystyle C_0:\,=\widetilde C^*(\cM)-\min_\cM \widetilde C^*(\cM)$. Finally, a simple application of the union bound (over $R$ models) leads to the claimed inequality in the theorem.

\subsection{Model Selection Consistency with BIC}
\label{pf::bic}
Note that we have
$$
\bic(\cM)-\bic(\cM_0)=-2\ell_n(\tmle_{\cM};X^n)+2\ell_n(\tmle_{\cM_0};X^n)+(d_\cM-d_{\cM_0})\log n.
$$
In Appendix \ref{pf::pred}, we showed that
$$
\left|\frac{1}{n}\ell_n(\tmle_{\cM};X^n)-\E_{\P_0}[\ell(\theta_{\cM}^*;X)]\right|\le  C\alpha_n
$$
holds with probability $1-CK^{-2}$. Therefore, we have
$$
\left|\bic(\cM)-\bic(\cM_0)-2n D\left(\P_0\verts \P_{\theta_\cM^*}\right)-(d_\cM-d_{\cM_0})\log n\right|\le C\alpha_n.
$$
When $D(\P_0\verts \P_{\theta_\cM^*})\ge \epsilon>0$, we can see that $\bic(\cM)-\bic(\cM_0)$ has order $O(2\epsilon n)$ with high probability.

Suppose $\cM$ contains the true model and $d_{\cM}>d_{\cM_0}$. Then we have
$$
\ell_n(\tmstar;X^n)=\ell_n(\theta^*_{\cM_0};X^n)\le\ell_n(\tmle_{\cM_0};X^n)\le \ell_n(\tmle_{\cM};X^n)
$$
Therefore,
$$
\left|\ell_n(\tmle_{\cM_0};X^n)-\ell_n(\tmle_{\cM};X^n)\right|\le \left|\ell_n(\theta_{\cM}^*;X^n)-\ell_n(\tmle_{\cM};X^n)\right|\le \sum_i|Z(X_i)|\cdot\|\tmle_{\cM}-\theta_{\cM}^*\|^2,
$$
which is bounded by $K^3\log^\epsilon(n)$ with probability at least $1-CK^{-2}$ for some $\epsilon>0$ small enough, by applying the Markov inequality, as we have that $\|\tmle_{\cM}-\theta_{\cM}^*\|\le K\sqrt{\frac{\log^\epsilon(n)}{n}}$ holds with high probability. Therefore, we have with high probability
$$
\bic(\cM)-\bic(\cM_0)\ge (d_{\cM}-d_{\cM_0})\log n-C\log^\epsilon(n)\gtrsim \log n.
$$
This implies that $\cM_0$ minimizes BIC across all model candidates.

\subsection{Proof of Lemma \ref{lem::gauss_cavi}}
\label{pf::gauss_cavi}
Note that for the Gaussian posterior, we have $\pi_n=\phi_n$ and $\widehat q =\phi^*$. Recall that the corresponding ``regret'' value is defined in Section~\ref{sec::thmcavi} as $D^{(t)}=D(p^{(t)}\verts \pi_n)-D(\widehat q\verts \pi_n)$, where $p^{(t)}$ dentoes the $t$-th iteration of CAVI for the Gaussian posterior.
It is easy to verify that 
\begin{align}\label{eqn:regret_KL}
    D^{(t)}=\frac{n}{2}\,(b^{(t)})^T V b^{(t)} \quad \mbox{and}\quad D(p^{(t)}\verts\phi^*) = \frac{n}{2}\,(b^{(t)})^T S b^{(t)},
\end{align}
where $b^{(t)}=\theta^{(t)} -\widehat\theta$. Therefore, there exists some positive constant such that $D(p^{(t)}\verts\phi^*) \leq C\,D^{(t)}$ for all $t\geq 0$. Now we consider the two update schemes separately.
\begin{itemize}
    \item Sequential update (randomized): Since the coordinate to be updated is chosen uniformly over $\{1,\ldots,d\}$, we obtain by using the updating formula~\eqref{eq::bconv} with a random $\ell\sim$Unif$(1,\ldots,d)$ that after one update, 
\begin{align*}    
&D^{(t)}-\mb E[D^{(t+1)}\mid D^{(t)}]=  \frac{1}{d}\sum_{\ell=1}^{d}\frac{n\gamma(2-\gamma)}{2V_{\ell\ell}}\left(V_{\ell\ell} b_\ell^{(t)}+\sum_{k \neq \ell} b_{k}^{(t)} V_{k\ell}\right)^{2}\\
&\qquad\qquad=\frac{n\gamma(2-\gamma)}{2d}\,b^{(t)T}V S^{-1}V b^{(t)}
\geq\frac{c_s n\gamma(2-\gamma)}{2d}\,b^{(t)T}V  b^{(t)}=\frac{c_s\gamma(2-\gamma)}{d}D^{(t)},
\end{align*}
where 
\begin{equation}
c_s=\max_{\|b\|=1}\frac{b^TV S^{-1}Vb}{b^TVb}.
\label{gauss_c}
\end{equation}
Therefore, by taking another expectation, we can obtain
$\Exp[D^{(t+1)}]\leq (1-c_s\gamma(2-\gamma)/d)\,\Exp[D^{(t)}]$ and therefore $\Exp[D^{(t)}]\leq (1-c_s\gamma(2-\gamma)/d)^t\, D^{(0)}$. We can always take $\gamma=1$ to achieve largest decrease in $D^{(t)}$ for sequential update.

\item Parallel update with $d$ coordinates: Using the updating formula~\eqref{eq::bconv}, we obtain
$$
D^{(t+1)}=\frac{n}{2}\, (b^{(t+1)})^TVb^{(t+1)}= \frac{n}{2}\, (b^{(t)})^TA_\gamma V A_\gamma b^{(t)}.
$$
Note that the matrix $c_p(\gamma)\,V- A_\gamma V A_\gamma$ is positive semidefinite for constant
\begin{equation}
c_p(\gamma)=\max_{\|b\|=1}\frac{b^TA_\gamma VA_\gamma b}{b^TVb}=\max_{\|b\|=1}\frac{b^T(I-\gamma S^{-1}V)V(I-\gamma S^{-1}V)b}{b^TVb}.
\label{jacobi_c}
\end{equation}
Therefore, we have $D^{(t+1)}\le c_p(\gamma)\,D^{(t)}$. By taking $\gamma$ sufficiently small, we can make $c_p(\gamma)\in(0,1)$, which leads to the geometric convergence of the parallel update as claimed in the theorem.
\end{itemize}

\subsection{Proof of Theorem \ref{thm::caviconv}}
\label{pf::caviconv}
The proof of Theorem \ref{thm::caviconv} is based on standard perturbation analysis. Recall we have shown the exponential convergence for Gaussian posteriors as in Lemma \ref{lem::gauss_cavi}. It remains to extend the results to general non-Gaussian posteriors. We will apply perturbation analysis to bound the difference introduced by the normal approximation to the general posterior. As before, we omit all the terms involving $L$ in our assumptions to simplify the notations. We only prove the theorem for the parallel update scheme; the randomized sequential update scheme involves only one coordinate per iteration, so its analysis is even simpler and therefore omitted. 

To proceed from the Gaussian posterior, we state some useful lemmas. The first lemma extends Lemma \ref{lem::kltri} to a general distribution $q$ rather than the mean-field solution $\widehat q$. While both lemmas apply to those $q$ having sub-Gaussian tails, Lemma \ref{lem::kldif_to_normal} makes no assumption about its center. When the center is $\theta^*$, it reduces to Lemma \ref{lem::kltri}. 
\begin{lemma}
\label{lem::kldif_to_normal}
Under Assumption \ref{ass::prior} and \ref{ass::l}, for any $1\le K=O(\sqrt{n})$, it holds with at least probability $1-CK^{-2}$ that for any probability measure $q$ in the mean-field family satisfying the sub-Gaussian tail decay, i.e.~satisfying inequality~\eqref{eqn:sub_Ga} with center $\theta^\ast$ therein being replaced with any other vector, we have with probability at least $1-CK^{-2}$,
$$
\left|D\left(q\verts \pi_n\right)-D\left(q \verts \phi_n\right)\right| \leq Cn\Exp_q\|\theta-\tmle\|^3+\frac{C K^{3}(\log n)^{3}}{\sqrt{n}}.
$$
\end{lemma}
Lemma \ref{lem::kldif_to_normal} provides a bridge for perturbation analysis that connects previous results for Gaussian posteriors with general posteriors. We defer its proof to Appendix \ref{pf::kldif_to_normal}.

Recall that we assume the initialization $q^{(0)}$ to be supported in a small but constant radius neighborhood of $\theta^*$. This enables us to apply local quadratic approximation to approximate the log-posterior function, as each iteration $Q^{(t)}$ tends to concentrate around its center (or mean) with a typical radius of order $\sqrt{\log n/n}$, which is ensured by Lemma~\ref{lem::onedlocalexpansion} below. Before stating this lemma, we first make some definitions to simplify the notations. 
Define the log-posterior function $U_n(\theta)=\log \pi_n(\theta)$ and its quadratic approximation $G_n(\theta)=-n(\theta-\tmle)^TV(\theta-\tmle)/2$, which is also the log-density of its approximating Gaussian posterior analyzed in Lemma~\ref{lem::gauss_cavi}. We also use $\mu^{(t)}=\Exp_{q^{(t)}}[\theta]$ to denote the mean of $q^{(t)}$. Recall that $\theta^{(t)}=E_{p^{(t)}}[\theta]$ denotes the mean of $p^{(t)}$, where $p^{(t)}$ is the $t$-th iteration of CAVI for the Gaussian posterior. We will conduct perturbation analysis to show that $q^{(t)}$ is close to $p^{(t)}$ and $\mu^{(t)}$ is close to $\theta^{(t)}$. 

For each coordinate index $\ell\in[d]$, we define the linear operator 
\begin{align}\label{eqn:lo}
    T_{\gamma,\ell}(\theta)=\theta_\ell-\gamma\sum_{k}\frac{V_{k\ell}}{V_{\ell\ell}}(\theta_k-\widehat\theta_k)\deltaeq\widetilde\theta_\ell(\gamma),
\end{align}
which characterizes the updating rule of the $\ell$-th coordinate of the expectation $\theta^{(t)}=E_{p^{(t)}}[\theta]$ under the Gaussian posterior using the CAVI algorithm with step size $\gamma$; for example, see equation~\eqref{eq::bconv}. 
For $\gamma=1$, we will write $\widetilde\theta_\ell=\widetilde\theta_\ell(1)$ for short. Note that these notation leads to $\widetilde\theta_\ell(\gamma)=\gamma\widetilde\theta_\ell+(1-\gamma)\theta_\ell$. Lastly, we also define two functions of $\theta$: $\widetilde U_{n,\ell} (\theta)=U_n(\widetilde\theta_\ell,\theta_{-\ell})$ and $\widetilde G_{n,\ell}(\theta)=G_n(\widetilde\theta_\ell,\theta_{-\ell})$. It is easy to check that we can simplify $\widetilde G_{n,\ell}(\theta)$ into $\widetilde G_{n,\ell}(\theta)=G_n(\theta)+nV_{\ell\ell}(\theta_\ell-\widetilde\theta_\ell)^2/2$. According to equation~\eqref{eqn:regret_KL}, we can also write $G_n(\theta^{(t)})=-\big[D(p^{(t)}\verts \phi_n)-D(\widehat q\verts \phi_n)\big]$, which is the negative objective function of CAVI for the normal posterior evaluated at the $t$-th iteration up to a constant; hence $\widetilde G_{n,\ell}(\theta)$ is the maximal objective value that we can obtain by updating the $\ell$-th coordinate in the CAVI for the Gaussian posterior with $N(\theta, (nS)^{-1})$ as the previous iteration, that is, $\widetilde G_{n,\ell}(\theta) = \max_{\theta_{\ell}} G_n(\theta_{\ell},\theta_{-\ell})$. According to this property, we can view $\widetilde U_{n,\ell}(\theta)$ as an approximation to $\widetilde G_{n,\ell}(\theta)$, where the later approximates the maximal objective value corresponding to updating the $\ell$-th coordinate in the CAVI for the true posterior with a distribution centered at $\theta$ as the previous iteration, by plugging in the (coordinate-wise) maximizer $\widetilde\theta_\ell$ derived from the CAVI for the Gaussian posterior. The following lemma (second part) essentially makes this intuition formal by showing that $Q_\ell^{(t+1)}$ almost centers at $T_{\gamma,\ell}(\mu^{(t)})$ with dispersion (or effective radius) of order $O(1/\sqrt{n})$, where recall that $\mu^{(t)}$ is the expectation of $Q_\ell^{(t)}$ in the previous iteration. Similar to the Gaussian posterior setting, we may assume that the initial distribution $Q^{(0)}$ is obtained by applying a full step size parallel update (i.e.~using~\eqref{cavi_jacobi} or~\eqref{cavi_jacobi_step} with $\gamma =1$) to any distribution supported in a constant neighborhood of $\theta^\ast$. The first part of the following lemma shows that similar to the Gaussian case, $Q^{(0)}$ is then guaranteed to concentrate at its expectation with dispersion (or effective radius) $O(1/\sqrt{n})$. A proof of this lemma is deferred to Appendix  \ref{pf::onedlocalexpansion}.

\begin{lemma}
\label{lem::onedlocalexpansion}
There exists some constant $\delta>0$ that only depends on $V(\theta^*)$, such that if $Q^{(0)}$ is obtained by applying a full step size parallel update to any distribution supported in $B_{\widehat\theta}(\delta)$, then $Q^{(0)}$ has sub-Gaussian tail with dispersion $O(1/\sqrt{n})$, that is, for each $\ell\in[d]$,
\begin{align*}
    Q_\ell^{(0)}(|\theta_\ell-\mu^{(0)}|>\epsilon) \le \exp\{-Cn\epsilon^2\},\quad\forall \epsilon>C\sqrt{\log n /n}.
\end{align*}
Moreover, for such a $Q^{(0)}$, we have that for any $t\geq 0$ and each $\ell\in[d]$, 
$$
Q_\ell^{(t+1)}(|\theta_\ell-T_{\gamma,\ell}(\mu^{(t)})|>\epsilon) \le \exp\{-Cn\epsilon^2\},\quad\forall \epsilon>C\sqrt{\log n /n},
$$
where recall that $\mu^{(t)}$ denotes the expectation of $Q^{(t)}$.
\end{lemma}
In the rest of the proof, we assume that the initialization $Q^{(0)}$ is obtained as in Lemma~\ref{lem::onedlocalexpansion}.
A direct corollary of Lemma \ref{lem::onedlocalexpansion} is that
 $$|\mu^{(t+1)}_\ell-T_{\gamma,\ell}(\mu^{(t)})| \le C\sqrt{\log n/n}.$$
If we denote the bias of the expectation of $q^{(t)}$ as $\tb^{(t)}=\mu^{(t)}-\tmle$, then the preceding display and equation~\eqref{eqn:lo} together imply
\begin{equation}
\|\tb^{(t+1)}-A_\gamma \tb^{(t)}\|\le C'\sqrt{\log n/n},
\label{eq::bmuconb}
\end{equation} 
where we recall that $A_\gamma=I-\gamma S^{-1}V$ is defined after equation~\eqref{eq::bconv} and we may take $C'=C\sqrt{d}$.
As a result, $\tb^{(t+1)}$ converges to a neighborhood around zero with radius $O(\sqrt{\log n/n})$ exponentially fast due to the exponential convergence of the CAVI for the Gaussian posterior (Lemma \ref{lem::gauss_cavi}), i.e.,
\begin{equation}
\|\widetilde b^{(t)}\| \le C_V\alpha^{t/2}\|\widetilde b^{(0)}\|+C''\sqrt{\log n/n},
\label{eq::exp_perturb}
\end{equation}
where $C_V$ is the conditional number of $V$ and $\alpha$ is the contraction factor defined in Lemma \ref{lem::gauss_cavi}.
Moreover, the sub-Gaussian tail from Lemma \ref{lem::onedlocalexpansion} also implies a moment bound: using the formula $\mb E[X^k]=\int_0^\infty k\,t^{k-1}\,\mb P(X\geq t)\,\dd t$ for any nonnegative random variable $X$, it is straightforward to show that for any  $k\ge 1$,
$$\E_{q_\ell^{(t+1)}}\big[|\theta_\ell-\mu^{(t+1)}_\ell|^k\big]\le C(\log n/n)^{k/2},$$
which implies 
\begin{align*}
\E_{q_\ell^{(t+1)}}\big[|\theta_\ell-\theta^*_\ell|^k\big]
\le &\, 2^{k-1}\left(\E_{q_\ell^{(t+1)}}\big[|\theta_\ell-\mu^{(t+1)}|^k\big]+|\mu_\ell^{(t+1)}-\theta^*_\ell|^k\right)\\
\le &\, C(\log n/n)^{k/2}+2^{k-1}|\tb^{(t+1)}_\ell|^k.
\end{align*}
Combining this with inequality~\eqref{eq::bmuconb},we have that for any $k\ge 1$,
\begin{equation}
\label{eq::exp_perturb_E}
\E_{q^{(t)}}\big[\|\theta-\theta^*\|^k\big] \le 2^{k-1} C_V\alpha^{kt/2}\|\tb^{(0)}\|^k+C(\log n/n)^{k/2}.
\end{equation}


Now let us proceed to the perturbation analysis for proving the first claimed inequality~\eqref{mainineq3} about $D^{(t)} =D(q^{(t+1)}\verts \pi_n)-D(\widehat q\verts \pi_n)$. Without loss of generality, we assume 
$\gamma=1$ for the remainder of this proof for the sake of simplicity.
Note that due to the factorization of $q^{(t+1)}$, we have 
\begin{align}
&\,D(q^{(t+1)}\verts \pi_n)=\int q^{(t+1)}\log q^{(t+1)}-\int q^{(t+1)} \log\pi_n  \notag\\
=&\,\sum_{\ell=1}^d\int q_\ell^{(t+1)}\log q_\ell^{(t+1)}-\int q^{(t+1)}\log \pi_n   \notag\\
\overset{(i)}{=}&\,\sum_{\ell=1}^d\left(\int q_\ell^{(t+1)}\log\left(\frac{\exp\{\int q_{-\ell}^{(t)}\,(U_n-\widetilde U_{n,\ell})\}}{\int_{\R}\exp\{\int_{\R^{d-1}} q_{-\ell}^{(t)}(\theta_{-\ell})\,[U_n(\theta)-\widetilde U_{n,\ell}(\theta)]\,\dd\theta_{-\ell}\}\,\dd\theta_\ell}\right)\right)-\int q^{(t+1)}\log \pi_n   \notag\\
=&\,\sum_{\ell=1}^d\int q_\ell^{(t+1)}\int q_{-\ell}^{(t)}\,(U_n-\widetilde U_{n,\ell})-\int q^{(t+1)}\log \pi_n   \notag\\
&\qquad\qquad\qquad-\sum_{\ell=1}^d\log\left(\int_{\R}\exp\left\{\int_{\R^{d-1}} q_{-\ell}^{(t)}(\theta_{-\ell})\,\big[U_n(\theta)-\widetilde U_{n,\ell}(\theta))\big]\,\dd\theta_{-\ell}\right\}\dd\theta_\ell\right), \label{eqn:DL_dec}
\end{align}
where we have used the CAVI updating formula for $q_\ell^{(t+1)}$, and in step (i) we added a $\theta_{\ell}$ independent term $\int q_{-\ell}^{(t)}\,\widetilde U_{n,\ell}$ (since $\widetilde U_{n,\ell}(\theta)$ only depends on $\theta_{-\ell}$) to both the exponents in the denominator and the numerator. Recall that $\phi_n$ is the density of $N(\tmle, (nV)^{-1})$, which approximates the posterior $\pi_n$ and has form $C_V\cdot\exp\{-n(\theta-\tmle)^TV(\theta-\tmle)/2\}$ with $C_V$ being some normalizing constant depending only on $V$. We first analyze the first term next to the last equality of the preceding display. By \eqref{eq::pi_taylor} in the proof of Lemma \ref{lem::kldif_to_normal}, we have with probability at least $1-CK^{-2}$,
\begin{equation}\label{eqn:t1}
\begin{aligned}
&\int_{\R^{d-1}} q_{-\ell}^{(t)}(\theta_{-\ell})\,U_n(\theta)\dd\theta_{-\ell}=\int_{\R^{d-1}} q_{-\ell}^{(t)}(\theta_{-\ell})\,\log\pi_n(\theta)\dd\theta_{-\ell}\\
=&\int_{\R^{d-1}} q_{-\ell}^{(t)}(\theta_{-\ell})\,\log\phi_n(\theta)\dd\theta_{-\ell}+O\left(n\E_{q_{-\ell}^{(t)}}\|\theta_{-\ell}-\theta_{-\ell}^*\|^3\right)+O\left(n|\theta_\ell-\theta_\ell^*|^3\right)+\alpha_n \\
=&\log C_V-n\E_{q_{-\ell}^{(t)}}[(\theta-\tmle)^TV(\theta-\tmle)]/2+O\left(n\E_{q_{-\ell}^{(t)}}\|\theta_{-\ell}-\theta_{-\ell}^*\|^3\right)+O\left(n|\theta_\ell-\theta_\ell^*|^3\right)+\alpha_n.
\end{aligned}
\end{equation}
Using the definition of $\widetilde U_{n,\ell}(\theta)=U_n(\widetilde\theta_\ell,\theta_{-\ell})$, we also have
\begin{equation}\label{eqn:t2}
\begin{aligned}
&\int_{\R^{d-1}} q_{-\ell}^{(t)}(\theta_{-\ell})\,\widetilde U_{n,\ell}(\theta)\dd\theta_{-\ell}\\
=&\log C_V-n\E_{q_{-\ell}^{(t)}}[((\widetilde\theta_\ell,\theta_{-\ell})-\tmle)^TV((\widetilde\theta_\ell,\theta_{-\ell})-\tmle)]/2+O\left(n\E_{q_{-\ell}^{(t)}}\|\theta_{-\ell}-\theta_{-\ell}^*\|^3\right)+\alpha_n.
\end{aligned}
\end{equation}
Similar to the Gaussian posterior case, we can obtain the following by straightforward algebras,
\begin{equation}
    \label{eqn:t3}
\begin{aligned}
&\E_{q_{-\ell}^{(t)}}[(\theta-\tmle)^TV(\theta-\tmle)]-\E_{q_{-\ell}^{(t)}}[((\widetilde\theta_\ell,\theta_{-\ell})-\tmle)^TV((\widetilde\theta_\ell,\theta_{-\ell})-\tmle)]\\
=&\, V_{\ell\ell}\,\E_{q_{-\ell}^{(t)}}[(\theta_\ell-\widetilde\theta\,)^2]=V_{\ell\ell}\,[\theta_\ell-T_{1,\ell}(\mu^{(t)})]^2+V_{\ell\ell}\,\var_{q_{-\ell}^{(t)}}(\widetilde\theta\,).
\end{aligned}
\end{equation}
Therefore, an application of Lemma \ref{lem::onedlocalexpansion} with equations~\eqref{eqn:t1}-\eqref{eqn:t3} yields
\begin{align*}
\sum_{\ell=1}^d\int q_\ell^{(t+1)}\int q_{-\ell}^{(t)}\,(U_n-\widetilde U_{n,\ell})
= &\,-\frac{n}{2}\sum_{\ell=1}^dV_{\ell\ell}\left(\var_{q_\ell^{(t+1)}}(\theta_\ell)+\var_{q_{-\ell}^{(t)}}(\widetilde\theta\,)\right)\\
&+O\left(n\E_{q^{(t)}}\big[\|\theta-\theta^*\|^3\big]\right)+O\left(n\E_{q^{(t+1)}}\big[\|\theta-\theta^*\|^3\big]\right)+O(\alpha_n).
\end{align*}
We next analyze the third term of sum of log-normalizing constants in the last line of equation~\eqref{eqn:DL_dec}. Note that we can decompose
\begin{align*}
&\int_{\R}\exp\left\{\int_{\R^{d-1}} q_{-\ell}^{(t)}(\theta_{-\ell})\,(U_n-\widetilde U_{n,\ell})\dd\theta_{-\ell}\right\}\dd\theta_\ell  \\
=&\,\int_{|\theta_\ell-T_{1,\ell}(\mu^{(t)})|\le c_1\sqrt{\frac{\log n}{n}}}\exp\left\{\int_{\R^{d-1}} q_{-\ell}^{(t)}(\theta_{-\ell})\,(U_n-\widetilde U_{n,\ell})\dd\theta_{-\ell}\right\}\dd\theta_\ell\\
&\qquad+\int_{|\theta_\ell-T_{1,\ell}(\mu^{(t)})|> c_1\sqrt{\frac{\log n}{n}}}\exp\left\{\int_{\R^{d-1}} q_{-\ell}^{(t)}(\theta_{-\ell})\,(U_n-\widetilde U_{n,\ell})\dd\theta_{-\ell}\right\}\dd\theta_\ell,
\end{align*}
with some constant $c_1$ to be decided later. Using equations~\eqref{eqn:t1}-\eqref{eqn:t3}, we have
\begin{equation}
 \begin{aligned}
&\int_{|\theta_\ell-T_{1,\ell}(\mu^{(t)})|\le c_1\sqrt{\frac{\log n}{n}}}\exp\left\{\int_{\R^{d-1}} q_{-\ell}^{(t)}(\theta_{-\ell})\,(U_n-\widetilde U_{n,\ell})\dd\theta_{-\ell}\right\}\dd\theta_\ell\\
=&\int_{|\theta_\ell-T_{1,\ell}(\mu^{(t)})|\le c_1\sqrt{\frac{\log n}{n}}}\exp\left\{-nV_{\ell\ell}(\theta_\ell-T_{1,\ell}(\mu^{(t)}))^2/2+O(n|\theta_\ell-\theta_\ell^*|^3)\right\}\dd\theta_\ell\\
&\cdot O\left(\exp\left\{n\E_{q_{-\ell}^{(t)}}\|\theta_{-\ell}-\theta_{-\ell}^*\|^3+\alpha_n\right\}\right)\\
=&\left(\sqrt{2\pi/(nV_{\ell\ell})}+O(n^{-c_1^2V_{\ell\ell}})\right)\cdot O\left(\exp\left\{n\E_{q_{-\ell}^{(t)}}\|\theta_{-\ell}-\theta_{-\ell}^*\|^3+c_1^3\alpha_n\right\}\right).
\end{aligned}  
\label{eq::Z_l_interior}
\end{equation}
Moreover, by equation~\eqref{eq::un_quad_bound} in the proof of Lemma \ref{lem::kldif_to_normal}, we have
$$
\int_{|\theta_\ell-T_{1,\ell}(\mu^{(t)})|> c_1\sqrt{\frac{\log n}{n}}}\exp\left\{\int_{\R^{d-1}} q_{-\ell}^{(t)}(\theta_{-\ell})\,(U_n-\widetilde U_{n,\ell})\dd\theta_{-\ell}\right\}\dd\theta_\ell\\
=O(n^{-c_1^2V_{\ell\ell}}),$$
where we have used the inequality $\int_x^\infty \exp\{-y^2\} \dd y=O(\exp\{-x^2\})$ for $x>1$.
Now by putting pieces together with $c_1$ large enough such that $O(n^{-c_1^2V_{\ell\ell}})$ is $o(\alpha_n n^{-1})$, we can get
\begin{equation}
 \begin{aligned}
&\log\left(\int_{\R}\exp\left\{\int_{\R^{d-1}} q_{-\ell}^{(t)}(\theta_{-\ell})\,(U_n(\theta)-\widetilde U_{n,\ell}(\theta))\dd\theta_{-\ell}\right\}\dd\theta_\ell\right)\\
&\qquad\qquad =\frac{1}{2}\log \frac{2\pi}{nV_{\ell\ell}}-nV_{\ell\ell}\var_{q_{-\ell}^{(t)}}(\widetilde\theta)/2+O\left(n\E_{q_{\ell}^{(t)}}\|\theta_{\ell}-\theta_{-\ell}^*\|^3\right)+O(\alpha_n).
\end{aligned}   
\label{eq::logZ_l}
\end{equation}
Lastly, we analyze the second term in the last step of equation~\eqref{eqn:DL_dec}. Using inequality~\eqref{eq::pi_taylor} in the proof of Lemma \ref{lem::kldif_to_normal}, we obtain
\begin{align*}
&\int q^{(t+1)}\log \pi_n
=\int q^{(t+1)}\log \phi_n+n\E_{q^{(t+1)}}\|\theta-\tmle\|^3+O(\alpha_n) \\
=&\,\log C_V-n\E_{q^{(t+1)}}[(\theta-\tmle)^TV(\theta-\tmle)]/2+O(\alpha_n)\\
=&\,\log C_V-\frac{n}{2}(\mu^{(t+1)}-\tmle)^TV(\mu^{(t+1)}-\tmle)-\frac{n}{2}\sum_{\ell=1}^nV_{\ell\ell}\var_{q_\ell^{(t+1)}}(\theta_\ell)+O(\alpha_n).
\end{align*}
Putting all pieces together, we have
\begin{align}
D(q^{(t+1)}\verts\pi_n)=&\frac{1}{2}\sum_{\ell=1}^d \log \frac{nV_{\ell\ell}}{2\pi}-\log C_V+\frac{n}{2}(\mu^{(t+1)}-\tmle)^TV(\mu^{(t+1)}-\tmle) \notag\\
&+O\left(n\E_{q^{(t)}}\big[\|\theta-\theta^*\|^3\big]\right)+O\left(n\E_{q^{(t+1)}}\big[\|\theta-\theta^*\|^3\big]\right)+O(\alpha_n).\label{eqn:regret_bound}
\end{align}
The first two constant terms are independent of $t$, and the third and fourth moment terms converge to zero exponentially fast up to $O(\alpha_n)$ by \eqref{eq::exp_perturb_E}. It suffices to show the exponential convergence of the quadratic term, which requires a sharper result than \eqref{eq::exp_perturb}. To achieve this, we apply finer analysis similar to the derivation of \eqref{eq::logZ_l}, utilizing the fact that density function of $q_{\ell}^{(t)}(\theta_\ell)$ for each $\ell$ is locally symmetric around its mode (expectation), leading to the following lemma. A proof of this lemma is provided in Appendix~\ref{proof_lem:mean_perturbation}.
\begin{lemma}
\label{lem:mean_perturbation}
Under the assumptions of the theorem, there exists some constant $C>0$, such that for each $\ell\in[d]$,
\begin{align*}
  \big|\mu^{(t+1)}_\ell-T_{1,\ell}(\mu^{(t)}) \big|\leq C\,\big(n\E_{q_{-\ell}^{(t)}}\big[\|\theta_{-\ell}-\theta_{-\ell}^*\|^3\big]+\alpha_n\big)\,n^{-1/2}(\log n)^{3/2},
\end{align*}
where recall that $\alpha_n=K^3(\log n)^3/\sqrt{n}$.
\end{lemma}
Since $\E_{q_{-\ell}^{(t)}}\big[\|\theta_{-\ell}-\theta_{-\ell}^*\|^3\big]$ converges exponentially fast to $O([\log(n)/n]^{3/2})$ by \eqref{eq::exp_perturb_E}, we can conclude that $\mu^{(t)}-\tmle$ converges exponentially quickly to be within an  order of $O(n^{-1/2}\alpha_n\log^{3/2} n)$, using a similar argument that leads to inequality~\eqref{eq::exp_perturb}.
This result refines \eqref{eq::exp_perturb} and concludes the first statement. In particular, by sending $t\to\infty$, this bound leads to an error bound between the variational mean estimator $\widehat\theta_{VB}=\E_{\widehat q_\theta}[\theta]$ and the MLE $\tmle$as
\begin{equation}
\|\widehat\theta_{VB}-\tmle\|=O(n^{-1}(\log n)^{9/2}). 
\label{eq::VB_mean_error_bound}
\end{equation}
Now by using the above result, inequality~\eqref{eq::exp_perturb_E} and equation~\eqref{eqn:regret_bound}, we obtain
\begin{align*}
    D(q^{(t+1)}\verts\pi_n)\leq \frac{1}{2}\sum_{\ell=1}^d \log \frac{nV_{\ell\ell}}{2\pi}-\log C_V+ C\alpha^{t} + CK^3(\log n)^3/\sqrt{n},
\end{align*}
for some sufficiently large constant $C>0$. Since equation~\eqref{eqn:regret_bound} also applies to $\widehat q$ (by letting $t\to\infty$), we obtain
\begin{align*}
    D(\widehat q\verts\pi_n)\geq \frac{1}{2}\sum_{\ell=1}^d \log \frac{nV_{\ell\ell}}{2\pi}-\log C_V - CK^3(\log n)^3/\sqrt{n}.
\end{align*}
Now taking the difference between the two preceding displays leads to the first claimed inequality~\eqref{mainineq3}.

\medskip
Next, let us prove the second claimed inequality~\eqref{eqn:mainineq4} by comparing $q^{(t)}$ and $p^{(t)}$.
Define $\epsilon_\ell^{(t)}\deltaeq D(p^{(t)}_\ell\verts q^{(t)}_\ell)$, and $\epsilon^{(t)}=D(p^{(t)}\verts q^{(t)})\deltaeq\sum_{\ell=1}^d\epsilon_\ell^{(t)}$. Now we show that $\epsilon^{(t)}$ converges exponentially quickly to zero up to an $O(\alpha_n)$ term. In fact, we have
\begin{align*}
    \epsilon_\ell^{(t+1)}=&D(p_\ell^{(t+1)}\verts q_\ell^{(t+1)})\\
    =&\Exp_{p_\ell^{(t+1)}}\Big[\log p_\ell^{(t+1)}-\log q_\ell^{(t+1)}\Big]\\
    =&\Exp_{p_\ell^{(t+1)}}\Big[\Exp_{p_{-\ell}^{(t)}}[F_n-\widetilde G_{n,\ell}]-\Exp_{q_{-\ell}^{(t)}}[U_n-\widetilde U_{n,\ell}]+\log Z_\ell^{(t+1)}-\log Z_{\ell,0}^{(t+1)}\Big]\\
    =&\Exp_{p_\ell^{(t+1)}}\Exp_{q_{-\ell}^{(t)}}[(F_n-\widetilde G_{n,\ell})-(U_n-\widetilde U_{n,\ell})]+\Exp_{p_\ell^{(t+1)}}(\Exp_{p_{-\ell}^{(t)}}-\Exp_{q_{-\ell}^{(t)}})[F_n-\widetilde G_{n,\ell}]\\
    &+\left[\log Z_\ell^{(t+1)}-\log Z_{\ell,0}^{(t+1)}\right]\deltaeq \I(\ell)+\II(\ell)+\III(\ell).
\end{align*}
where $$ Z_\ell^{(t+1)}=\int\exp\{\Exp_{q_{-\ell}^{(t)}}[U_n-\widetilde U_{n,\ell}]\}\quad\mbox{and}\quad Z_{\ell,0}^{(t+1)}=\int\exp\{\Exp_{p_{-\ell}^{(t)}}[F_n-\widetilde G_{n,\ell}]\}$$ are normalizing constants. We will bound the three terms $\I(\ell)$, $\II(\ell)$ and $\III(\ell)$ separately.

For $\I(\ell)$, using inequality~\eqref{eq::un_expand} we have 
\begin{align*}
    |\I(\ell)|\le 2C_\delta n\left(\Exp_{p_\ell^{(t+1)}}\big[|\theta_\ell-\tmle_\ell|^3\big]+\Exp_{q_{-\ell}^{(t)}}\|\theta_{-\ell}-\tmle_{-\ell}\|^3\right).
\end{align*}
which converges exponentially quickly to be within $O(\alpha_n)$ according to inequality~\eqref{eq::exp_perturb_E}.

For $\II(\ell)$, we have 
\begin{align*}
\II(\ell)=&\,-nV_{\ell\ell}\Exp_{p_\ell^{(t+1)}}(\Exp_{p_{-\ell}^{(t)}}-\Exp_{q_{-\ell}^{(t)}})[(\theta_\ell-\widetilde\theta_\ell)^2/2]\\
=&\,-nV_{\ell\ell}\Exp_{p_\ell^{(t+1)}}\Exp_{p_{-\ell}^{(t)}}[(\theta_\ell-\widetilde\theta_\ell)^2/2]+nV_{\ell\ell}\Exp_{p_\ell^{(t+1)}}\Exp_{q_{-\ell}^{(t)}}[(\theta_\ell-\widetilde\theta_\ell)^2/2] \\
=&\,\frac{nV_{\ell\ell}}{2}\left[-\var_{p_{-\ell}^{(t)}}(\widetilde\theta_\ell)+\var_{q_{-\ell}^{(t)}}(\widetilde\theta_\ell)+(\Exp_{p_\ell^{(t+1)}}\theta_\ell-\Exp_{q_{-\ell}^{(t)}}\widetilde\theta_\ell)^2\right].
\end{align*}
Note that 
\begin{align*}
 &\,nV_{\ell\ell}\var_{p_{-\ell}^{(t)}}(\widetilde\theta_\ell)/2+\log Z_{\ell,0}^{(t+1)}\\
=&\,\log\int\exp\{\Exp_{p_{-\ell}^{(t)}}[-nV_{\ell\ell}(\theta_\ell-\widetilde\theta_\ell)^2/2]\}\dd \theta_{-\ell} +nV_{\ell\ell}\var_{p_{-\ell}^{(t)}}(\widetilde\theta_\ell)/2\\
=&\,\log\int\exp\{-nV_{\ell\ell}(\theta_\ell-\Exp_{p_{-\ell}^{(t)}}\widetilde\theta_\ell)^2/2\}\dd \theta_{-\ell}=\frac{1}{2}\log\frac{2\pi}{nV_{\ell\ell}}.  
\end{align*}
From \eqref{eq::un_expand} and \eqref{eq::exp_perturb_E}, we have 
\begin{equation}
\Exp_{p_\ell^{(t+1}}\Exp_{q_{-\ell}^{(t)}}[U_n(\theta)-\widetilde U_{n,\ell}] = -\frac{nV_{\ell\ell}}{2}\left[(\Exp_{p_\ell^{(t+1)}}\theta_\ell-\Exp_{q_{-\ell}}^{(t)}\widetilde\theta_\ell)^2+\var_{q_{-\ell}^{(t)}}(\widetilde\theta_\ell)\right]
+O(\alpha_n).
 \label{localonedexpansion}
\end{equation}
On the other hand, by \eqref{eq::logZ_l} we have
\begin{align*}
 &\,\frac{nV_{\ell\ell}}{2}\left[\var_{q_{-\ell}^{(t)}}(\widetilde\theta_\ell)+(\Exp_{p_\ell^{(t+1)}}\theta_\ell-\Exp_{q_{-\ell}^{(t)}}\widetilde\theta_\ell)^2\right]+\log Z_\ell^{(t+1)}\\
=&\,\Exp_{p_\ell^{(t+1)}}\log\int\exp\{\Exp_{q_{-\ell}^{(t)}}[U_n-\widetilde U_{n,\ell}]\}+\frac{nV_{\ell\ell}}{2}\left[\var_{q_{-\ell}^{(t+1)}}(\widetilde\theta_\ell)+(\Exp_{p_\ell^{(t+1)}}\theta_\ell-\Exp_{q_{-\ell}^{(t)}}\widetilde\theta_\ell)^2\right]\\
=&\,\frac{1}{2}\log \frac{2\pi}{nV_{\ell\ell}}+R_{n,\ell}+O(\alpha_n)+nV_{\ell\ell}(\Exp_{p_\ell^{(t+1)}}\theta_\ell-\Exp_{q_{-\ell}^{(t)}}\widetilde\theta_\ell)^2/2\\
=&\,\frac{1}{2}\log \frac{2\pi}{nV_{\ell\ell}}+R_{n,\ell}+O(\alpha_n)+nV_{\ell\ell}\left((\E_{p_\ell^{(t+1)}}\theta-\tmle)^2+(\E_{q_{-\ell}^{(t)}}\widetilde\theta_\ell-\tmle_\ell)^2\right),
\end{align*}
where the remainder term $R_{n,\ell}\le 2C_\delta n\E_{q_{-\ell}^{(t)}}\big[|\theta_{-\ell}-\tmle_{-\ell}|^3\big]$ by \eqref{eq::un_expand}.
Putting all pieces together, we obtain
\begin{equation}
\begin{aligned}
 \epsilon^{(t)}\le &\,2C_\delta n\Exp_{p^{(t+1)}}\|\theta-\tmle\|^3+4C_\delta dn\Exp_{q^{(t)}}\|\theta-\tmle\|^3\\
 &\,+\sum_{\ell=1}^dnV_{\ell\ell}\left((\E_{p_\ell^{(t+1)}}\theta-\tmle)^2+(\E_{q_{-\ell}^{(t)}}\widetilde\theta_\ell-\tmle_\ell)^2\right)+O(\alpha_n),
\end{aligned}
 \label{klepsilont}
\end{equation}
and hence $\epsilon^{(t)}$ converges exponentially quickly to be within $O(\alpha_n)$, i.e.,
$$
\E[D(p^{(t)}\verts q^{(t)})]\le \alpha^t \widetilde C+C\alpha_n,
$$
where
\begin{align*}
\widetilde C=&\,2C_\delta\, n\,\Exp_{p^{(1)}}\big[\|\theta-\tmle\|^3\big]+4C_\delta \,dn\,\Exp_{q^{(0)}}\big[\|\theta-\tmle\|^3\big]\\
 &\,+\sum_{\ell=1}^dnV_{\ell\ell}\left((\E_{p_\ell^{(1)}}\theta-\tmle)^2+(\E_{q_{-\ell}^{(0)}}\widetilde\theta_\ell-\tmle_\ell)^2\right)+O(\alpha_n).
\end{align*}
Pinsker's inequality implies $H(p,q)\le \sqrt{D(p\verts q)/2}$ for any distributions $p$ and $q$; therefore, we can apply Jensen's inequality and the triangle inequality to finally conclude 
\begin{align*}
\E[H(q^{(t)},\widehat q \,)]
\le &\, \E[H(p^{(t)},q^{(t)})]+\E[H(p^{(t)},\phi^*)]+H(\phi^*,\widehat q \,)\\
\le&\, \E\left[\sqrt{D(p^{(t)}\verts q^{(t)})/2}\right]+\E\left[\sqrt{D(p^{(t)}\verts \phi^*)/2}\right]+\sqrt{D(\widehat q\verts\phi^*)/2}\\
\le&\, \sqrt{\E\left[D(p^{(t)}\verts q^{(t)})\right]/2}+\sqrt{\E\left[D(p^{(t)}\verts \phi^*)\right]/2}+\sqrt{D(\widehat q\verts\phi^*)/2}\\
\le &\,\alpha^{t/2}\sqrt{\widetilde C}+\alpha^{t/2}\sqrt{D(p^{(0)}\verts \phi^*)}+C\sqrt{\alpha_n},
\end{align*}
where we have used Theorem \ref{thm::main1} to bound the $D(\widehat q\verts\phi^*)$ term. This proves the second claimed inequality~\eqref{eqn:mainineq4}.

\subsection{Proof of Theorem~\ref{thm::caviconv2}}
\label{sec::cavi_latent}
This subsection includes the theoretical analysis for the CAVI algorithm with latent variables. In this case, the one-step update from $q_{\theta} \otimes q_{S^n}$ to $q^\ast_{\theta} \otimes q^\star_{S^n}$ can be summarized as follows: for each $i \in [n]$, we have
\begin{equation}
\qs i ^{*}\left(s_i\right) \propto \exp \left\{\int_{\R^{d}} q_\theta\left(\theta\right) \log p\left(s_i \mid \theta, X_i\right) \,\dd\theta\right\},
\label{cavi_s}
\end{equation}
and each $j\in[d]$, we have
\begin{equation}
\qt j^{*}\left(\theta_{j}\right) \propto \exp \left\{\int_{\R^n}\int_{\R^{d-1}} q_{S^n}^\ast(s^n)\qt{-j}\left(\theta_{-j}\right) \log p\left(\theta_{j} \mid \theta_{-j}, \,X^n,s^n\right) \,\dd\theta_{-j}\dd s^n\right\}.
\label{cavi_theta}
\end{equation}
Recall that we used the shorthand $V=V(\theta^\ast)$ to denote the observed data Fisher information, $V_s = V_s(\theta^\ast)$ the missing data Fisher information, and $V_c=V+V_s$ the complete data Fisher information.
Let $U_n = \sum_{i=1}\sum_{s\in\m S} \nabla \ell_S(\theta^\ast, X_i, s)\, p(S_i=s\,|\,X_i,\theta^\ast)$. It is standard and straightforward to show that under Assumptions~\ref{ass::prior}-\ref{ass::ls}, we have that $n^{-1/2} \,U_n$ converges in distribution to $N(0, V)$, and the MLE $\tmle$ of $\theta$ satisfies $\|\tmle - \theta^\ast-(n V)^{-1} U_n\| =O_p(n^{-1})$ so that $\sqrt{n}\,(\tmle-\theta^\ast)$ converges in distribution to $N(0,V^{-1})$. 
By using this property and following a similar type of analysis as in \cite{wei2019} based on Taylor expansions (first analyze $\qs i ^{*}$ and then plug-in it into $\qt j^{*}$), we can show (see Appendix \ref{sec::cavi_latent_gauss} for a proof) that
\begin{equation}
\begin{aligned}
    \qt j^{*}(\theta_j) \propto &\,\exp\Big\{  O(n^{-1/2}) +O(\E_{q_\theta}\|\theta-\theta^* \|) + O(n^{1/2} \E_{q_{\theta}}\|\theta-\theta^*\|^2) +O(n\E_{q_{\theta}}\|\theta-\theta^*\|^3) \\
    &\,\qquad -\frac{n}{2}\, [V_c]_{jj}\, (\theta_j-\theta_j^\ast)^2 + n(\theta_j-\theta_j^\ast) \sum_{k=1}^d V_{jk} \,(\tmle_k - \theta_k^\ast) \\
    &\, + n (\theta_j-\theta_j^\ast) \sum_{k=1}^d [V_s]_{jk} \,\big(\mb E_{q_{\theta_k}}[\theta_k] -\theta^\ast_k\big) -  n (\theta_j-\theta_j^\ast) \sum_{k\neq j} [V_c]_{jk} \,\big(\mb E_{q_{\theta_k}}[\theta_k] -\theta^\ast_k\big)\Big\}.
\label{cavi_latent_gauss}
\end{aligned}
\end{equation}
Note that without the first four high-order remainder terms in the first line in the preceding expansion, the corresponding $\qt j^{*}(\theta_j)$ will be a Gaussian distribution. In fact, we can remove the two terms $O(\E_{q_\theta}\|\theta-\theta^* \|)$ and $O(n^{1/2} \E_{q_{\theta}}\|\theta-\theta^*\|^2)$ from the above expression as the basic inequality implies that they can be bounded by $O(n^{-1/2}+n\E_{q_{\theta}}\|\theta-\theta^*\|^3)$, which coincides with the remainder terms in the previous perturbation analysis for CAVI without latent variables.

In the rest of the proof, we will primarily focus on this Gaussian (or quadratic) case and derive the corresponding convergence rate of the CAVI. Such an analysis is similar to that in the proof of Theorem \ref{thm::caviconv} for CAVI without latent variables, where we analyzed the CAVI algorithm for the Gaussian posterior, resulting in equation~\eqref{eqn:lo} that governs the convergence of CAVI. A subsequent perturbation analysis, which extends from the Gaussian case to the general case by keeping track of the higher-order remainder terms, can be carried out in a completely analogous manner (as in the proof of Theorem \ref{thm::caviconv}); however, since it is extremely involved and tedious, and does not provide much additional insight, we omit the details of such a perturbation analysis.

\noindent {\bf Gaussian case.}
As we discussed above, we will focus on analyzing the convergence of CAVI for the following Gaussian case, where $\qt \ell^{(t+1)}$ is updated from $\qt \ell^{(t)}$ according to
\begin{align*}
    \qt \ell^{(t+1)}(\theta_\ell) \propto \exp\Big\{ &\, -\frac{n}{2}\, [V_c]_{\ell\ell}\, (\theta_\ell-\theta_\ell^\ast)^2 + n(\theta_\ell-\theta_\ell^\ast) \sum_{k=1}^d V_{\ell k} \,(\tmle_k - \theta_k^\ast) \\
    &\, + n (\theta_\ell-\theta_\ell^\ast) \sum_{k=1}^d [V_s]_{\ell k} \,\big(\mb E_{q_{\theta_k}^{(t)}}[\theta_k] -\theta^\ast_k\big) -  n (\theta_\ell-\theta_\ell^\ast) \sum_{k\neq \ell} [V_c]_{jk} \,\big(\E_{q_{\theta_k}^{(t)}}[\theta_k] -\theta^\ast_k\big)\Big\}\\
    &\, :\,= \exp\Big\{-\frac{n}{2}\, [V_c]_{\ell\ell}\, \theta_\ell^2 + L^{(t)} \theta_\ell + C^{(t)}\Big\}.
\end{align*}
The expression inside the exponential is a quadratic form of $\theta_\ell$ with the leading term equal to $-\frac{n}{2}[V_c]_{\ell\ell}\theta_\ell^2$. Therefore, similar to the analysis of CAVI for Gaussian posterior without latent variables, after a warm-up during which every component has been updated at least once, we have $q^{(t)}_{\theta}$ following a multivariate normal distribution with diagonal variance matrix $[nS_c]^{-1}$, where $S_c=\diag(V_c)$. Similarly, we denote $S=\diag(V)$ and $S_s=\diag(V_s)$, which will be used afterwards. It is also easy to see that the coefficient $L^{(t)}$ for the linear term in the preceding display is
\begin{align*}
&n[V_c]_{\ell\ell}\,\theta_\ell^*+n\sum_{k=1}^dV_{\ell k}\,(\tmle_k-\theta^*_k)+n\sum_{k=1}^d[V_s]_{\ell k}\,\big(\E_{q^{(t)}_{\theta_k}}[\theta_k]-\theta^*_k\big)-n\sum_{k\not=\ell}[V_c]_{\ell k}\,\big(\E_{q^{(t)}_{\theta_k}}[\theta_k]-\theta^*_k\big)\\
=&\,n[V_c]_{\ell \ell}\,\E_{q^{(t)}_{\theta_\ell}}[\theta_k]-n\sum_{k=1}^dV_{\ell k}\,\big(\E_{q^{(t)}_{\theta_k}}[\theta_k]-\tmle_k\big).
\end{align*}
Correspondingly, the center (mean) of $q^{(t+1)}_{\theta_\ell}$ (as a normal distribution) is given by $(n[V_c]_{\ell\ell})^{-1}L^{(t)}$, so that $q^{(t+1)}_{\theta_\ell} = N\big(L^{(t)}/(n[V_c]_{\ell\ell}), \,[nS_c]^{-1}\big)$.

Similar to the analysis of CAVI for Gaussian posteriors without latent variables, we define $b^{(t)}=\E_{q^{(t)}_\theta}[\theta]-\tmle$. The updating formula for the $\ell$-th component $b^{(t+1)}_\ell$ of $b^{(t)}_\ell$ using a full step size can be accordingly written as
$$
b^{(t+1)}_\ell=b^{(t)}_\ell-\sum_{k=1}^{d}[V_c]_{\ell\ell}^{-1}V_{\ell k}\,b_\ell^{(t)}.
$$
Generally, for a partial step size $\gamma\in (0, 1]$, we can adapt the above updating formula into
\begin{align}\label{eqn:leading_up}
  b^{(t+1)}=b^{(t)}-\gamma S_c^{-1}Vb^{(t)}:\,=A_\gamma b^{(t)}.
\end{align}
It is interesting to note that this updating formula reduces to the updating formula~\eqref{eq::bconv} for Gaussian posteriors without latent variables by replacing $S_c$ with $S$; correspondingly, $A_\gamma=(I-\gamma \,S_c^{-1}V)$ simplifies to the $(I-\gamma \,S^{-1}V)$ therein. Now we consider the two update schemes separately as before.

\smallskip
\begin{itemize}
    \item Sequential update (randomized): This case is similar to the earlier analysis without latent variables, by simply replacing the step size $\gamma$ therein with $\gamma V_{\ell \ell}/[V_c]_{\ell \ell}$ throughout. In particular, we can obtain that after one iteration,
\begin{align*}    
&D^{(t)}-\mb E[D^{(t+1)}\mid D^{(t)}]=  \frac{1}{d}\sum_{\ell=1}^{d}\frac{n\gamma(2-\gamma\frac{V_{\ell \ell}}{[V_c]_{\ell \ell}})}{2[V_c]_{\ell \ell}}\left(V_{\ell\ell} b_\ell^{(t)}+\sum_{k \neq \ell} b_{k}^{(t)} V_{k\ell}\right)^{2}\\
&\qquad\qquad=\frac{n\gamma}{2d}\,b^{(t)T}V S_c^{-1}(2S_c-\gamma S) S_c^{-1}V b^{(t)}
\geq\frac{c_{s} (\gamma) n\gamma}{2d}\,b^{(t)T}V  b^{(t)}=\frac{c_{s}(\gamma)\gamma}{d}D^{(t)},
\end{align*}
where 
$$
c_s(\gamma)=\max_{\|b\|=1}\frac{b^TVS_c^{-1}(2S_c-\gamma S) S_c^{-1}V b}{b^TVb}.
$$
By taking another expectation, we can finally obtain
$\Exp[D^{(t+1)}]\leq (1-c_s(\gamma)\gamma/d)\,\Exp[D^{(t)}]$ and therefore $\Exp[D^{(t)}]\leq (1-c_s(\gamma)\gamma/d)^t\, D^{(0)}$. 
Again, we can always take $\gamma=1$ to achieve largest decrease in $D^{(t)}$ for the sequential update. It is interesting to compare this result to the earlier result without latent variables at $\gamma=1$. Specifically, we have
$$c_{s}(1)=\max_{\|b\|=1}\frac{b^TVS_c^{-1}(2S_c-S)S_c^{-1}V b}{b^TVb}\ge \beta(2-\beta)\cdot\max_{\|b\|=1}\frac{b^TVS^{-1}V b}{b^TVb}= \beta (2-\beta) \,c_s,$$
where $\beta=\min_{1\le\ell\le d}\frac{V_{\ell\ell}}{[V_c]_{\ell\ell}}\in [0,1]$ and $c_s$ is the corresponding factor given by~\eqref{gauss_c} for the no latent variable case. Therefore, the presence of latent variables will slow down the reduction of the regret $D^{(t)}$ by a factor of $\beta (2-\beta)\in[0,1]$.

\item Parallel update with $d$ coordinates: similar to the earlier analysis without latent variables and the sequential update analysis, we have that for the parallel update,
$$
D^{(t+1)}=\frac{n}{2}\, (b^{(t+1)})^TVb^{(t+1)}= \frac{n}{2}\, (b^{(t)})^TA_\gamma V A_\gamma b^{(t)}.
$$
Therefore, we can choose $\gamma$ sufficiently small to guarantee the matrix $c_p(\gamma)\,V- A_\gamma V A_\gamma$ to be positive semidefinite, where 
$$
c_p(\gamma)=\max_{\|b\|=1}\frac{b^TA_\gamma VA_\gamma b}{b^TVb}=\max_{\|b\|=1}\frac{b^T(I-\gamma S_c^{-1}V)V(I-\gamma S_c^{-1}V)b}{b^TVb} \in [0,1].
$$
This will lead to the geometric convergence of the parallel update, or $D^{(t+1)} \leq c_p(\gamma) D^{(t)}$, as claimed in the theorem.
\end{itemize}

\section{Proof of Technical Lemmas}
\subsection{Proof of Lemma \ref{lem::kltri}}
\label{pf::kltri}
We omit the terms of $\|\theta-\theta^*\|^L$ to simplify the notations of the proof, as the argument trivially generalizes due to the sub-Gaussian properties (these terms only contribute a higher-order remainder term).

Denote the density function of $N(\tmle,[nV(\theta^*)]^{-1})$ as $\phi_n$.
Note that the posterior can be written as
$$
\pi_n(\theta)=\frac{\exp\{\ell(\theta;X^n)-\ell(\tmle;X^n)\}\pi(\theta)}{\int\exp\{\ell(\theta;X^n)-\ell(\tmle;X^n)\}\pi(\theta)\di\theta}.
$$
The proof is based on the following lemma which slightly refines Lemma 10 in \cite{wei2019} and characterizes the denominator of $\pi_n$. Note that the we have removed the unnecessary power $d$ in the original lemma which turns out to be unnecessary.
\begin{lemma}
\label{lem::post_expansion}
Under Assumption, for any $1\le K=O(\sqrt{n})$, we have
$$
\left|\frac{\int\exp\{\ell(\theta;X^n)-\ell(\tmle;X^n)\}\pi(\theta)\di\theta}{(2\pi/n)^{d/2}|V(\theta^*)|^{-1/2}\pi(\theta^*)}-1\right|\le \frac{CK^3(\log n)^{3}}{\sqrt{n}}
$$
holds with probability at least $1-CK^{-2}$.
\end{lemma}
 Using this lemma, we can write the log ratio between $\pi_n$ and $\phi_n$ as
\begin{align*}
\log\frac{\pi_n}{\phi_n}=&\ell(\theta;X^n)-\ell(\tmle;X^n)-\frac{n}{2}(\theta-\tmle)^TV(\theta^*)(\theta-\tmle)\\
&+\log\frac{\pi(\theta)}{\pi(\theta^*)}+\log(1+C\alpha_n|V(\theta^*)|^{1/2}),
\end{align*}
where recall that $\alpha_n=\frac{K^3(\log n)^{3}}{\sqrt{n}}$.

Note that from Assumption~\ref{ass::l}, we can apply a Taylor expansion to get
\begin{align*}
&\ell(\theta;X^n)-\ell(\tmle;X^n)-\frac{n}{2}(\theta-\tmle)^TV(\theta^*)(\theta-\tmle)\\
=&\frac{1}{2}(\theta-\tmle)^T[\nabla^2\ell(\tmle;X^n)-nV(\theta^*)](\theta-\tmle)+R_n,
\end{align*}
with
$$
R_n\le \sum_i |Z(X_i)|\cdot\|\theta-\tmle\|^3\le K^2 \|\theta-\tmle\|^3,
$$
under the same Markov-inequality-based tail event as in the proof of Lemma~\ref{lem::post_expansion}.
In addition, we also have
$$
|\nabla^2\ell(\tmle;X^n)-nV(\theta^*)|\le |\tmle-\theta^*| \sum_i |Z(X_i)|\le\frac{CK^3(\log n)^{1/2}}{\sqrt{n}},
$$
as $\|\tmle-\theta^*\|\le\frac{CK(\log n)^{1/2}}{\sqrt{n}}$ with probability at least $1-CK^{-2}$.

Since Assumption \ref{ass::prior} essentially implies that $|\log\pi(\theta)-\log\pi(\theta^*)|\le C(\|\theta-\theta^*\|+\|\theta-\theta^*\|^2)$ for some constant $C$, by putting all pieces together we have
\begin{equation}
\begin{aligned}
\log\frac{\pi_n}{\phi_n}\le&\Big|\ell(\theta;X^n)-\ell(\tmle;X^n)-\frac{n}{2}(\theta-\tmle)^TV(\theta^*)(\theta-\tmle)\Big|\\
&+C(\|\theta-\theta^*\|+\|\theta-\theta^*\|^2)+C\alpha_n\\
\le&\frac{CK^3(\log n)^{1/2}}{\sqrt{n}}\|\theta-\tmle\|^2+K^2\|\theta-\tmle\|^3+C(\|\theta-\theta^*\|+\|\theta-\theta^*\|^2)+C\alpha_n\\
\le&\frac{CK^3(\log n)^{1/2}}{\sqrt{n}}\|\theta-\theta^*\|^2+K^2\|\theta-\theta^*\|^3+C(\|\theta-\theta^*\|+\|\theta-\theta^*\|^2)+C\alpha_n
\end{aligned}
\label{eq::post_ratio}
\end{equation}
Taking the integral with respect to $Q_\theta$ and using (\ref{eq::mmt_bd}), we can get
\begin{align*}
&\left|D(Q_\theta \verts \Pi_n)-D(Q_\theta\verts N(\tmle,[n V(\theta^*)]^{-1}))\right|\\
=&\left|\int \log \left(\frac{\pi_n}{\phi_n}\right)(\theta) \di Q_\theta(\theta)\right| \le \frac{CK^3(\log n)^3}{\sqrt{n}}+\frac{CK^5(\log n)^3}{n^{3/2}}\le \frac{CK^3(\log n)^3}{\sqrt{n}},
\end{align*}
as we have assumed $K=O(\sqrt{n})$.

\subsection{Proof of Lemma \ref{lem::kldif_to_normal}}
\label{pf::kldif_to_normal}
Using the AM-GM inequality, we have
$\|\theta-\theta^*\|\le \frac{2}{3\sqrt{n}}+\frac{n}{3}\|\theta-\theta^*\|^3\le \alpha_n+\frac{n}{3}\|\theta-\theta^*\|^3$, and $\|\theta-\theta^*\|^2\le \frac{1}{3n^2}+\frac{2n}{3}\|\theta-\theta^*\|^3\le \alpha_n+\frac{2n}{3}\|\theta-\theta^*\|^3$. Combining these inequalities with (\ref{eq::post_ratio}), we can get
\begin{equation}
\log(\pi_n/\phi_n)\le C(n\|\theta-\theta^*\|^3+\alpha_n).
\label{eq::pi_taylor}
\end{equation}
We conclude the proof by taking expectation with respect to $q$.

\subsection{Proof of Lemma \ref{lem::onedlocalexpansion}}
\label{pf::onedlocalexpansion}
By Assumption \ref{ass::l}, there exists some constant $C$ that depends on $\delta$ such that for all $\theta\in \mbox{supp}(q)$,
$$
|U_n(\theta)-U_n(\tmle)- G_n(\theta)|\le C_\delta n\|\theta-\tmle\|^3,
$$
As $C_\delta$ is decreasing in $\delta$, we may find a constant $\delta$ small enough such that for any $\theta$ in $B_{\tmle}(\delta)$ we have
$$\|\theta-\tmle\|^3\le (\theta-\tmle)^TV(\theta-\tmle)/(4C_\delta)=-G_n(\theta)/(2nC_\delta).$$
As a result, for any $\theta\in B_{\tmle}(\delta)$ we have
\begin{equation}
|U_n(\theta)-U_n(\tmle)-  G_n(\theta)|\le nC_\delta\|\theta-\tmle\|^3 \le -G_n(\theta)/2,
\label{eq::un_expand}
\end{equation}
or equivalently
$$
G_n(\theta)/2\le U_n(\theta)-U_n(\tmle)-  G_n(\theta)\le  - G_n(\theta)/2.
$$
It is easy to see that the construction of $\widetilde \theta_\ell$ is coming from the convergence analysis for the Gaussian posterior as in the proof of Lemma \ref{lem::gauss_cavi}, implying that $(\widetilde\theta_\ell(\gamma),\theta_{-\ell})=(T_{\gamma,l}(\theta), \theta_{-\ell})$ is a contraction (after re-centering at $\tmle$). As a result, we have $(\widetilde\theta_\ell(\gamma),\theta_{-\ell})\in B_{\tmle}(\delta)$.
For simplicity of notation, we mainly consider $\gamma=1$, although the following argument holds for any $\gamma$ satisfying the condition of Lemma \ref{lem::gauss_cavi}. Plugging-in $\theta_\ell=\widetilde\theta_\ell$ and using the definition of $\widetilde U_{n,\ell}$ and $\widetilde G_{n,\ell}$, we get
\begin{equation}
\widetilde G_{n,\ell}(\theta)/2\le\widetilde U_{n,\ell}(\theta)-U_n(\tmle)- \widetilde G_{n,\ell}(\theta)\le  -\widetilde G_{n,\ell}(\theta)/2.
\label{eq::un_tilde_expand}
\end{equation}
Combine \eqref{eq::un_expand} and \eqref{eq::un_tilde_expand}, we have
\begin{equation}
|U_n(\theta)-\widetilde U_{n,\ell}(\theta) - (G_n(\theta)-\widetilde G_{n,\ell}(\theta))|\le \widetilde G_{n,\ell}(\theta)/2-G_n(\theta)/2.
\label{eq::perturb}
\end{equation}
Finally, by taking the expectation with respect to $q_{-\ell}$, and recalling $\widetilde G_{n,\ell}(\theta)=G_n(\theta)+nV_{\ell\ell}(\theta_\ell-\widetilde\theta_\ell)^2/2$, we obtain
\begin{equation}
\begin{aligned}
&\left|\Exp_{q_{-\ell}}[U_n(\theta)-\widetilde U_{n,\ell}(\theta)]+nV_{\ell\ell}(\theta_\ell-\Exp_{q_{-\ell}}\widetilde\theta_\ell)^2/2+nV_{\ell\ell}\var_{q_{-\ell}}(\widetilde\theta_\ell)/2\right|\\ 
\le & \, nV_{\ell\ell}(\theta_\ell-\Exp_{q_{-\ell}}\widetilde\theta_\ell)^2/4+nV_{\ell\ell}\var_{q_{-\ell}}(\widetilde\theta_\ell)/4.    
\end{aligned}
\label{eq::un_quad_bound}
\end{equation}
This implies that as long as $\ell$-th component of $q^{(t)}$ has been updated in CAVI algorithm with full step size, we have that $\sqrt{n}\theta_\ell$ follows some non-degenerate distribution, and hence $
n\var_{q_\ell^{(t)}}(\theta_\ell)=O(1).$ In particular, this property applies to the initialization $q_{\ell}^{(0)}$ constructed in the statement of the lemma (based on a full step size CAVI). Similarly, if every component has been updated in CAVI algorithm with full step size, we conclude that $n\var_{q_{-\ell}^{(t)}}(\widetilde\theta_\ell)=O(1)$, which also applies to the constructed $q_{\ell}^{(0)}$. Moreover, as long as $q^{(t)}$ satisfies these properties, all future iterations will satisfy the same properties due to~\eqref{eq::un_quad_bound}.
Therefore, we conclude that for each $t\geq 0$, $q_\ell^{(t+1)}$ is sub-Gaussian and centers at $\Exp_{q_{-\ell}^{(t)}}\widetilde\theta_\ell=\E_{q_{-\ell}^{(t)}}[T_{1,\ell}(\theta)]=T_{1,\ell}(\E_{q^{(t)}}[\theta])$, as in the lemma.

\subsection{Proof of Lemma~\ref{lem:mean_perturbation}}\label{proof_lem:mean_perturbation}
By definition, we have
\begin{align*}
\mu^{(t+1)}_\ell-T_{1,\ell}(\mu^{(t)})=&\frac{\int_{\R}(\theta_\ell-\tmle_\ell)\exp\{\int_{\R^{d-1}} q_{-\ell}^{(t)}(U_n(\theta)-\widetilde U_{n,\ell}(\theta))\dd\theta_{-\ell}\}\dd\theta_\ell}{\int_{\R}\exp\{\int_{\R^{d-1}} q_{-\ell}^{(t)}(U_n(\theta)-\widetilde U_{n,\ell}(\theta))\dd\theta_{-\ell}\}\dd\theta_\ell}.
\end{align*}
The denominator is already analyzed in~\eqref{eq::logZ_l}.
It remains to analyze the numerator. We make the same decomposition as the proof of~\eqref{eq::logZ_l} with some constant $c_2$ to be decided later,
\begin{align*}
&\int_{\R}(\theta_\ell-T_{1,\ell}(\mu^{(t)}))\exp\left\{\int_{\R^{d-1}} q_{-\ell}^{(t)}(\theta_{-\ell})\,(U_n-\widetilde U_{n,\ell})\dd\theta_{-\ell}\right\}\dd\theta_\ell  \\
=&\int_{|\theta_\ell-T_{1,\ell}(\mu^{(t)})|\le c_2\sqrt{\frac{\log n}{n}}}(\theta_\ell-T_{1,\ell}(\mu^{(t)}))\exp\left\{\int_{\R^{d-1}} q_{-\ell}^{(t)}(\theta_{-\ell})\,(U_n-\widetilde U_{n,\ell})\dd\theta_{-\ell}\right\}\dd\theta_\ell\\
&+\int_{|\theta_\ell-T_{1,\ell}(\mu^{(t)})|> c_2\sqrt{\frac{\log n}{n}}}(\theta_\ell-T_{1,\ell}(\mu^{(t)}))\exp\left\{\int_{\R^{d-1}} q_{-\ell}^{(t)}(\theta_{-\ell})\,(U_n-\widetilde U_{n,\ell})\dd\theta_{-\ell}\right\}\dd\theta_\ell.
\end{align*}
We have by \eqref{eq::un_quad_bound},
$$
\int_{|\theta_\ell-T_{1,\ell}(\mu^{(t)})|> c_2\sqrt{\frac{\log n}{n}}}(\theta_\ell-T_{1,\ell}(\mu^{(t)}))\exp\left\{\int_{\R^{d-1}} q_{-\ell}^{(t)}(\theta_{-\ell})\,(U_n-\widetilde U_{n,\ell})\dd\theta_{-\ell}\right\}\dd\theta_\ell\\
=O(n^{-\frac{1}{2}-c_2^2V_{\ell\ell}}),$$
as $\int_x^\infty y\exp\{-y^2\} \dd y=O(\exp\{-x^2\})$, and
\begin{align*}
&\int_{|\theta_\ell-T_{1,\ell}(\mu^{(t)})|\le c_2\sqrt{\frac{\log n}{n}}}(\theta_\ell-T_{1,\ell}(\mu^{(t)}))\exp\left\{\int_{\R^{d-1}} q_{-\ell}^{(t)}(\theta_{-\ell})\,(U_n-\widetilde U_{n,\ell})\dd\theta_{-\ell}\right\}\dd\theta_\ell\\
=&\frac{1}{\sqrt{n}}\int_{|\theta_\ell-T_{1,\ell}(\mu^{(t)})|\le c_2\sqrt{\frac{\log n}{n}}}\sqrt{n}(\theta_\ell-T_{1,\ell}(\mu^{(t)}))\exp\left\{-nV_{\ell\ell}(\theta_\ell-T_{1,\ell}(\mu^{(t)}))^2/2+O(n|\theta_\ell-\theta_\ell^*|^3)\right\}\dd\theta_\ell\\
&\cdot O\left(\exp\left\{n\E_{q_{-\ell}^{(t)}}\|\theta_{-\ell}-\theta_{-\ell}^*\|^3+\alpha_n\right\}\right)\\
=&n^{-1/2}\cdot O(\exp\{c_2^3\log^{3/2}n/\sqrt{n}\})\cdot O\left(\exp\left\{n\E_{q_{-\ell}^{(t)}}\big[\|\theta_{-\ell}-\theta_{-\ell}^*\|^3\big]+\alpha_n\right\}\right)\\
=& O\left(n^{-1}c_3^3(n\E_{q_{-\ell}^{(t)}}\big[\|\theta_{-\ell}-\theta_{-\ell}^*\|^3\big]+\alpha_n)\log^{3/2}n\right)
\end{align*}
where note that the exponential convergence of $n\E_{q_{-\ell}^{(t)}}\big[\|\theta_{-\ell}-\theta_{-\ell}^*\|^3\big]$ up to a higher-order term from~\eqref{eq::exp_perturb_E} implies that it is bounded and hence we can take away the exponentially small term therein. Finally, we take $c_2$ large enough so that $O(n^{-\frac{1}{2}-c_2^2V_{\ell\ell}})$ is $o(n^{-2})$.
Combining with \eqref{eq::logZ_l}, we can conclude that
$$
\big|\mu^{(t+1)}_\ell-T_{1,\ell}(\mu^{(t)})\big|=O\left(n^{-1/2}(n\E_{q_{-\ell}^{(t)}}\big[\|\theta_{-\ell}-\theta_{-\ell}^*\|^3\big]+\alpha_n)\log^{3/2}n\right).
$$

\subsection{Proof of Equation~\eqref{cavi_latent_gauss}}
\label{sec::cavi_latent_gauss} 
The proof is mainly based on repeatedly applying Taylor expansions with respect to $\theta$.
In the following presentation, we use $C$ to denote a generic constant, whose value may vary from line to line.
We will also use the shorthands $\ell_{is}(\theta)=\log p(S_i=s\mid X_i,\theta)$ and $h_{is}(\theta)=\log p(X_i, S_i=s\mid \theta)$ to denote the conditional distribution of latent variable $S_i$ given $(X_i,\theta)$ and the joint distribution of $(X_i, S_i)$, respectively. 

\smallskip
\noindent{\bf Analysis of $q_{S_i}^*$.}
For a given $q_\theta$, we can apply Taylor expansion to the updating formula~\eqref{cavi_s} for the latent distribution $q_{S_i}^*$ to obtain
\begin{align*}
\log q_{S_i}^\ast(s)= C+\int\ell_{is}(\theta)q_\theta(\theta)=C+\ell_{is}(\theta^*)+\langle \nabla\ell_{is}(\theta^*), \mu_{q_\theta}-\theta^*\rangle +O(\E_{q_\theta}\|\theta-\theta^*\|^2),
\end{align*}
where we recall that $\mu_{q_\theta}=\E_{q_\theta}[\theta]$.
To simplify the presentation, we use the shorthand $f_{is}=\langle\nabla \ell_{i s}(\theta^*),\mu_{q_\theta}-\theta^*\rangle$.
Since $1=\sum_{s'}p(S_i=s' \mid X_i, \theta^*)=\sum_{s'} \exp\{\ell_{is'}(\theta^\ast)\}$, we can further get
\begin{align*}
q^\ast_{S_i}(s)=&\, \frac{p(S_i=s \mid X_i, \theta^*)(1+f_{is})}{\sum_{s'}p(S_i=s' \mid X_i, \theta^*)(1+f_{is'})}+O(\E_{q_\theta}\|\theta-\theta^*\|^2)\\
=&\,p(S_i=s \mid X_i, \theta^*)(1+f_{is})\Big[1-\sum_{s'}p(S_i=s' \mid X_i, \theta^*)f_{is'}\Big]+O(\E_{q_\theta}\|\theta-\theta^*\|^2)\\
=&\,p(S_i=s \mid X_i, \theta^*)\Big[ 1+f_{is}-\sum_{s'}p(S_i=s' \mid X_i, \theta^*)f_{is'}\Big]+O(\E_{q_\theta}\|\theta-\theta^*\|^2).
\end{align*}

\smallskip
\noindent{\bf Analysis of $q_{\theta_j}^*$.}
For each $j\in[d]$, the updating formula~\eqref{cavi_theta} for $q^\ast_{\theta_j}$ can be written as
\begin{align*}
    \log q^\ast_{\theta_j}(\theta_j) = C + \E_{q_{S^n}} \E_{q_{\theta_{-j}}}[\log p(\theta, X^n, S^n)].
\end{align*}
By applying a Taylor expansion to $\theta$, we can write
\begin{align*}
&\, \log p(\theta, X^n, S^n)
=C+\log \pi(\theta)+\sum_i h_{i s}(\theta)=C+\log \pi(\theta)\\
&\qquad +\sum_i \Big(h_{i s}(\theta^*)+\langle\nabla h_{is}(\theta^*), \theta-\theta^*\rangle +\frac{1}{2}(\theta-\theta^*)^T \nabla^2 h_{i s}(\theta^*)(\theta-\theta^*)\Big)+O(n\|\theta-\theta^*\|^3).
\end{align*}
Throughout the proof, we will use $\nabla_j$ to denote the partial derivative relative to $\theta_j$. By reordering the terms and absorbing $\theta_j$ independent terms to the leading constant, we obtain
\begin{equation}
\begin{aligned}
&\E_{q_{S^n}} \E_{q_{\theta_{-j}}}[\log p(\theta, X^n, S^n)]
=\\
& \qquad\qquad C+\E_{q_{\theta_{-j}}}[\log \pi(\theta))+O(n\|\theta-\theta^*\|^3)+\sum_i \sum_s\Big[\nabla_j h_{i s}(\theta^*)(\theta_j-\theta_j^*) \\
& \qquad \qquad +\frac{1}{2} \nabla_{j j} h_{i s}(\theta^*)(\theta_j-\theta_j^*)^2+\sum_{k \neq j} \nabla_{j k} h_{i s}(\theta^*)([\mu_{\theta_q}]_k-\theta_k^*)(\theta_j-\theta_j^*)\Big]q^\ast_{S_i}(s).
\end{aligned}
\label{eq::latent_exp_logp}
\end{equation}
Note that by the central limit theorem and the definition of various Fisher information matrices, we have
\begin{align*}
& \sum_i \sum_s \nabla^2 h_{i s}(\theta^*) p(S_i=s \mid X_i, \theta^*)=-n V_c+O_p(\sqrt{n}),\\
& \sum_i \sum_s \nabla^2 \ell_{i s}(\theta^*) p(S_i=s \mid X_i, \theta^*)=-n V_s+O_p(\sqrt{n}),
\end{align*}
and also recall $V_c=V+V_s$. 
Now we can plug-in the expansion of $q^*_{S_i}(s)$ into~\eqref{eq::latent_exp_logp} and utilize the preceding display to obtain
\begin{align*}
&\E_{q_{S^n}} \E_{q_{\theta_{-j}}}[\log p(\theta, X^n, S^n)] = C+E_{q_{\theta_{-j}}}[\log \pi(\theta)]+O(n\E_{q_\theta}\|\theta-\theta^* \|^3)\\
&\qquad\qquad +(\theta_j-\theta_j^*)\sum_i \sum_s p(S_i=s \mid X_i, \theta^*)\Big[\nabla_j h_{i s}(\theta^*)\Big(1+\langle\nabla \ell_{i s}(\theta^*), \mu_{q_\theta}-\theta^*\rangle \\
&\qquad\qquad -\sum_{s'} p(S_i=s' \mid X_i, \theta^*)\langle\nabla \ell_{i s'}(\theta^*), \mu_{q_\theta}-\theta^*\rangle\Big) +\sum_{k \neq j} \nabla_{j k} h_{i s}(\theta^*)([\mu_{q_\theta}]_k-\theta_k^*)\Big]\\
&\qquad\qquad\qquad\qquad +\frac{1}{2}  (\theta_j-\theta_j^*)^2\sum_i \sum_s \nabla_{j j} h_{i s}(\theta^*)p(S_i=s \mid X_i, \theta^*) \\
&\, = C+O(\E_{q_\theta}\|\theta-\theta^* \|)+O(n\E_{q_\theta}\|\theta-\theta^* \|^3)\\
& \qquad\qquad+(\theta_j-\theta_j^*)\left(\sum_i \sum_s \nabla_j h_{is}(\theta^*) p(S_i=s \mid X_i, \theta^*)\right) +(\theta_j-\theta_j^*)\cdot (\mu_{q_\theta}-\theta^*)^T \\
& \ \cdot \left\{\sum_i \sum_s \nabla_j h_{i s}(\theta^*) p(S_i=S\mid X_i, \theta^*)\left[\nabla \ell_{i s}(\theta^*)-\sum_{s'}p(S_i=s' \mid X_i, \theta^*) \nabla \ell_{i s'} (\theta^*) \right]\right\} \\
& \qquad\qquad -n(\theta_j-\theta_j^*)\sum_{k\not= j}[V_c]_{j k}([\mu_{q_\theta}]_k-\theta_k^*)-\frac{1}{2} n[V_c]_{jj}(\theta_j-\theta_j^*)^2+O(\sqrt{n}\E_{q_\theta}\|\theta-\theta^*\|^2).
\end{align*}
To further simplify some terms, we note that by using integration by parts,
\begin{align*}    
&\E\left[\sum_s \nabla h_{is}(\theta^*) \nabla \ell_{i s}(\theta^*)^T p(S_i=s \mid X_i, \theta^*) \right]\\
=&\, \int\sum_s \frac{\nabla p(x, s\mid \theta^*)}{p(x, s \mid \theta^*)} \frac{\nabla p(s \mid x, \theta^*)^T}{p(s \mid x, \theta^*)}p(s \mid x, \theta^*)p(x\mid\theta^*)\di x \\
=&\,\int \sum_s \nabla p(x, s \mid \theta^*) \nabla \log p(s \mid x, \theta^*)^T\di x 
=-\int \sum_s p(x, s \mid \theta^*) \nabla^2 \log p(s \mid x, \theta^*)\di x 
=V_s,
\end{align*}
and
\begin{align*}
&\E\left[\sum_s \sum_{s'} \nabla h_{i s}(\theta^*) p(S_i=s \mid X_i, \theta^*)\nabla \ell_{i s'}(\theta^*)^T p(S_i=s'\mid X_i, \theta^*)\right]\\ 
=&\,\int\sum_s \sum_{s'}\frac{\nabla p(x, s\mid \theta^*)}{p(x, s \mid \theta^*)} \frac{\nabla p(s' \mid x, \theta^*)^T}{p(s' \mid x, \theta^*)}p(s \mid x, \theta^*)p(s' \mid x, \theta^*)p(x\mid\theta^*)\di x \\
=&\,-\int \sum_s \sum_{s'} p(x, s \mid \theta^*) \nabla^2 p(s' \mid x, \theta^*) 
=-\int \sum_{s'}p(x \mid \theta^*) \nabla^2 p(s' \mid x, \theta^*) \di x \\
=&\,-\int p(x \mid \theta^*) \nabla^2 \sum_{s'} p(s' \mid x, \theta^*) \di x
=0.
\end{align*}
Therefore, we can once again applying the central limit theorem to replace some of the summations over $i$ with their corresponding expectations to obtain
\begin{align*}
&\E_{q_{\theta_{-j}}} \E_{S^n}[\log p(\theta, X^n, S^n)]\\
=&\, C+O(\E_{q_\theta}\|\theta-\theta^* \|)+O(\sqrt{n}\E_{q_\theta}\|\theta-\theta^*\|^2)+O(n\E_{q_\theta}\|\theta-\theta^* \|^3) +(\theta_j- \theta_j^*) U_j \\
&\, +n(\theta_j-\theta_j^*)[V_s(\mu_{q_\theta}-\theta^*)]_j-n(\theta_j-\theta_j^*) \sum_{k \neq j}[V_c]_{k j}([\mu_{q_\theta}]_k-\theta_k^*)-\frac{1}{2} n[V_c]_{jj}(\theta_j-\theta_j^*)^2,
\end{align*}
where $U_j:\,=\sum_i \sum_s \nabla_j h_{i s}(\theta^*) \,p(S_i=s \mid X_i, \theta^*)$.
Let random vector $U=\left(U_1, \cdots, U_d\right)^T$. Since
$$E[U]=\int \sum_s \frac{\nabla p(x, s \mid \theta^*)}{p(x, s \mid \theta^*)} p(s \mid x, \theta^*)p(x \mid \theta^*)\di x=0,
$$
and
\begin{align*}
& E[UU^T]=\int \sum_s \sum_{s'} \frac{\nabla p(x, s \mid \theta^*)}{p(x, s \mid \theta^*)}p(s \mid x, \theta^*) \frac{\nabla p(x, s' \mid \theta^*)^T}{p(x, s' \mid \theta^*)} p(s'\mid x,\theta^*)p(x\mid \theta^*) \di x \\
=&\,\int \sum_s\sum_{s'}p(x, s \mid \theta^*)\frac{\nabla p(x, s' \mid \theta^*)^T}{p(x \mid \theta^*)}\di x 
=\int \frac{\nabla p(x \mid \theta^*) \nabla p(x \mid \theta^*)^T}{p(x \mid \theta^*)}\di x =V,
\end{align*}
we can apply the central limit theorem to get $n^{-1/2} U \cod N(0, V) $ as $n\to\infty$, and it is also standard to verify that the MLE $\tmle$ satisfies
$\tmle=\theta^*+ n^{-1} V^{-1} U+O_p(n^{-1})$, implying $\sqrt{n} (\tmle - \theta^\ast) \cod N(0, V^{-1})$.
Putting all pieces together, we finally reach
\begin{align*}
\log q_{\theta_j}(\theta_j)=&\, C+O(n^{-1/2})+O(\E_{q_\theta}\|\theta-\theta^* \|)+O(\sqrt{n}\E_{q_\theta}\|\theta-\theta^*\|^2)+O(n\E_{q_\theta}\|\theta-\theta^* \|^3)  \\
&-\frac{1}{2} n[V_c]_{j j}(\theta_j-\theta_j^*)^2+n(\theta_j-\theta_j^*) \sum_k V_{k j}(\tmle_k-\theta_k^*) \\
& +n(\theta_j-\theta_j^*) \sum_k[V_s]_{k j}([\mu_{q_\theta}]_k-\theta_k^*) -n(\theta_j-\theta_j^*) \sum_{k \neq j}[V_c]_{kj}([\mu_{q_\theta}]_k-\theta_k^*), 
\end{align*}
which is equation~\eqref{cavi_latent_gauss}.
\end{document}